\documentclass[3p]{elsarticle}
\pdfoutput=1
\usepackage{amsmath} % AMS Math Package
\usepackage{amsthm} % Theorem Formatting
\usepackage{amssymb}
\usepackage{mathrsfs}
\usepackage{xcolor}

\usepackage{cellspace}

\newcommand{\ket}[1]{\left| #1 \right>} % for Dirac bras
\newcommand{\braket}[2]{\left< #1 \vphantom{#2} \right|
	\left. #2 \vphantom{#1} \right>}
\newcommand{\ketbra}[2]{\left| #1 \vphantom{#2}\right> 
	\left< #2 \vphantom{#1} \right|} % for Dirac
\newcommand{\vv}[1]{\ensuremath{\mathbf{#1}}} % for vectors
\newcommand{\gv}[1]{\ensuremath{\boldsymbol{#1}}}

\newcommand{\abs}[1]{\left| #1 \right|} % for absolute value
\newcommand{\avg}[1]{\left< #1 \right>} % for average

\newcommand{\pd}[2]{\frac{\partial #1}{\partial #2}} 
\newcommand{\pdd}[2]{\frac{\partial^2 #1}{\partial #2^2}}

\newtheorem{thm}{Theorem}
\newtheorem{prop}[thm]{Proposition}
\newtheorem{cor}[thm]{Corollary}
\newtheorem*{lem}{Lemma}

\newtheorem*{prp}{Proposition}

\newdefinition{rmk}{Remark}
\theoremstyle{remark}
\newtheorem*{dfn}{\underline{\textsc{def}}}

\begin{document}

\begin{frontmatter}
	
	\title{Quantum Hall Ground States and Regular Graphs}
	
	\author{Hamed Pakatchi}
	\ead{pakatchi@alum.mit.edu}
	\affiliation{organization={Massachusetts Institute of Technology}, addressline={77 Massachusetts Avenue}, 	city={Cambridge}, postcode={02139}, state={MA}, country={USA}}

\begin{abstract}
We show that every uniform state on the sphere is essentially a superposition of regular graphs. In addition, we develop a graph-based ansatz to construct trial FHQ ground states sharing the local properties of Jack polynomials. In particular, our graphic states have the $(k,r)$ clustering property. Moreover, a subclass of the construction is realizable as the densest zero-energy state of a model that modifies the projection Hamiltonian. 
\end{abstract}

\end{frontmatter}

%\tableofcontents

\section{Introduction}
The fractional quantum Hall effects (FQHE) are exotic phases of matter that emerge from two-dimensional electronic systems exposed to a large uniform transverse magnetic field. In these systems, the filling fraction $\nu$, and thus the Hall conductance $\sigma_H = (e^2/h)\nu$, are fractional. The theory behind the FQHE arguably starts with Laughlin's famous variational states \cite{Laughlin_Original}, which successfully captures the physics of $\nu=1/m$ filling fractions. Jain's composite fermion approach \cite{jain, jain2007composite}, in which, through an adiabatic process, some flux tubes are attached to the electrons, subsequently led to the Jain states describing the family $\nu= \frac{p}{2mp+1}$ ($m,p$ integers). The scope of our understanding of FQH states vastly widened when Moore \& Read related them to conformal blocks in certain conformal field theories \cite{Moore-Read}. Based on the Ising CFT, they constructed the celebrated Pfaffian state ($\nu=1/2$) and initiated the study of non-Abelian FQH states. In turn, the Moore-Read state was itself generalized to Read-Rezayi \cite{Read-Rezayi} (also see  \cite{blok1992many}). Nowadays, the ideas behind and around Read-Rezayi states lay the foundation of the theory of non-Abelian FQHE. There are several generalizations of Read-Rezayi states as well. The most prominent ones are the negative rational Jack polynomials \cite{Feigin, Bernevig_Haldane_Model, Bernevig_Haldane_Cluster}, which are currently the subject of intense study. Evidently, the construction and analysis of trial FQH states have been central to studying FQHE. Inspired by the local properties of Jack polynomials, the current paper provides a novel graph-based method for designing and characterizing trial FQH ground states.

To begin the analysis of FQH states on a given surface $M\subset \mathbb{R}^3$, the first step is to find a basis for the Landau levels. These are the energy levels for a charged particle moving on the surface $M$ in the presence of a uniform transverse magnetic field. Every Landau level is highly degenerate, with an energy gap proportional to the magnetic field $B$. Since the magnetic fields required for FQHE are typically large, focusing on those states confined in the lowest Landau level (LLL) is reasonable. The LLL restriction leads to FQH states having a remarkable holomorphic structure. For example, in the plane geometry, with $z=x+iy$ the complex coordinates of an electron, the $N$-body wavefunctions have the following form:
\[
\label{intro_wf_eq}
\Phi(z_1, \bar{z}_1,\cdots,z_N, \bar{z}_N)
=
\Psi(z_1,\cdots,z_N) \exp\left(-\frac{eB}{4\hbar}\sum_i z_i\bar{z}_i\right)
\]
Here, $\Psi$ is a symmetric (resp. antisymmetric) \emph{polynomial} for a bosonic (resp. fermionic) system. Our focus in this paper is the study of (trial) bosonic FQH \emph{ground} states. Such polynomials have additional symmetries (e.g., translational invariance) and certain local properties. We will explore some of these local properties in this section and throughout the paper. Henceforth, if there is no chance for confusion, the term `wavefunction' will refer to the polynomial part $\Psi$ alone.

In principle, one can design a model FQH Hamiltonian and solve for its ground state. That would be the most direct approach to characterizing a model FQH ground state. The simplest and most popular examples are the projection Hamiltonians $H_{k+1}^{r-1}$ \cite{Gaffnian, Simon_Pseudopotentials}, which are the $(k+1)$-body generalizations of Haldane's (two-body) pseudopotential formalism \cite{Haldane_Sphere_Pseudo}. This operator projects a wavefunction into the sector where no cluster of $k+1$ particles can have relative angular momentum $r$ or more. The densest (i.e., lowest total angular momentum) zero-energy eigenstates of $H_{k+1}^{r-1}$ is called a model FQH ground state. Laughlin states \cite{Laughlin_Original}, Moore-Read state \cite{Moore-Read}, Read-Rezayi states \cite{Read-Rezayi, blok1992many}, and Gaffnian \cite{Gaffnian} are all the \emph{unique} ground states of $H_{k+1}^{r-1}$ Hamiltonians for an appropriate $(k,r)$.  Unfortunately, for most pairs $(k,r)$, the null space of the Hamiltonian $H^{r-1}_{k+1}$ (restricted to lowest possible angular momentum) is not necessarily one-dimensional \cite{Simon_Pseudopotentials}. In other words, most $H^{r-1}_{k+1}$ operators do not possess a \emph{unique} ground state. A finer characterizing method and a modified Hamiltonian are required to resolve the degeneracy. We should also mention that, as the first step, constructing a ``nice'' wavefunction is often more practical. One then uses the local properties of that state to design a Hamiltonian that \emph{realizes} it as (hopefully, the unique) ground state.

The \emph{pattern of zeros} \cite{Pattern_of_Zeros} is another characterizing method for FQH states. This method captures the relative behavior of clusters of electrons and boils it down to a few integers. Given a bosonic (i.e. symmetric) ground state polynomial $\Psi(z_1, \cdots, z_N)$ with $N\gg 1$, the integer $S_a$ is defined as the minimal relative angular momentum of clusters of $a$ particles. The sequence $S = (S_1, S_2, \cdots, S_a, \cdots)$ is called the pattern of zeros of $\Psi$. In Ref. \cite{Pattern_informal}, a wavefunction is said to satisfy a ``$k$-cluster condition'' if
\[
S_{pk+q}=\frac{p(p-1)}{2}kr+pqr+pS_k+S_q
\label{kcluster}
\]
If this reasonable relation holds, then the pattern of zeros truncates to a finite set $(\frac{k}{r};  S_2, S_3, \cdots, S_k)$. This finite set is a characteristic of the FQH phase that $\Psi$ represents. However, the pattern of zeros is not a constructive method of classification. Given an \emph{allowed} pattern of zeroes, there is no systematic way to reverse-engineer any wavefunction that presumably led to the data. Moreover, this method is only a partial classification; there can be multiple distinct FQH systems with the same pattern of zeros.

Describing FQH ground states as correlators in certain conformal field theories(CFT) \cite{Moore-Read} is yet another approach for characterizing these many-body states. In this picture, FQH ground states are (neutral) correlators of $N$ electrons vertex operators $V_e(z) = \psi(z) :e^{i\phi(z)/\sqrt{\nu}}:$, where $:\: :$ is normal order, $\phi(z)$ is a free boson, $\nu$ is the filling fraction, and $\psi_1(z)$ is a parafermion \cite{Zamolodchikov_Fateev_Parafermion}. The underlying parafermionic CFT has $k$ primary fields $\psi_0\equiv 1, \psi_1, \psi_2, \cdots, \psi_{k-1}$. Denoting the scaling dimension of $\psi_a$ by $h_a$, the operator product expansions are given by (addition done in $\mathbb{Z}_k$):
\begin{align*}
	&\psi_a(z)\psi_b(w)(z-w)^{h_a+h_b-h_{a+b}}= C_{a,b}\psi_{a+b}(w)+\cdots\quad(a+b\neq 0)
	\\
	&\psi_a(z)\psi_{k-a}(w)(z-w)^{2h_a}= 1+\frac{2h_a}{c}(z-w)^2T(w)+\cdots
\end{align*}
Here, $T$ is the energy-momentum tensor, $c$ is the central charge, and $C_{a,b}$ are the structure constants. The corresponding trial FQH droplet (in the plane or sphere geometry) is thus:
\[
\label{CFT_WF_eq}
\Psi(z_1, \cdots, z_{nk})=\avg{\psi_1(z_1)\cdots \psi_1(z_{nk})}\prod_{i<j}(z_i-z_j)^{2h_1-h_2}
\]
In particular, $\nu^{-1}=2h_1-h_2\equiv r/k\in \mathbb{Q}$ is the inverse of the filling fraction. The characterizing data of this wavefunction is a set $\{c, h_a, C_{a,b}, \cdots\}$ that results in associative OPEs. For example, when $h_a=a(k-a)/k$, associativity fixes all of the structure constants, and $c=2(k-2)/(k+2)$ \cite{Zamolodchikov_Fateev_Parafermion}. In this case, the corresponding FQH ground states are the $\nu=k/2$ Read-Rezayi states \cite{Read-Rezayi}.

As mentioned before, in the pursuit of classifying the FQH phases of matter, the explicit construction of trial FQH wavefunctions and analyzing their properties has been an invaluable source of insight. To that end, we believe developing further constructive machinery to design FQH trial ground states is necessary. Arguably, as powerful as the CFT description of FQH states is, it is not entirely computation friendly. To expand on that, while the correlation functions are constructive, the notorious difficulty of computing conformal blocks hides some of the internal structure of these wavefunctions. As for the pseudopotential formalism, it is still an open question to determine model Hamiltonians with unique ground states. Unfortunately, even if we had access to the perfect model Hamiltonian, \emph{explicitly} solving for null states is a non-starter. 
Moving on to composite fermions, while of utmost theoretical value, the Jain states \cite{jain} are also tricky to manipulate explicitly. To our knowledge, the only existing \emph{constructive} program to produce trial FQH states is the Jack polynomial approach. We briefly describe this construction next.

Jack polynomials $J_\lambda^{(\alpha)}$ are a family of multivariate homogeneous symmetric polynomials that depend on a formal parameter $\alpha$ and are labeled by a partition $\lambda$. They are the eigenfunctions of a Calegro-Sutherland Hamiltonian \cite{sutherland1971quantum, sutherland1971exact1, sutherland1972exact2}:
\begin{align*}
	H_{CS}J^{(\alpha)}_\lambda&:=\Big[\sum_i \Big(z_i \pd{}{z_i}\Big)^2+\frac{1}{\alpha}
	\sum_{i<j}\frac{z_i+z_j}{z_i-z_j}
	\Big(z_i \pd{}{z_i}-z_j \pd{}{z_j}\Big)
	\Big]J^{(\alpha)}_\lambda = \varepsilon_\lambda J^{(\alpha)}_\lambda\\
	\varepsilon_\lambda &= \sum_i \lambda_i (\lambda_i + (n+1-2i)/\alpha)
\end{align*}
Endowing the space of symmetric functions with a particular inner product, the Jacks constitute a triangular orthogonal basis (see \cite[Theorem 1.1.]{stanleyjack}). To elaborate on triangularity, the Jacks have the following decomposition in the symmetric monomial $m_{\lambda}$ basis: 
\[
J_{\lambda}^{(\alpha)} = m_{\lambda} + \sum_{\mu \prec \lambda} v_{\lambda\mu}(\alpha)m_{\mu}
\]
Here,  $\mu \prec \lambda$ signals that $\mu_1+\cdots+\mu_a\leq \lambda_1+\cdots+\lambda_a$ for all $a$ (i.e., $\lambda$ dominates $\mu$), and $v_{\lambda\mu}(\alpha)$ are certain coefficients. Building on the findings of Ref. \cite{Feigin}, Bernevig and Haldane realized that the following Jacks, with specialized parameter $\alpha_{k,r}=-\frac{k+1}{r-1}$ (with $k+1, r-1$ relatively prime), are suitable as trial FQH ground states:
\begin{align*}
	\mathcal{J}_n^{(k,r)}&:= J_{\Lambda(n,k,r)}^{(\alpha_{k,r})}\\
	\Lambda(n,k,r)&=([(n-1)r]^{\times k}, \cdots, [2r]^{\times k}, r^{\times k}, 0^{\times k})
\end{align*}
Here, the number of particles and flux quanta are respectively $N=nk$ and $N_\phi=(n-1)r$. The Laughlin [$(k,r)=(1,2m)$], Read-Rezayi [$(k,r)=(k,2)$], and Gaffnian [$(k,r)=(2,3)$] states all belong to the large family of Jack trial wavefunctions. In addition to triangularity, another attractive feature of these Jacks is the so-called $(k,r)$ clustering property:
\[
\mathcal{J}_{n+1}^{(k,r)}(k,r)(\overbrace{w, \cdots, w}^{\times k}, z_1, z_2, \cdots, z_{nk})=
\prod_{i=1}^{nk} (z_i-w)^{r}\mathcal{J}_n^{(k,r)}(z_1, z_2, \cdots, z_{nk})
\]
This property was first conjectured in Refs. \cite{Bernevig_Haldane_Model, Bernevig_Haldane_Cluster} and later on proved using CFT techniques \cite{estienne1, estienne2}. There is also a secondary proof \cite{zamaere}, which uses the representation theory of Cherednik algebras. The local properties of the Jacks inspire much of the graph-based construction of trial states that we introduce in this paper. That being said, we do not use Jack polynomials to develop any of the formalism.

Regardless of the approach, model FQH ground states are presumably classified (at least partially) by a finite set of intrinsic data. For example, this role is fulfilled by $(\frac{k}{r}; S_2, S_3, \cdots, S_k)$ in pattern of zeros approach, while it is the set $\{c, h_a, C_{a,b}\}$ that encodes the model FQH ground state in CFT picture. The current paper will introduce a new point of view on the characterization and construction of model FQH ground states. Verbally, our findings can be summarized as follows:
\emph{One may think of bosonic FQH ground states as (a finite collection of) regular graphs. The intrinsic data of such a state is also a (small size-independent) graph.} What is more, our construction is entirely constructive. One can use it in parallel with the existing approaches, e.g., composite fermions, CFT, and Jack polynomials. However, as this approach is still very young, at this stage, we do not know what new physical results can come from it.

We emphasize that the current paper will concentrate on \emph{bosonic} systems. The so-called `bosonic electrons' play the role of electrons in the bosonic systems. In the language of composite particles \cite{jain}, a `bosonic electron' is an electron with a single flux quantum attached to it. In that sense, the filling fraction of the actual electronic system $\nu_f$ is given by $\nu_f = \frac{\nu}{1+\nu}<1$, with $\nu$ being the bosonic filling fraction. We assume all electrons, and thus `bosonic electrons,' are spinless and confined to their respective lowest Landau level. For bosonic FQH systems, the polynomials $\Psi(z_1, \cdots, z_N)$  are always symmetric. It is understood that the corresponding fermionic polynomial $\Psi_f$ can be obtained by multiplying with a Laughlin-Jastrow factor: $\Psi_f(z_1, \cdots, z_N) = \Psi (z_1, \cdots, z_N)\prod_{i<j}(z_i-z_j)$.

%%%%%%%%%%%%%%%%%%%%%%%%%%%%%%%%%%%

\section{Outline and Summary}
This paper introduces a graph-based method to construct trial FQH ground states systematically. Our construction has two stages. We devise a graphic method in the first stage to produce states enjoying the $(k,r)$ clustering property. We then investigate which wavefunctions are realizable as the densest zero-energy state of a model Hamiltonian. In the second stage, we filter out the trial wavefunctions that are unlikely to be realizable. The filtration is implemented by enforcing additional properties on the underlying graphs. The entirety of the construction is heavily influenced by the local properties of negative rational Jack polynomials. We aim to produce realizable graphic states that share the same local properties as Jacks. We now outline and summarize the content of the paper.

We are primarily interested in FQH ground states on the \emph{sphere}. Let $\nu, \mathcal{S}$ be the filling fraction and shift of the FQH phase, respectively. Suppose there are $N$ electrons and $N_\phi=N\nu^{-1}-\mathcal{S}$ flux quanta pass through the sphere. Aside from a geometric factor $\prod_{i=1}^N (1+z_i\bar{z}_i)^{-N_\phi/2}$, the ground state is a symmetric polynomial $\Psi$ where the degree of $z_1$ does not exceed $N_\phi$. Moreover, $\Psi$ has total angular momentum zero in the induced $SO(3)$ structure (see \ref{HaldaneSphere_app} for a review). The latter condition translates to the following:
\begin{align*}
	L_-\Psi &:= \sum_i \partial_i\Psi=0\\
	L_z\Psi &:= \Big(-\frac{NN_\phi}{2}+\sum_i z_i\partial_i\Big)\Psi=0\\
	L_+\Psi &:= 
	\Big(\sum_i N_\phi z_i-z_i^2\partial_i\Big)\Psi=0
\end{align*}
Such $\Psi$ is called an $(N, N_\phi)$ uniform state. In particular, $L_-\Psi=0$ and $L_z\Psi=0$ mean that $\Psi$ is translational invariant (i.e., a function of differences $z_i-z_j$) and homogeneous of degree $NN_\phi/2$ respectively.

The link between regular graphs and quantum Hall states is the uniform states. We explore uniform states and their graphic interpretation in section \ref{uniform_sec} (also in \ref{uniform_appendix} and \ref{cayley_appendix}). We may formulate the key observation in terms of a question: A typical wavefunction with translational and rotational symmetry is of the form:
\[
\Psi(z_1, \cdots, z_N) = \mathscr{S}\Big[\prod_{i, j}(z_i-z_j)^{\mu(i,j)}\Big]
\]
Here, $\mathscr{S}$ is symmetrization, and $\mu$ is some matrix. What should the conditions on $\mu$ be so that $\Psi$ is an $(N, N_\phi)$ uniform state? As it turns out, it is enough for $\mu$ to be the adjacency matrix of a regular graph in $N$ nodes and degree $N_\phi$. To elaborate, a graph $G$ is a set of $N$ nodes, say $\{1,2,\cdots, N\}$, and a set of (possibly multiple) \emph{arrows} between nodes. The \emph{adjacency matrix} $\mu_G(i,j)$ records the number of arrows \emph{from} $i$ \emph{to} $j$. A graph $G$ is regular of degree $N_\phi$ if each node has $N_\phi$ arrows incident to it (i.e., $\sum_{j}\mu_G(i,j)+\mu_G(j,i)=N_\phi$ for all $i$). Given such a graph, the following is called its \emph{symmetrized graph polynomial (SGP)}:
\[
\mathrm{SGP}[G]=\mathscr{S}\Big[\prod_{i, j}(z_i-z_j)^{\mu_G(i,j)}\Big]
\]
The SGP of a regular graph is either zero or a uniform state. Conversely, suppose $\Psi$ be any $(N, N_\phi)$ uniform state. Then one can find a (finite) number of regular graphs $G_p$ ($N$ nodes, degree $N_\phi$) such that $\Psi$ is a superposition of $\mathrm{SGP}[G_p]$'s.

\begin{figure}
	\centering
	\includegraphics[scale=.65]{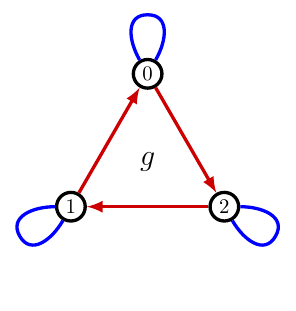}
	\hspace{10pt}
	\includegraphics[scale=.65]{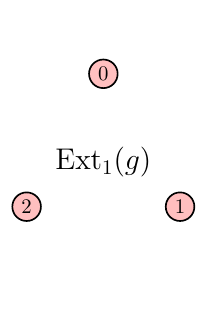}
	\hspace{10pt}
	\includegraphics[scale=.65]{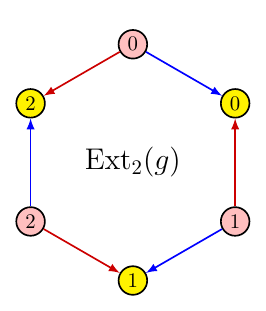}
	
	\includegraphics[scale=.65]{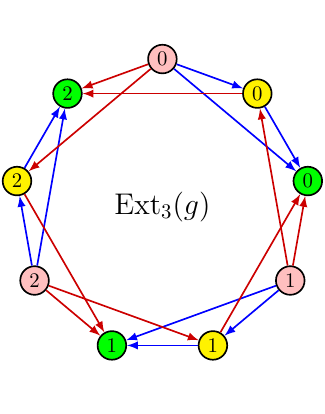}
	\hspace{10pt}
	\includegraphics[scale=.65]{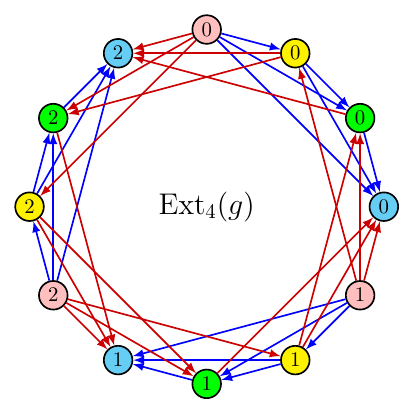}
	\caption{The $(k,r)=(3,2)$ intensive graph $g$ for $\mathbb{Z}_3$ Read-Rezayi state with $\nu=3/2$, and its first four extensions. Note that, with $x,y=0,1,2$, there is an arrow of the form $x\to y$ in $\mathrm{Ext}_n(g)$ if and only if $g$ has that arrow as well. For clarity, we have colored the loops of $g$ (and the corresponding arrows in the extensions) in blue. We paint all other arrows of $g$ (and the corresponding arrows in extensions) in red. The graph $\mathrm{Ext}_n(g)$ has nodes of $n$ different colors. The vertex colors have nothing to do with $g$ and are how the size of the system $n$ is reflected in the extension graphs $\mathrm{Ext}_n(g)$.}
	\label{intro_fig}
\end{figure}

The SGPs allow us to regard FQH ground states as (a superposition of) regular graphs. We want to design infinite sequences of regular graphs $\vv{G}=(G_n)_{n\geq 1}$, $G_n$ having $nk$ nodes and degree $(n-1)r$, such that $\Psi_n=\mathscr{N}_n\mathrm{SGP}[G_n]$ ($\mathscr{N}_n$ is an appropriate normalization) satisfies the so-called $(k,r)$ clustering property
\[
\Psi_{n+1}(\underbrace{w, \cdots, w}_{\times k}, z_1, \cdots, z_{nk})=\prod_{i<j}(z_i-w)^r \Psi_n(z_1, \cdots, z_{nk})
\]
To do so, in section \ref{construction_sec}, we devise an ansatz which we now describe. We need two graphic ingredients: The graph $T_n$ (related to Laughlin states) and the tensor product.
\begin{enumerate}
	\item $T^r_n$ is a graph in nodes $1,2,\cdots, n$. If $i<j$, then there is an arrow  $i\to j$ of multiplicity $r$, but no arrow $j\to i$. The SGP of $T_n^{2m}$ is the Laughlin $2m$-state. We use the notation $T_n\equiv T_n^1$.
	\item Let $G$ (resp. $H$) be a graph in $N_G$ (resp. $N_H$) nodes $x_i$ (resp. $y_j$) and adjacency matrix $\mu_G$ (resp. $\mu_H$). Then $G\otimes H$ has $N_GN_H$ nodes $x_i\otimes y_i$, and its adjacency matrix is the Kronecker product $\mu_G\otimes \mu_H$.
\end{enumerate}
With this terminology, in our ansatz, the graphs $G_n$ are of the form $G_n=T_n\otimes g$. The graph $g$ is size-independent and is part of the classifying data of the respective FQH phase of matter. For $\Psi_n\propto \mathrm{SGP}[T_n\otimes g]$ to be $(k,r)$ clustering, the graph $g$ needs to satisfy the following conditions:
\begin{enumerate}
	\item Every node in $g$ has at
	least one loop. In other words, the diagonal entries of $\mu_g$ are non-zero.
	\item Each connected component of $g$ has an even
	number of arrows.
	\item $g$ has $k$ nodes. Each node has $r$ outgoing arrows and $r$ incoming ones. In other words, $\sum_{j}\mu_g(i,j)=\sum_{j}\mu_g(j,i)=r$.
\end{enumerate}
We call $g$ satisfying these properties a \emph{$(k,r)$ intensive} graph. The graph $G_n:=\mathrm{Ext}_n(g):=T_n\otimes g$ is called the $n$th extension of $g$.  The filling fraction and shift are $\nu=k/r$ and $\mathcal{S}=r$. In Fig. \ref{intro_fig}, we have drawn the intensive graph and the first few extensions for $\mathbb{Z}_3$ Read-Rezayi state to demonstrate the graphic procedure. The Laughlin $2m$-states, Read-Rezayi, Gaffnian, Haffnian, and possibly the Jacks $\mathcal{J}^{(2k,3)}$ all fit into our construction. In Table \ref{table:intensive}, we have gathered their intensive graphs.

The discussion regarding realizability begins in section \ref{discussion_sec}. We first examine if the projection Hamiltonians $H_{k+1}^{r-1}$ are viable models. It is instructive to describe $H_{k+1}^{r-1}$ formally. If $\mathcal{P}_{k+1}^m$ denote the projection to the sector where particles at $z_1, \cdots, z_{k+1}$ have a relative angular momentum $m$, then $H_{k+1}^{r-1}$ is given by
\[
H_{k+1}^{r-1}=\sum_{i_1<\cdots<i_{k+1}}
\sum_{m=0}^{r-1}V_m\mathcal{P}_{k+1}^m(i_1, \cdots, i_{k+1}), \qquad (V_m>0)
\]
These Hamiltonians have shown great success for Laughlin \cite{Haldane_Sphere_Pseudo}, Read-Rezayi \cite{Read-Rezayi}, and Gaffnian states \cite{Gaffnian}. However, it is known that they do not have a \emph{unique} ground state if $k$ or $r$ are large enough \cite{Simon_Pseudopotentials}. Therefore, an appropriate modification is needed to resolve the non-uniqueness issue. We have made limited progress toward such a resolution elsewhere \cite{pakatchi2021CFT}. In that paper, using the framework of $\mathbb{Z}_k^{(r)}$-algebras, we have found a natural candidate for a modified Hamiltonian. The new Hamiltonian relies on a local property that we dubbed \emph{separability}. To describe it, let us introduce a bit of terminology. Let $\mathcal{T}_{k+1}^r$ be the Bargmann space \cite{girvin1984formalism} of symmetric homogeneous translational invariant polynomials with $k+1$ variables and degree $r$. Suppose $\ket{\chi; z_1, \cdots, z_{k+1}}\in \mathcal{T}_{k+1}^r$ is such that
\[
|\chi; 1, \underbrace{0, \cdots, 0}_{\times k}\rangle =1
\]
Also, let $\gv{\Psi}=(\Psi_n)_{n\geq 1}$ be a sequence satisfying the $(k,r)$ clustering property. Then, we say $\gv{\Psi}$ is \emph{separable} with \emph{minimal polynomial} $\chi$ if (for all $n$)
\[
(\mathcal{P}_{k+1}^r\Psi_{n+1})(z_1, \cdots, z_{k+1}, z_{k+2}, \cdots) =
\ket{\chi;z_1, \cdots, z_{k+1}}\prod_{i>k+1 }(z_i-z_{\mathrm{cm}})^r\Psi_n(z_{\mathrm{cm}},z_{k+2}, \cdots)
\]
Here, $z_{\mathrm{cm}}=(z_1+\cdots+z_{k+1})/(k+1)$ is the cluster center-of-mass. As an example, the Jacks $\mathcal{J}_n^{(k,r)}$ are all separable. This is because they are a special $\mathbb{Z}_{k}^{(r)}$ wavefunction \cite{estienne1, estienne2}, and all $\mathbb{Z}_{k}^{(r)}$ wavefunctions are separable \cite{pakatchi2021CFT}. Now, separability immediately suggests a modification to $H_{k+1}^{r-1}$, parametrized by $\chi$, as follows
(Notation: $z_I\equiv z_{i_1}, \cdots, z_{i_{k+1}}$):
\[
H_\chi= H_{k+1}^{r-1}+
V_r 
\sum_{i_1<\cdots< i_{k+1}}
\left(1-\frac{\ketbra{\chi;z_I}{\chi;z_I}}{\braket{\chi}{\chi}}\right)	\mathcal{P}_{k+1}^r(i_1, \cdots, i_{k+1})
\label{eq_Hamiltonian}
\]
The densest null states of $H_\chi$ are those null states of $H_{k+1}^{r-1}$ that are $(k,r)$ clustering and separable with minimal polynomial $\chi$. For the current paper, we need to find intensive graphs that lead to separable states to get a realizable graphic state. Finding such intensive graphs is the subject of section \ref{Separability_sec}, and we will present a simple example shortly. We should also mention that when $\dim \mathcal{T}_{k+1}^{r}=1$ the new Hamiltonian reduces to $H_{k+1}^{r-1}$. For example, in the case of Laughlin, Read-Rezayi, and Gaffnian states, we may use either $H_{k+1}^{r-1}$ or $H_\chi$ (see \ref{Zk2_appendix}). In Ref. \cite{pakatchi2021CFT}, we argue that when $r=4$, $k\geq 3$, the Hamiltonians $H_\chi$ are quite likely to have a unique ground state. The possibly unique ground state is the droplet wavefunction of the unique $\mathbb{Z}_k^{(2)}$-algebra that is fixed by $\chi$. An abridged version of our arguments can be found in \ref{Zk2_appendix}.

The final local property we will consider is periodicity. In a sense, the job of periodicity is to filter out the pathological clustering states. A $(k,r)$ clustering state $\Psi$ is \emph{periodic} if the minimal relative angular momentum of a cluster of $a=pk+q$ particles ($0\leq q<k$) is given by
\[S_{pk+q}=\left\{\frac{p(p-1)}{2}k+pq\right\}r
\]
The quantity $S_a$ is also the minimal total power the variables $z_1, \cdots, z_a$ can have in the wavefunction. Alternatively, we may describe periodicity in terms of the free boson expansion of $\Psi$. Let $m_\lambda$ be the symmetric monomial and $\lambda\prec \Lambda$ symbolize $\lambda_1+\cdots+\lambda_i\leq \Lambda_1+\cdots+\Lambda_i$ for all $i$ (i.e., $\Lambda$ dominates $\lambda$). Then a $(k,r)$ clustering state is periodic if and only if
\[
\Psi = m_{\Lambda} + \sum_{\mu\prec \Lambda} c_\mu m_\mu\\
\]
Here, $\Lambda$ is the partition corresponding to the periodic orbital occupation given by:
\[
n = (k, \underbrace{0, \cdots, 0}_{\times (r-1)}, k, \underbrace{0, \cdots, 0}_{\times (r-1)}, \cdots, k, \underbrace{0, \cdots, 0}_{\times (r-1)}, k)
\]
Almost by definition, the Jacks $\mathcal{J}_n^{(k,r)}$ are periodic. We first discuss periodicity and pattern of zeroes in section \ref{discussion_sec}. Then, in section \ref{periodicity_sec}, we continue the discussion and derive a condition on the intensive graphs to ensure periodicity.

Let us present an example of intensive graphs leading to separability and periodicity. A \emph{$(k,l,m)$ saturated} graph $g$ has $k$ nodes, $l$ loops for each node, and an arrow $x\to y$ of multiplicity $m>0$ for any distinct pair of nodes $x\neq y$. Moreover, we demand that $kl=$even and $m\leq l$. With this definition, the wavefunctions $\Psi^{(g)}_n=\mathscr{N}_n\mathrm{SGP}[T_n\otimes g]$ are periodic and separable. Moreover, if $h=g^c$ is the disjoint union of $c$ copies of $g$, then $\Psi^{(h)}_n=\mathscr{N}_n\mathrm{SGP}[T_n\otimes g]$ is also separable and periodic. In particular, $\Psi_n^{(h)}$ are realizable by $H_\chi$, where $\chi$ is explicitly given by:
\[
\ket{\chi; z_0, z_2, \cdots, z_{kc+1}} =\frac{1}{(1+\delta_{1,k})(kc)!} \mathscr{S}\Big[(z_0-z_1)^l \prod_{i=2}^{k+1}(z_0-z_i)^m\Big]
\]
Let us present a few special cases:
\begin{enumerate}
	\item The $(k=2, l=2r, m=-)$ saturated graph $\ell_r$ is the intensive graph of the Laughlin $2r$-state (i.e. a single loop of multiplicity $2r$).
	\item Using the representation of Ref. \cite{cappelli}, the $k$-fold disjoint union $\ell_1^{k}$ is an intensive graph for $\mathbb{Z}_k$ Read-Rezayi states.
	
	\item The $(k=2, l=2,m=1)$ saturated graph, denoted $\jmath$, is the intensive graph for Gaffnian.
	\item The $(k=2, l=2,m=2)$ saturated graph is the intensive graph for Haffnian \cite{Haffnian}.
	\item One can show that $\mathcal{J}_2^{(2c,3)}\propto \mathrm{SGP}[T_2\otimes \jmath^c]$ (for all $c$). This loosely suggests that $\jmath^c$ is the intensive graph for these Jacks.
\end{enumerate}

We finish this section by pointing out some advantages and avenues of generalization of the graphic methodology. Aside from visualization benefits, our graph-based formalism's advantages are primarily computational at the moment. Various graphic concepts, some of which we discuss here, can be employed to make studying and manipulating the model wavefunctions easier. There is also a close connection between regular graphs and the invariants of binary forms. In Ref. \cite{pakatchi2021CFT}, we explore this connection and demonstrate the computational prowess that comes with it. 

As for generalization, it would be interesting if the graphic language could be extended to include quasi-holes. Naively speaking, due to the combinatorical nature of braiding statistics, graph theory seems to be an excellent framework for studying the exchange properties of quasi-holes. However, developing a graph-based formalism for quasi-holes is a work in progress and has many subtleties.

Yet another generalization path is to alter the graphic ansatz $\Psi_n\propto \mathrm{SGP}[T_n\otimes g]$. Let us give a concrete example of such an alteration. In Ref. \cite{simon2008correlators}, Simon computes the $2n$ point correlation of $\mathcal{N}=1$ superconformal currents $G(z)$:
\[
\Psi_n(z_1, \cdots, z_{2k})=\mathscr{N}_n\avg{G(z_1)G(z_2)\cdots G(z_{2n})}\prod_{i<j}(z_i-z_j)^3
\]
With $A=(c/3)-1$ (with $c$ the central charge), we have $\Psi\propto \mathscr{S}\prod_{1\leq a<b\leq n}f(z_{2a-1}, z_{2a}, z_{2b-1},z_{2b})$, where $f$ is given by
\[
\begin{aligned}
	f(z_1, z_2, z_3, z_4)&=A(z_1-z_3)^3(z_2-z_4)^3(z_1-z_4)^3 (z_2-z_3)^3\\
	& \quad+(z_1-z_3)^4(z_2-z_4)^4(z_1-z_4)^2 (z_2-z_3)^2
\end{aligned}
\]
In terms of our intensive graphs, we may symbolically write this as
\[
\Psi_n = \mathscr{N}_n \mathrm{SGP}\Big[T_n\otimes \Big\{
A g_1
+
g_2
\Big\}\Big], \qquad 
g_1=\vcenter{\hbox{
		\includegraphics[scale=.5]{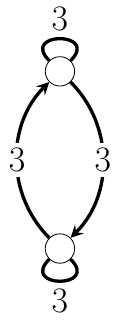}
}}, \qquad
g_2=
\vcenter{\hbox{
		\includegraphics[scale=.5]{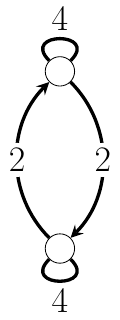}
}}
\]
Let us expand on the meaning of this symbol. Let's call an arrow in $T_n$ a \emph{bond} for clarity. We begin by reinterpreting the tensor product $T_n\otimes g$. A procedural way to obtain $T_n\otimes g$ is by making $n(n-1)/2$ copies of the \emph{shard} $\sigma=T_2\otimes g$, and patching them along the bonds. We understand the symbol $T_n\otimes (Ag_1+g_2)$ via the procedural interpretation. Concretely, the symbol
$\mathrm{SGP}[T_n\otimes (Ag_1+g_2)]$ is the superposition of the SGPs of $2^{n(n-1)/2}$ graphs $G_f$. The label `$f$' is an edge-coloring of $T_n$ by two colors $1,2$. To obtain $G_f$, each bond colored $1$ (resp. 2) is replaced by the \emph{shard} $\sigma_1=T_2\otimes g_1$ (resp. $\sigma_2=T_2\otimes g_2$). Once all the shards are in place, we glue them together to get $G_f$. If the resulting graph $G_f$ has $m$ bonds colored by $1$, then the polynomial $\mathrm{SGP}[G_f]$ is accompanied by weight $A^m$ in the superposition. An ansatz of this form will be quite valuable for computing wavefunctions that are conformal blocks. However, formalizing this ansatz in the general case has some difficulties. The main issue is how to patch the individual shards. In highly symmetric cases, like the example above, there is no possibility of ambiguity. Nevertheless, in the general case, the situation is quite subtle and under current scrutiny.

\section{Uniform States \& Cayley Decomposition}
\label{uniform_sec}
\begin{figure}
	\centering
	\includegraphics[width=.5\linewidth]{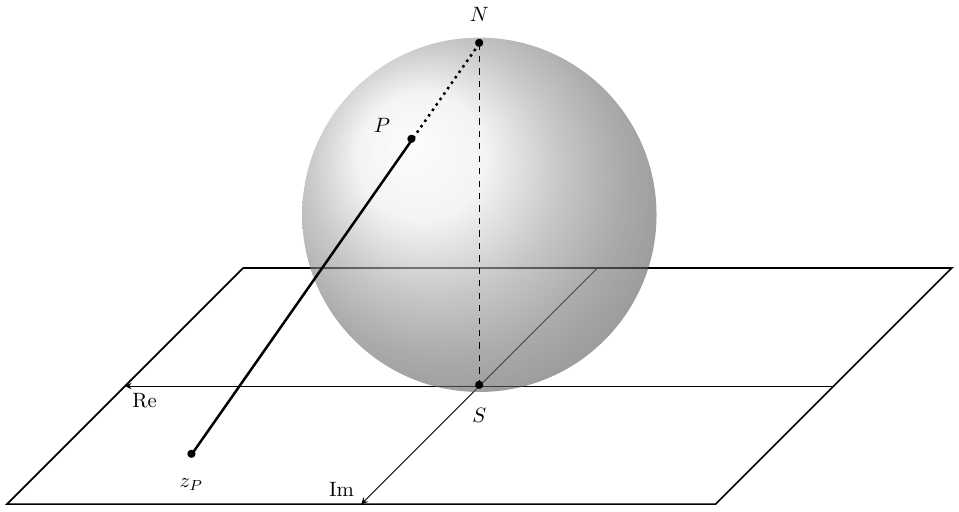}
	\includegraphics[width=.4\linewidth]{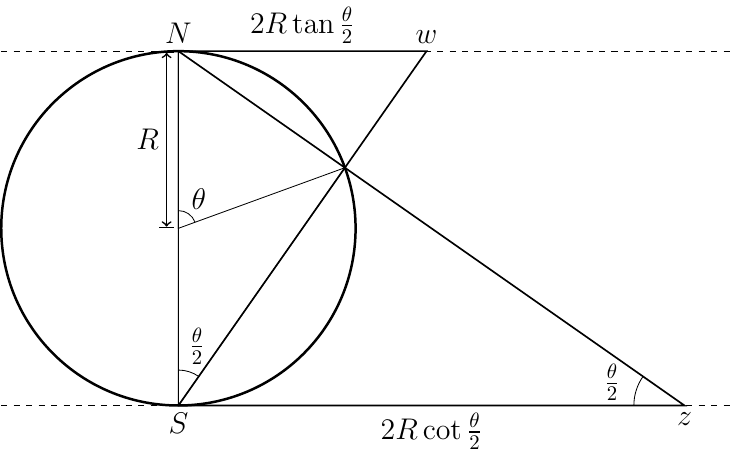}
	\label{stereo_fig}
	\caption{(Left) Illustration of stereographic projection. We make the complex plane $\mathbb{C}$ to be tangential to the sphere at the south pole (Let $\hat{\vv{x}},\hat{\vv{y}}$ be the direction of the real and imaginary axis, and $\gv{\Omega}$ be the outward unit normal to the sphere. We require $\hat{\vv{x}}\times \hat{\vv{y}}=-\gv{\Omega}$). The complex coordinate $z_P$ of a point $P$ is the intersection of the ray $NP$ with the complex plane. (Right) Geometric illustration of the reparametrization $z=2R\cot(\theta/2)e^{i\phi}$ and $w=2R\tan(\theta/2)e^{-i\phi}=1/z$.}
\end{figure}

We begin by reviewing the relevant features of the lowest Landau level in spherical geometry (see \ref{HaldaneSphere_app} for a complete treatment). We denote the radius by $R$. A magnetic monopole at the center causes a transverse uniform magnetic field $B$ on the surface. Let $\ell_B=(\hbar/e B)^{1/2}$ be the magnetic length. The radius $R$ is such that the number of flux quanta $N_\phi = 4\pi R^2 B/(h/e) =2(R/\ell_B)^2$ is an integer. Under these circumstances, the lowest Landau level (LLL) is $N_\phi+1$ dimensional. Using the stereographic coordinates $z=2R\cot(\theta/2)e^{i\phi}$ for the position of particles (Fig. \ref{stereo_fig}), the following wavefunctions constitute a basis for the LLL ($0\leq l\leq N_\phi$):
\begin{equation}
	\varphi_l(z) = z^lK(z,\bar{z}), \qquad K(z,\bar{z})=(1+z\bar{z}/4R^2)^{-N_\phi/2}
\end{equation}
[Note that in the limit $R\to \infty$, i.e. for an infinite plane, we have $N_\phi\to \infty$ and $\varphi_l = z^l \exp(-z\bar{z}/4\ell_B^2)$.] Let us describe the inner product. Let $P_i(z)$ (with $i=1,2$) be polynomials of degree at most $N_\phi$, and $\psi_i(z,\bar{z})=P_i(z)K(z,\bar{z})$ the corresponding LLL wavefunctions. The $L^2$ inner product is then given by
\begin{equation}
	\braket{\psi_1}{\psi_2} = \int R^2\sin \theta d\theta d\phi \:\overline{\psi_1(\theta, \phi)}\psi_2(\theta, \phi)=
	\int \frac{dxdy}{(1+z\bar{z}/4R^2)^{N_\phi+2}} \:\overline{P_1(z)}P_2(z)
\end{equation}
where $x,y$ are defined via $z=x+iy$. Next, we utilize the single-body states to build many-body wavefunctions.

We place $N$ spin-polarized bosonic electrons on the spherical setup described above. We assume that the gap between the Landau levels is large enough that all particles live in the LLL. The complex coordinate of $i$th particle is denoted by $z_i = 2R\cot(\theta_i/2)\exp(i\phi_i)$ (Fig. \ref{stereo_fig}). Due to the confinement to the LLL, the most general many-body wavefunction is of the form:
\begin{equation}
	\Phi(z_1, \cdots, z_N)= \Psi(z_1, \cdots, z_N) \prod_{i=1}^N K(z_i, \bar{z}_i)
\end{equation}
Here, $\Psi=\Psi(z_1, \cdots, z_N)$ is a \emph{symmertic polynomial} of degree no higher than $N_\phi$ in any of its variables. It is useful to call the degree of $\Psi$ in $z_1$ (or any other variable) the \emph{local degree} of $\Psi$. We denote the local degree by $\deg_\ell \Psi$ and recognize the LLL condition as $\deg_\ell \Psi\leq N_\phi$. From now on, we will ignore the kernels $K$ and call $\Psi$ the `wavefunction' or the `state.'

If $\Psi$ is the ground state of a bosonic fractional quantum Hall (FQH) system in the LLL, then it will be a liquid state satisfying the symmetries of the sphere: the wavefunction $\Psi$ must be an $SO(3)$ singlet. Concretely, the presence of the magnetic flux induces a particular representation of $SO(3)$ (see \ref{HaldaneSphere_app}). With $L_z, L_{\pm}$ denoting the angular momentum operator in this representation, we have
\begin{subequations}
	\begin{align}
		L_-\Psi&:=
		\sum_i \partial_i\Psi
		=0\\
		L_z\Psi&:= \sum_i \left(-\frac{N_\phi}{2}+z_i\partial_i\right)\Psi = 0\\
		L_+\Psi &:= \left(\sum_{i}N_\phi z_i-z_i^2\partial_i\right)\Psi = 0
	\end{align}
\end{subequations}
In other words, $L^2\Psi = 0$. As we explain in \ref{uniform_appendix}, this is equivalent to $\Psi$ being an $(N,N_\phi)$ uniform state: we call a polynomial $\Psi$ an $(N,N_\phi)$ \emph{uniform state} (on the sphere) if
\begin{enumerate}
	\item $\Psi=\Psi(z_1, \cdots, z_N)$ is a \emph{symmetric} polynomial in $N$ variables, of degree no higher than $N_\phi$ in $z_1$ ($\deg_\ell \Psi\leq N_\phi$).
	\item $\Psi$ is \emph{translational invariant}; i.e., for any $c\in \mathbb{C}$, we have $\Psi(z_1+c, \cdots, z_N+c)=\Psi(z_1, \cdots, z_N)$.
	\item $\Psi$ is \emph{homogeneous}; i.e., for any $0\neq \alpha \in \mathbb{C}$, we have $\Psi(\alpha z_1, \cdots, \alpha z_N)=\alpha^M \Psi(z_1, \cdots, z_N)$ for some $M$ called the total \emph{degree} (which is equal to the total angular momentum in the \emph{plane} geometry).
	\item Defining the \emph{adjoint} of a polynomial of degree no higher than $N_\phi$ in each variable as
	\begin{equation}
		\Psi^\dagger(z_1, \cdots, z_N):=\prod_{i=1}^N z_i^{N_\phi}\Psi\left(-\frac{1}{z_1}, \cdots, -\frac{1}{z_N}\right)
		\label{adjoint_eq}
	\end{equation}
	we have $\Psi=\Psi^\dagger$; i.e. $\Psi$ is \emph{self-adjoint}.
\end{enumerate}
An FQH ground state is necessarily (but not sufficiently) an $(N, N_\phi)$ uniform state. Including the above two interpretations, there are (at least) four equivalent formulations of uniform states. In \ref{uniform_appendix}, we have listed all four and proved their equivalence.

Most of the well-known examples of uniform states in the FQH literature are special cases of the family of negative rational Jack polynomials  \cite{Bernevig_Haldane_Model} ($k+1$ and $r-1$ are relatively prime):
\begin{subequations}
	\label{jack_eq}
	\begin{align}
		\mathcal{J}_n^{(k,r)}&:= J_{\Lambda(n,k,r)}^{(\alpha_{k,r})}, \qquad \alpha_{k,r} = - \frac{k+1}{r-1}\\
		\Lambda(n,k,r)&=([(n-1)r]^{\times k}, \cdots, [2r]^{\times k}, r^{\times k}, 0^{\times k})
		\label{Lambda}
	\end{align}
\end{subequations}
The Laughlin $2m$-state (with $k=1, r=2m$), Pfaffian state \cite{Moore-Read} (with $k=2, r=2$), $\mathbb{Z}_k$ Read-Rezayi states \cite{Read-Rezayi} (with appropriate $k$ and $r=2$), and Gaffnian \cite{Gaffnian} (with $k=2, r=3$) are special cases. We will return to these examples a few times in due course of the paper. A non-Jack example would be Haffnian \cite{Haffnian}.

\begin{figure}
	\centering
	\includegraphics[width=.3\linewidth]{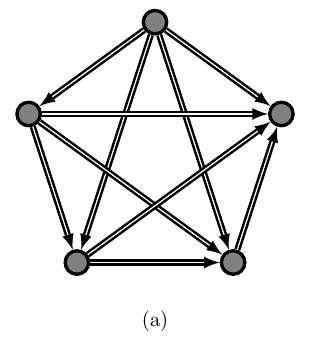} 
	\includegraphics[width=.3\linewidth]{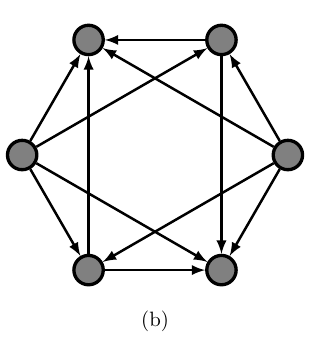}
	\includegraphics[width=.3\linewidth]{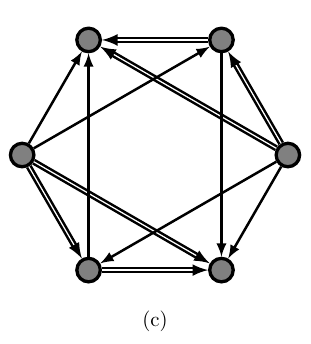} 
	
	\includegraphics[width=.3\linewidth]{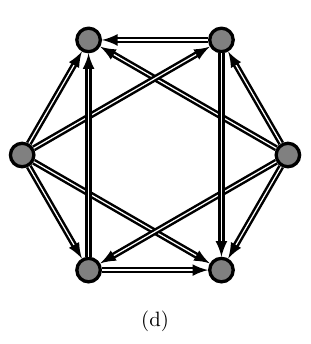}
	\includegraphics[width=.3\linewidth]{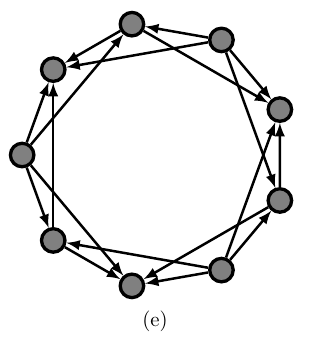} 
	\includegraphics[width=.3\linewidth]{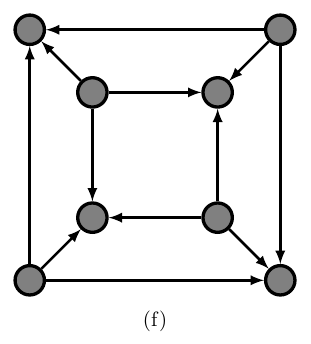}
	\caption{Six examples of extensive graphs. Single and double arrows are arrows of multiplicity one and two, respectively. The SGP of these graphs are the following (bosonic) states: (a) $N=5$ Laughlin 2-state ($N_\phi=8$); (b) $N=6$ Moore-Read state/Pfaffian ($N_\phi=4$); (c) $N=6$ Gaffnian ($N_\phi=6$); (d) $N=6$ Haffnian ($N_\phi=8$); (e) $N=9$, $\mathbb{Z}_3$ Read-Rezayi state ($N_\phi=4$); (f) Jack polynomial $\mathcal{J}_2^{(4,3)}$ ($N_\phi=3$).} 
	\label{extensive_graphs_fig}
\end{figure}

Uniform states are closely related to regular graphs, which we now describe. In this paper, ``graph'' is shorthand for ``directed weighted/multiple graph''. Intuitively, our notion of a graph is a bunch of nodes and arrows between nodes. More precisely, a graph $G$ is a pair $G=(V,\mu)$, with $V$ being the \emph{node set} and $\mu:V\times V\to \mathbb{Z}_{\geq}$ called the \emph{multiplicity function}. If $i\neq j\in V$, we understand $\mu(i,j)\neq 0$ as $G$ having an \emph{arrow} $i\to j$ of multiplicity $\mu(i,j)$. If $\mu(i,i)\neq 0$ we say $G$ has a \emph{loop} at $i$ of multiplicity $\mu(i,i)$. Furthermore, $\mu(i,j)=0$ means \emph{no arrow} starting at $i$ and ending on $j$. We will recite (and, at points, invent) graph-theoretic notions in the course of the paper. Some immediately relevant definitions are:
\begin{dfn}
	The \emph{order} of $G$, denoted by $|G|$, is the number of nodes that it has (i.e. $|G|=|V|$).
\end{dfn}
\begin{dfn}
	If $G$ has a loop nowhere/everywhere we say $G$ is \emph{loopless}/\emph{fully looped} respectively.
\end{dfn}
\begin{dfn}
	For each node $i$, we define the \emph{out-degree} ($+$)/\emph{in-degree} ($-$) as
	\begin{equation}
		d_i^{+} = \sum_{j\in V}\mu(i, j), \quad
		d^{-}_i = \sum_{j\in V}\mu(j, i)
		\label{outindeg_eq}
	\end{equation}
	Put differently, the out-degree (resp. in-degree) is the number of outgoing (resp. incoming) arrows incident to $i$.
\end{dfn}
\begin{dfn}
	$G$ is \emph{regular} of degree $\Delta$ if $d_i^++d_i^- = \Delta$ for all $i$. In other words, every node has $\Delta$ arrows incident to them.
\end{dfn}
We say a graph $G$ is $(N,N_\phi)$ \emph{extensive} if $G$ is a loopless regular graph of degree $N_\phi$ and order $N$. In subsection \ref{intensive_graphs_sec}, we will define the so-called intensive graphs as well. However, the graphs relevant to $(N, N_\phi)$ uniform states are all extensive. Six examples of extensive graphs, all `presenting' known FQH ground states, are depicted in Fig. \eqref{extensive_graphs_fig}.

The connection between uniform states and extensive graphs manifests through \emph{symmetrized graph polynomials (SGPs)}. Given a graph $G$ with nodes labeled as $1,2,\cdots, N$, we define the symmetric polynomial $\mathrm{SGP}(G)$ as
\begin{equation}
	\label{SGP_eq}
	\mathrm{SGP}(G) = \sum_{\sigma\in \mathfrak{S}_N}\prod_{i,j=1}^N(z_{\sigma(i)}-z_{\sigma(j)})^{\mu_G(i,j)}
\end{equation}
Here, $\mathfrak{S}_N$ is the symmetric group, and $\mu_G$ is the multiplicity function of $G$. Note that $\mathrm{SGP}$ vanishes if there is any loop in $G$. The definition of $\mathrm{SGP}$ does not require extensivity of $G$. Nevertheless, the SGP of an extensive graph is particularly significant:
\begin{thm}
	\label{cayley_1}
	If $G$ is an $(N,N_\phi)$ extensive graph, then $\mathrm{SGP}(G)$ is either identically zero or an $(N,N_\phi)$ uniform state.
\end{thm}
\noindent
A uniform state $\Psi$ is said to be \emph{presentable} if $\Psi=\alpha\mathrm{SGP}(G)$ for some extensive graph $G$ and $\alpha\in \mathbb{C}$. If so, $G$ is called a \emph{presentation} of $\Psi$. The examples in Fig. \eqref{extensive_graphs_fig} (see the caption) are all  presentations of well-known FQH states. Presentations are not unique. For example, with $\Psi$ being the $N=6$ Pfaffian, an alternative presentation to Fig. (\ref{extensive_graphs_fig}.b) is found via the equality
\[
\mathrm{SGP}\left(
\vcenter{
	\hbox{
		\includegraphics[scale=.75]{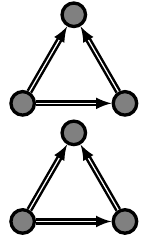} 
	}
}
\right)=
4\:
\mathrm{SGP}\left(
\vcenter{
	\hbox{
		\includegraphics[scale=.75]{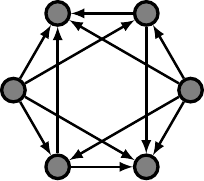} 
	}
}
\right)
\]
Although not every uniform state is presentable, one can always decompose (not necessarily uniquely) any uniform state as a superposition of presentable states:
\begin{thm}[Cayley Decomposition]\label{cayley_2} Let $\Psi$ be an $(N,N_\phi)$ uniform state. Then there exist a finite number of $(N,N_\phi)$ extensive graphs $G_1, \cdots, G_p$, together with $\mathbb{C}$-numbers $\alpha_1, \cdots, \alpha_p$, so that
	\begin{equation}
		\label{cayley_decomp_eq}
		\Psi = \sum_{l=1}^p \alpha_l \:\mathrm{SGP}(G_l)
	\end{equation}
\end{thm}
\noindent
To our knowledge, SGPs first appeared in communications between Petersen and Sylvester \cite{sylvesterA, petersen}. The original discovery of Theorem \ref{cayley_1} and \ref{cayley_2} is apparently due to Cayley. The proofs for Theorems \ref{cayley_1} and \ref{cayley_2} can be found in \ref{cayley_appendix}.

Much of our knowledge today about uniform states comes from the works of mathematical giants of the nineteenth century, like Sylvester, Cayley, Petersen, Clebsch, Gordan, Hermite, and Hilbert. The machinery introduced here is mostly a modern reformulation of the tools developed by them to determine and classify all \emph{binary invariants} (see Ref. \cite{olverinv} for an introduction). Since nothing in this paper explicitly needs invariant theory, we have refrained from discussing binary invariants and their \emph{equivalence} to uniform states in any capacity. We explore binary invariants in the context of quantum Hall states elsewhere \cite{pakatchi2021CFT}.

%%%%%%%%%%%%%%%%%%%

\section{Graph-Based Construction of Clustering States}
\label{construction_sec}
This paper aims to provide a systematic construction of trial FQH ground states. Although SGPs provide an opportunity to create trial wavefunctions out of regular graphs, they do not clarify which regular graphs are relevant to the FQH effect. To elaborate, the distinguishing characteristics of FHQ ground states must be in their local properties. For example, one the distinguishing features of the Jacks $\mathcal{J}_n^{(k,r)}$ is their so-called $(k,r)$ clustering property \cite{estienne1, estienne2, zamaere}:
\[
\mathcal{J}_{n+1}^{(k,r)}(k,r)(\overbrace{w, \cdots, w}^{\times k}, z_1, z_2, \cdots, z_{nk})=
\prod_{i=1}^{nk} (z_i-w)^{r}\mathcal{J}_n^{(k,r)}(z_1, z_2, \cdots, z_{nk})
\]
Since many of the most successful FQH ground states are special cases of Jacks, we are motivated to look for graphs leading to clustering states. Concretely, we are looking for sequences of extensive graphs $\vv{G}=(G_n)$ so that their SGP is  $(k, r)$ clustering:
\[
\gv{\Psi}\propto \mathrm{SGP(\vv{G})}=(\mathrm{SGP}(G_1), \mathrm{SGP}(G_2), \cdots, \mathrm{SGP}(G_n), \cdots)
\]
The graph $G_n$ will be regular of degree $(n-1)r$ and order $nk$. The philosophy behind the design of $ G_n$ is inspired by the classification ideas of the pattern of zeros formalism. In Ref. \cite{Pattern_of_Zeros}, a key idea of the authors is to reduce the large messy polynomials encountered in FQHE to a handful of \emph{classifying data}. For their purposes, the classifying data is the pattern of zeros. However, the classifying data for our construction is a size-independent graph $g$. All graphs $G_n$ are obtained from $g$ through a procedure called \emph{extension}.

The outline of this section is as follows. In subsection \ref{size_sec}, we open the section by commenting on the role of ``size'' for FQH ground states. We argue that if the goal is to construct trial ground states, one has to design a sequence of wavefunctions in \emph{all sizes}. In subsection \ref{clustering_sec}, we discuss the clustering property and its consequences. Building on that, in subsection \ref{heuristic_sec}, we heuristically present the design of the graphs $G_n$ and what motives the specifics of their construction. Subsection \ref{intensive_graphs_sec} formalizes the construction and introduces the intensive graphs $g$ (the classifying data). The proof that $\mathrm{SGP}(G_n)$ are clustering states is sketched in subsection \ref{localstructure_sec}.

\subsection{Size in FQH Systems}
\label{size_sec}
Consider a quantum Hall system consisting of $N$ electrons on a sphere. We measure the filling fraction $\nu$ (i.e., the Hall conductance $\sigma_H = (e^2/h)\nu$) and the Wen-Zee shift $\mathcal{S}$ \cite{wen1992shift} (i.e., the Hall viscosity $\eta_H = \frac{1}{4}\hbar \rho \mathcal{S}$, with $\rho$ being the mean density \cite{read2009non,read2011hall}). With $N_\phi$ the number of magnetic flux quanta, the following condition must hold:
\[
N_{\phi}^{\mathrm{eff}}:=N_{\phi} + \mathcal{S}= N\nu^{-1}
\]
The notation $N_{\phi}^{\mathrm{eff}}$ refers to the number of flux quanta of the `effective' magnetic field. To expand on that, since the sphere is not flat, the orbital motion also couples to the sphere's curvature \cite{wen1992shift}. This additional coupling changes the effective magnetic field. The Wen-Zee shift $\mathcal{S}$ is precisely the additional flux gain.

Now, the same FQH phase of matter can be prepared with a different number of electrons. For example, the new system can have $N(n)=nN$ particles and $N_{\phi}^{\mathrm{eff}}(n)=n N_{\phi}^{\mathrm{eff}}$ effective magnetic flux quanta. The constraint on the number of (actual) magnetic flux quanta becomes
\begin{equation}
	\label{Nphi_EQ}
	N_{\phi}(n) = N(n)\nu^{-1} - \mathcal{S}
\end{equation}
In this sense, we have prepared the phase at size $n$. The seemingly innocuous observation we make is that: although the phase of matter is insensitive to the size, it is meaningless to talk about the `ground state wavefunction' without specifying the size. To expand on that, let $\Psi_n$ be the symmetric polynomial that describes the FQH ground state at size $n$. To a physicist, the wavefunctions prepared at different sizes are all ``the same'', since they describe the same phase. However, to a mathematician interested in building these ground states, there is no obvious relation (as functions) between $\Psi_n$ and $\Psi_{n+1}$. After all, $\Psi_n$ and $\Psi_{n+1}$ do not even have the same number of variables. Additionally, the phase of matter is usually only meaningful in the thermodynamic limit $n\to \infty$. If we cannot say how $\Psi_n$ and $\Psi_{n+1}$ relate, we have no chance of describing the $n\to \infty$ limit of these functions. Due to these considerations, if the goal is to \emph{construct explicit polynomials}  representing a quantum Hall phase, it is more appropriate to work with the sequence of wavefunctions at `all sizes':
\begin{equation}
	\mathbf{\Psi} = (\Psi_1, \Psi_2, \cdots, \Psi_n, \cdots)
\end{equation}
To put it differently, any reasonable construction of trial FQH ground states should produce an infinite sequence of states $\Psi_n$. The number of particles at size $n$ must be $N(n)=nN(1)$, and the number of flux quanta given by \eqref{Nphi_EQ}.

%%%%%%%%%%%
\subsection{Clustering States}
\label{clustering_sec}
The simplest local feature of an FQH ground state $\Psi(z_1, \cdots, z_N)$ is how it vanishes as electrons are fused. Suppose $\Psi(z_1, \cdots, z_N)$ is an $(N, N_\phi)$ uniform state. For each $a\geq 1$, define the \emph{$a$-fusion} as (Notation: $w^{\times a}$ stands for $w$ repeated $a$ times)
\begin{equation}
	\Phi_a(w;z_1, \cdots, z_{N-a})=\Psi(w^{\times a}, z_1, \cdots, z_{N-a})
	\label{fusions_eq}
\end{equation}
Physically speaking, $\Phi_a$ is the result of bringing an $a$-cluster of bosonic electrons to a common point $w$. Due to $\Psi$ being symmetric and translational invariant, there exist integers $k,r>0$, so that 
\begin{subequations}
	\label{fusion_eqs}
	\begin{align}
		\Phi_a(w;z_1, \cdots, z_{N-a})&=0\quad \text{for}\quad  a>k\\
		\Phi_k(w;z_1, \cdots, z_{N-k})&=\prod_{i=1}^{N-k}(z_i-w)^rR(w;z_1, \cdots, z_{N-k})
		\label{kfusion_eq}
	\end{align}
\end{subequations}
where $R(w;z_i)$ is a symmetric polynomial in $z_i$'s and $\lim_{w\to z_1}R(w; z_1, \cdots)\neq 0$. In general, $R$ is a messy polynomial, and the only meaningful local information we may extract is the vanishing order $r$. An important exception is when $\deg_w R=0$, i.e., $R$ has no $w$ dependence.

In the above analysis, if $\deg_w R=0$, we say $\Psi$ is a $(k,r)$-\emph{clustering state}. It is straightforward to check that if $\Psi$ is $(k,r)$ clustering and $(N, N_\phi)$ uniform, then $R$ is an $(N-k, N_\phi-r)$ uniform state. In what follows, we will discuss additional properties of the clustering states.

Among all states that vanish of order \emph{at least} $r$ when $k+1$ particles are fused but do not vanish when $k$ particles are fused, the $(k,r)$-clustering states are the densest, i.e. have smallest local degree $N_\phi$ (also see Ref. \cite{Simon_projection}). To show this, consider a wavefunction $\Psi$ vanishing of order $d\geq r$ when $k+1$ particles are fused. Let us compute $\deg_w{\Phi_k}$. On the one hand, since $\Psi$ is self-adjoint, we have
\[
\Phi_k(w;z_i)=w^{kN_\phi}\prod_{i}z_i^{N_\phi}\Phi_k(-1/w;-1/z_i)
\]
Moreover, since $\Phi_k(w;z_i)$ is translational invariant, its `constant term' $\Phi_k(0;z_i)$ is non-zero (see the Lemma in \ref{uniform_appendix}). The two observation combine to give $\deg_w{\Phi_k}=kN_\phi$. On the other hand, using \eqref{kfusion_eq}, we find that $\deg_w{\Phi_k}=(N-k)d+\deg_w R$. Putting the two computations together, we obtain the following identity:
\begin{equation}
	\label{temp_id_Eq}
	kN_\phi = (N-k)d+\deg_w R
\end{equation}
Since $d\geq r$ and $\deg_w R\geq 0$, the minimum $N_\phi$ occurs when $\deg_w R=0$ and $d=r$. 

The filling fraction and the shift of a $(k,r)$ clustering state are respectively $\nu=k/r$ and $\mathcal{S}=r$. This can be obtained from \eqref{temp_id_Eq} (for $d=r, \deg_w R=0$) in conjunction with the constraint $N_{\phi}=N\nu^{-1}-\mathcal{S}$. Moreover, since $N_\phi$ has to be an integer, the numbers $N, N_\phi$ must be of the following form
($0\leq s<\mathrm{gcd}(k,r)$):
\begin{subequations}
	\label{nn}
	\begin{align}
		N^{(k,r,s)}(n)&=nk +\frac{k}{\mathrm{gcd}(k,r)}s
		\label{Nn}
		\\
		N^{(k,r,s)}_\phi(n)&=(n-1)r +\frac{r}{\mathrm{gcd}(k,r)}s
		\label{sizes_eq}
	\end{align}
\end{subequations}
The triple $(k,r,s)$ are characteristic of the FQH phase of matter. However, the index $n$ is the extensive parameter that determines the size of the system. For simplicity, this paper will only consider the $s=0$ case. Explicitly, we always have $N(n)=nk$ and $N_\phi(n)=(n-1)r$.

Let $\mathbf{\Psi} = (\Psi_1, \Psi_2, \cdots, \Psi_n, \cdots)$ be a sequence describing the ground state of an FQH phase. Suppose $\Psi_n$ is an $(n k, (n-1)r)$ uniform state, leading to $\nu=k/r$ and $\mathcal{S}=r$. In addition to these assumptions, if all $\Psi_n$ are $(k,r)$ clustering states, we say $\mathbf{\Psi}$ is a $(k,r)$ \emph{clustering sequence}. Such sequences enjoy the following relation:
\begin{equation}
	\Psi_{n+1}(w^{\times k}, z_1, \cdots, z_{nk})=
	\prod_{i=1}^{nk}
	(z_i-w)^r
	\Psi_n(z_{1}, \cdots, z_{nk})
	\label{krclustering_EQ}
\end{equation}
In other words, we may obtain $\Psi_n$ from $\Psi_{n+1}$ via a limit:
\begin{equation}
	\Psi_n(z_1, z_2, \cdots)=\lim_{w\to 0}
	w^{kN_\phi}\Psi_{n+1}\Big(\underbrace{-\frac{1}{w},\cdots, -\frac{1}{w}}_{\times k};z_1, z_2, \cdots, z_{nk}\Big)
	\label{reduction_alt_eq}
\end{equation}
This process can be interpreted as coalescing $k$ electrons \emph{at infinity}, creating a vortex there with flux $r$. If we remove this particle and vortex at infinity, we go from size $n$ to size $n-1$. The wavefunction resulting from this process is thus naturally identified with the ground state of the same phase at size $n-1$.

Let us finish by commenting on the relation between $(k,r)$ clustering states and Laughlin $kr$-states. First of all, $\Psi_1$ is a constant, we may choose $\Psi_1\equiv 1$. Equivalently, this normalization can be fixed via the \emph{refinement} equation:
\begin{equation}
	\Psi_{n}(w_1^{\times k}, w_{2}^{\times k},\cdots, w_n^{\times k}) = \prod_{1\leq i<j\leq n}(w_i-w_j)^{kr}
	\label{refinement_eq}
\end{equation}
In this sense, $(k,r)$ clustering states are refinements of Laughlin $kr$-state. In particular, $kr$ needs to be even. As a secondary consequence, observe that $\deg \Psi_n = \frac{1}{2}N(n)N_\phi(n)=\frac{n(n-1)}{2}kr$. We can use this to weaken the definition of a clustering sequence. Suppose $\Psi_n$ is symmetric, translational invariant, homogeneous, and $\deg_\ell \Psi_n\leq N_\phi$. A priori, we do not demand $\Psi_n$ to be self-adjoint. Nonetheless, suppose $\gv{\Psi}$ satisfies Eq. \eqref{krclustering_EQ}. According to part (4) of Theorem \ref{uniform_equivalent} in \ref{uniform_appendix}, since $\deg \Psi_n = \frac{1}{2}N(n)N_\phi(n)$, the wavefunctions $\Psi_n$ are automatically $(nk, (n-1)r)$ uniform states.

\subsection{The Ideas Behind the Graphic Construction}
\label{heuristic_sec}
We use an analogy between a ground state sequence $\gv{\Psi}=(\Psi_n)$ and a thermodynamic system. A thermodynamic system is controlled by a single extensive variable and a collection of intensive ones. The intensive variables (e.g., temperature, pressure) describe the phase, while the extensive variable determines the size. We should be able to identify an FQH system similarly. The number of particles $N$ is a choice for the extensive variable, while topological invariants, e.g., filling fraction $\nu$ and shift $\mathcal{S}$, are characteristics of the phase (playing a similar role to intensive variables). We want to apply this point of view to the level of FQH ground states $\mathbf{\Psi}=(\Psi_n)$ representing a phase. We expect that $\Psi_n$ can be characterized by a set of `intensive' variables and a single extensive parameter $n$ identifying the size. The intensive data, whatever it may be, must be a characteristic of the FQH phase of matter. At the level of polynomials,  while the number of variables is a good indicator for the extensive variable, it is unclear what qualifies as ``intensive'' data of a state $\Psi_n$. The best we can say is that ``intensive'' data depends solely on the local properties of $\Psi_n$. We believe the situation can be drastically clarified if we regard FQH ground states as graphs instead of polynomials. We demonstrate this idea by explicitly constructing a large family of trial ground states.

Let $\vv{G}=(G_n)$ be a placeholder for a sequence of $(n k, (n-1)r)$ extensive graphs, and $\gv{\Psi}\propto\mathrm{SGP}(\vv{G})$. We would like to design $\mathbf{G}$ in a way that $\gv{\Psi}$ is $(k,r)$ clustering. If we manage to do that, then $\Psi_n$ will satisfy Eq. \eqref{refinement_eq}:
\[
\Psi_{n}(w_1^{\times k}, w_{2}^{\times k},\cdots, w_n^{\times k}) = \prod_{1\leq i<j\leq n}(w_i-w_j)^{kr}
\]
In this sense, the polynomials $\Psi_n$ are a refinement of Laughlin $kr$-states. Now, let $T_n^{2m}$ denote the presentation of Laughlin $2m$-state (exact definition will follow shortly). The core intuition behind the construction is then as follows: \emph{If $\Psi_n$ is a refinement of Laughlin $kr$-state, then it is natural to presume $G_n$ (i.e., the presentation of $\Psi_n$) is a refinement of $T_n^{kr}$ in some sense.} To properly define what is meant by ``refinement'' in the context of graphs, we need the following notions:
\begin{figure}
	\centering
	\includegraphics[width=\linewidth]{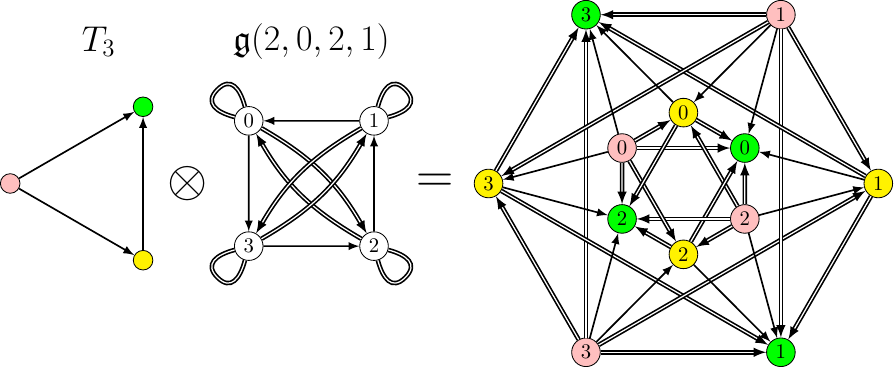} 
	\caption{Demonstration of the tensor product of $T_3$ with $g=\mathfrak{g}(2,0,2,1)$. Instead of labeling the nodes as $\mathrm{``color"}\otimes \mathrm{``number"}$, we have labeled them with the ``number'' and painted the node with the ``color''.}
	\label{tensorEXP}
\end{figure}

\begin{dfn}
	The $n$th \emph{transitive tournament}  $T_n$ is a graph with node set $V=\{1,2,\cdots, n\}$ and multiplicity function
	$$
	\mu_1(i,j)=\begin{cases}
		1 & i<j\\
		0 &\mathrm{otherwise}
	\end{cases}
	$$
	By convention, $T_1$ is a single node with no loop. The graph $T_n^{2m}$  is defined as the pair $T_n^{2m}=(V, 2m\mu_1)$; i.e. each arrow in $T_n$ is replaced with an arrow of multiplicity $2m$. Note that $T_n^{2m}$ is the unique presentation of the Laughlin $2m$-state in $n$ nodes (cf. Fig. \ref{extensive_graphs_fig}.a).
\end{dfn}
\begin{dfn}
	Let $G_1=(V_1, \mu_1)$ and $G_2=(V_2, \mu_2)$ be two graphs. We define the \emph{tensor product} $G_1\otimes G_2=(V,\mu)$ as follows. The node set of $G_1\otimes G_2$ is the Cartesian product $V=V_1\times V_2$. The nodes are denoted as $x\otimes y$ with $x\in V_1, y\in V_2$. The multiplicity is defined as follows
	\begin{equation}
		\mu(x_1\otimes y_1, x_2\otimes y_2)=\mu_1(x_1, x_2)\mu_2(y_1,y_2)
	\end{equation}
	We have drawn an example of a tensor product in Fig. \ref{tensorEXP}.	
\end{dfn}
Let us go back to the thermodynamic analogy to finish the heuristic construct. We have the naive expectation that the intensive data should somehow ``decouple'' from the size of the system. We symbolically write this decoupling as $G_n = \mathrm{Size}(n) \otimes \mathrm{Intensive}$. At the same time, $G_n$ is supposedly a ``refinement'' of $T_n^{kr}$. The most natural way to formalize these ideas is by defining
\[
G_n = T_n\otimes g
\]
Here, $g$ is a ``small nice'' graph independent of $n$ and encodes all of the intrinsic/intensive data of the FQH state. The graph $g$ must have $k$ nodes and $kr$ arrows. To see how $T_n\otimes g$ is a refinement of $T_n^{kr}$, note that if one identifies all nodes of $g$ into one, i.e. replaces $g$ with a single loop of multiplicity $kr$, then $G_n=T_n\otimes g$ reduces to $T_n^{kr}$.

\subsection{Intensive Graphs \& Wavefunctions $\Psi^{(g)}_n$}
\label{intensive_graphs_sec}
In this subsection, we will formalize the heuristic construction in subsection \ref{heuristic_sec}. 

Define the \emph{extension} functor $\mathrm{Ext}: \mathrm{Graphs}\to \text{Graph Sequences}$ via
\begin{equation}
	\mathrm{Ext}(g)=(\mathrm{Ext}_n(g))_{n\geq 1}, \qquad
	\mathrm{Ext}_n(g)=T_n\otimes g
\end{equation}
Calculating the symmetrized graph polynomial of this sequence yields a wavefunction sequence $\mathbf{\Psi}^{(g)}=(\Psi_n^{(g)})$, where $\gv{\Psi}^{(g)}\propto\mathrm{SGP}(\mathrm{Ext}(g))$. It remains to choose an appropriate normalization. Let $c(g)$ be the number of components of $g$ (see \ref{graph_appendix}). Then the complete definition should be:
\begin{equation}
	\Psi_n^{(g)}:=\frac{1}{n!^{c(g)}|g|!^{n}}\mathrm{SGP}(T_n\otimes g)
	\label{gsg}
\end{equation}
We need to impose criteria on $g$ so that $\Psi_n^{(g)}$ is $(k,r)$ clustering. Two graph-theoretic notions are needed to achieve this [Recall that $d_i^+$ (resp. $d_i^-$) counts the number of outgoing (resp. incoming) arrows for node $i$]:
\begin{dfn}
	$g=(V,\mu)$ is \emph{even} if $
	\sum_{i\in V} d_i^+=
	\sum_{i\in V} d_i^-=$even. In other words, if the number of arrows is even.
\end{dfn}
\begin{dfn}
	A graph $g=(V,\mu)$ is called \emph{flow-regular} of degree $r$ if $d_i^+=d_i^-=r$ for all $i\in V$. 
\end{dfn}
We say a graph $g$ is \emph{$(k,r)$ intensive} if $g$ is fully looped, has order $k$, is flow-regular of degree $r$, and each of its components is even. As it turns out (subsection \ref{localstructure_sec}), these restrictions on $g$ make $\mathbf{\Psi}^{(g)}$ a $(k,r)$ clustering sequence. In Table \ref{table:intensive}, we have gathered the intensive graphs of the well-known (presentable) FQH states.

\begin{table}
	\centering
	\begin{tabular}{|c|c|c|c|c|}
		\hline
		Laughlin & {\Large $\ell\equiv \vcenter{\hbox{\includegraphics{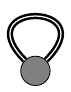}}}$} & {\Large $\ell_2\equiv \vcenter{\hbox{\includegraphics{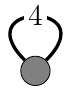}}}$} & $\cdots$ & $\vcenter{\hbox{\includegraphics{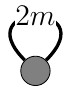}}}$ \\
		($k=1$) & $(r=2)$ & $(r=4)$ & $\cdots$ & $(r=2m)$  \\
		\hline
		Paired States & $\vcenter{\hbox{\includegraphics{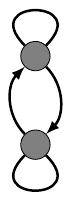}}}$ & {\Large$\jmath\equiv\vcenter{\hbox{\includegraphics{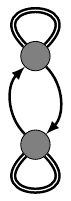}}}$} & $\vcenter{\hbox{\includegraphics{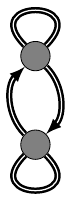}}}$ &  ---\\
		\textit{Alternative:}& {\Large$\ell^2$} & & {\Large $\ell_2^2$} &  \\
		&  Pfaffian & Gaffnian & Haffnian &  \\
		($k=2$)&  ($r=2$) & ($r=3$) & ($r=4$) &  \\
		\hline
		Read-Rezayi& $\vcenter{\hbox{\includegraphics{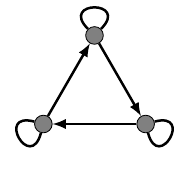}}}$ & $\vcenter{\hbox{\includegraphics{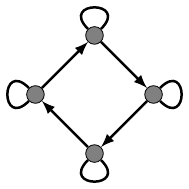}}}$ & $\cdots$ & $\vcenter{\hbox{\includegraphics{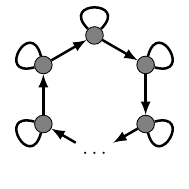}}}$ \\
		\textit{Alternative:}& {\Large $\ell^3$} & {\Large $\ell^4$} & $\cdots$ & {\Large $\ell^k$} \\
		$(r=2)$& $(k=3)$  & $(k=4)$ & $\cdots$  & ($k$=arbitrary)  \\
		\hline
		Jacks $\mathcal{J}^{(2c,3)}$?& {\Large $\jmath^2$} & {\Large $\jmath^3$} & $\cdots$ & {\Large $\jmath^c$} \\
		$(r=3)$& $(k=4)$ & $(k=6)$ & $\cdots$ & $(k=2c)$ \\
		\hline
	\end{tabular}
	\caption{The intensive graphs of the well-known presentable FQH states. The word ``alternative'' refers to a secondary (but equivalent) intensive graph for the respective state. For later use, we named some of the graphs in the table (i.e., $\ell, \ell_2, \gamma$). The notation $g^c$ means the disjoint union of $c$ copies of $g$. Finally, in the case of the Jacks $\mathcal{J}^{(2c,3)}$, the intensive graphs provided are speculative.}
	\label{table:intensive}
\end{table}

\subsection{Local Structure of $\mathrm{Ext}_n(g)$ with $g$ Intensive}
\label{localstructure_sec}
Throughout, $g$ stands for a \emph{connected} $(k,r)$ intensive graph. In this subsection, we will sketch the proof that $\gv{\Psi}^{(g)}$ is non-vanishing, uniform, and $(k,r)$ clustering. The proof of the clustering property is a byproduct of the local structure of the graphs $\mathrm{Ext}_n(g)$. These local structures are the main topic of this subsection. In this section, we will assume $g$ is connected for simplicity
The complete proof for the (possibly) disconnected $g$ can be found in \ref{proof_cluster_appendix}.

We first reduce uniformity and non-vanishing property to clustering property. Note that $G_n=T_n\otimes g$ has order $N=nk$. Moreover, since $g$ is flow-regular of degree $r$, the graph $G_n$ is regular of degree $N_\phi = (n-1)r$. Therefore, by Theorem \ref{cayley_1}, we either have $\Psi_n^{(g)}=0$ or $\Psi_n^{(g)}$ is an $(N,N_\phi)$ uniform state. Thus, it is enough to show $\Psi_n^{(g)}$ is non-vanishing. Next, we show that non-vanishing property is a consequence of clustering property. Recall that if $\gv{\Psi}=(\Psi_n)$ is an arbitrary $(k,r)$ clustering sequence, then 
\[
\Psi_{n+1}(w^{\times k}, z_1, z_2, \cdots) = 
\prod_{i=1}^{nk}(z_i-w)^r\Psi_n(z_1, z_2, \cdots)
\]
Therefore, if $\Psi_n\neq 0$ we necessarily have $\Psi_{n+1}\neq 0$. Consequently, if $\Psi_1\neq 0$ we are done. Now observe that $T_1\otimes g$ has $k$ nodes and no arrows. As a result, $\mathrm{SGP}(T_1\otimes g)=k!$ and  $\Psi_1^{(g)}=1$. The clustering property will be trickier to show, and we need a few definitions beforehand.

To illustrate the local structure of $\mathrm{Ext}_n(g)$, we need to introduce the concept of maximum independent sets. Given a graph $G=(V,\mu)$, a subset of nodes $S\subset V$ is called an \emph{independent set} of $G$ if for any two nodes $x,y\in S$ we have $\mu(x,y)=0$. Verbally, there are no arrows in between the nodes in $S$. We call $|S|$ the \emph{size} of the independent set. The quantity $\alpha(G)$, called the \emph{independence number} of $G$, is the size of the largest independent set in $G$. An independent set $S$ satisfying $|S|=\alpha(G)$ is called \emph{maximum}.

Let us quote some facts about maximum independent sets of $\mathrm{Ext}_n(g)$. Firstly, the fully-looped condition results in $\alpha(G_n)=|g|=k$. Verbally, the maximum size of an independent set is $k$. Secondly, as a result of fully-looped and flow-regularity conditions, the graphs $G_n=T_n\otimes g$ have exactly $n$ maximum independent sets:
\begin{equation}
	[J] = \{J\}\otimes U=\{J\otimes u\mid u\in V_{g}\}
\end{equation}
Clearly, $[I]\cap [J]=\emptyset$ and the node set of $G_n$ is equal to $[1]\cup [2]\cup\cdots \cup [n]$. We call $[J]$'s the color classes of $G_n$. Note that each color class corresponds to a node in $T_n$ (see Fig. \ref{tensorEXP} for an illustration). Since $T_n$ has no loops, it is obvious that $[J]$ is an independent set in $G_n$. The non-trivial part of the above statement is that $[J]$ is \emph{maximum} and that $[1], [2], \cdots, [n]$ are the \emph{only} maximum independent set of $G_n$. The proofs can be found in \ref{maxind_appendix}.

\begin{figure}
	\centering
	\includegraphics[width=0.45\linewidth]{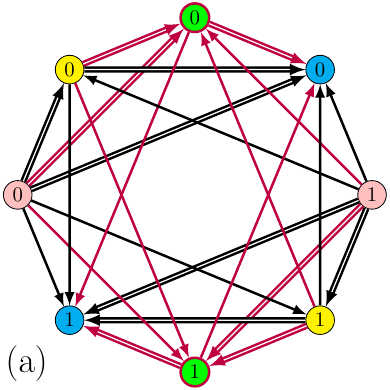} \hfill
	\includegraphics[width=0.45\linewidth]{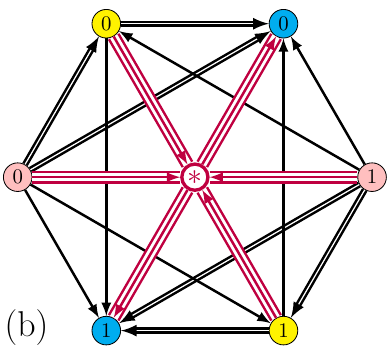}
	\caption{(a) This is the graph $G_n=T_4\otimes \mathfrak{g}(2,1)$, which is a presentation of Gaffnian state with $N=8$. Any arrow with an endpoint in the Green class is highlighted in purple. Note that all non-Green nodes have exactly $r=3$ purple arrows. (b) The Green class has coalesced into the special node $\circledast$. Disregarding the purple arrows, one can see the isomorphism $G_4|_{\mathrm{Green}^c}\simeq G_3=T_3\otimes \mathfrak{g}(2,1)$ (see Fig. (\ref{extensive_graphs_fig}.c)). Note that, for any color class $[J]$, the arrows between $\circledast$ and $[J]$ will either go entirely out of $\circledast$ or come entirely in $\circledast$. Moreover, there are $kr=6=$even arrows between $\circledast$ and any color class.}
	\label{clustering_fig}
\end{figure}

Restricted to a collection $m$ of maximum independent sets, $G_n$ reduces to $G_m$. In particular, we have the following isomorphisms ($[J]^c=$complement of $[J]$)
\begin{align}
	\label{reduction_graph_iso}
	G_n|_{[J]^c}&\simeq G_{n-1}=T_{n-1}\otimes g\\
	\label{patching_iso}
	G_n|_{[I]\cup [J]}&\simeq G_2=T_2\otimes g
\end{align}
We call the former isomorphism the \emph{reduction isomorphism}  (see Fig. (\ref{clustering_fig}.a)). We will use it to prove the clustering property shortly. The restriction of $G_n$ to a pair of distinct maximum independent sets, i.e. $G_n|_{[I]\cup [J]}$, is called a \emph{bond} of $G_n$. The graph $G_2=T_2\otimes g$ is called the \emph{shard} of the construction. The latter relation, called the \emph{patching isomrphism}, means: ``The bonds of $G_n$ are the same as the shard (for all $n$)''. This local structure is illustrated in Fig. (\ref{localfig}.b). One can think of $G_n=T_n\otimes g$ as a patching of $n(n-1)/2$ copies of the shard $G_2=T_2\otimes g$ [In this patching, each node of $T_n$ (e.g., $J$), is associated to a color class (e.g., $[J]$), and each arrow $I\to J$ corresponds to the bond $G_n|_{[I]\cup [J]}$.]

We now sketch the proof that $\gv{\Psi}^{(g)}$ is $(k,r)$ clustering. Let $V=\{1,2 \cdots, (n+1)k\}$ be the node set of $G_{n+1}$. We use the symbol $f$ for bijections $f:V\to \{z_1, \cdots, z_{(n+1)k}\}$. We can write the SGP in the form:
\begin{equation}
	\mathrm{SGP}(G_{n+1})=\sum_f \prod_{i,j}(f(i)-f(j))^{\mu(i,j)}
\end{equation}
Putting $z_{(n+1)k}=z_{(n+1)k-1}=\cdots=z_{nk+1}=w$, only those bijections $f$ that send a maximum independent set to $w$ will survive. The effect of such $f$ is essentially coalescing a maximum independent set into a special point $w$ (see Fig. (\ref{clustering_fig}.b)). With some labor, the reduction isomorphism can be used to show that:
\[
\Psi_{n+1}^{(g)}(w^{\times k},Z_n)=
\prod_{i}(z_i-w)^{r}\Psi^{(g)}_n(Z_n)
\frac{\sum_{J=1}^{n+1} \mathrm{sgn}(J)}{n+1}
\]
where $\mathrm{sgn}(J)=(-1)^{s(J)}$ with $s(J)=(n-J)kr$ the number of arrows going out of $[J]$ in $G_n$. The evenness condition now guarantees that $\mathrm{sgn}(J)=+1$, and the clustering condition is proven. The role of evenness, for the most part, is to make sure $\Psi_n^{(g)}$ is well-defined/non-vanishing.

\begin{figure}
	\centering
	\includegraphics[width=0.9\linewidth]{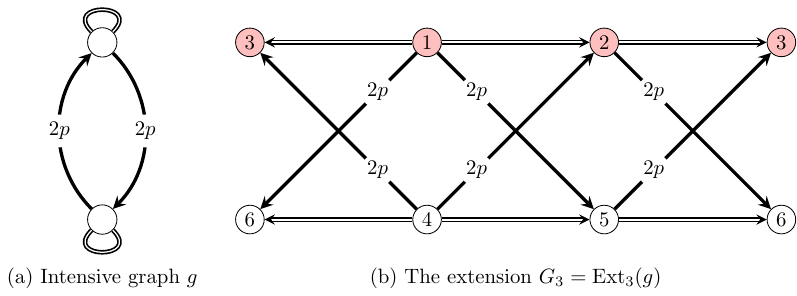}
	\caption{The graphic construction of a $(2,2p+2)$ clustering state that is not a null state of the projection Hamiltonian $H_{3}^{2p+1}$. (a) The intensive graph $g$ for this nonexample. The loops have multiplicity two, and the arrows between the distinct nodes have multiplicity $2p$. We may choose $p>2$ as large as needed. (b) For illustration purposes, the third extension is shown. For more visibility, we have duplicated the nodes $3$ and $6$ (at the left and right end). However, the nodes with the same label need to be identified. Note that the $3$-cluster 1,2,3 (shown in pink) has relative angular momentum $6$. If $p>2$, this leads to $S_3=6>2p+2$.}
	\label{fig:nonexample}
\end{figure}

\section{Discussion on Realizability}
\label{discussion_sec}
In this section, we will probe realizable clustering states. To elaborate, we look for sensible local properties besides $(k,r)$ clustering so that the wavefunctions are the ground state of some model Hamiltonian. Here, a `ground state' means a zero-energy eigenstate that is as dense as possible (i.e., has a minimal total degree). The new local properties will help us filter intensive graphs so that the graphic wavefunctions are realizable. All the Hamiltonians discussed in this section are within the pseudopotential formalism \cite{Simon_Pseudopotentials}.

\subsection{Null States of Projection Hamiltonians}
For the moment, we will limit ourselves to \emph{projection} Hamiltonians 
\begin{equation}
	H_{k+1}^{m-1} = \sum_{i_1<\cdots<i_{i+1}}\sum_{l=0}^{m-1} V_l \mathcal{P}_{k+1}^l(i_1, \cdots, i_{k+1}), \qquad V_l>0
\end{equation}
Here, $\mathcal{P}_{k+1}^l$ is the projection to the sector where the $k+1$-cluster $i_1, \cdots, i_{k+1}$ have relative angular momentum $l$. The Hamiltonian $H_{k+1}^{m-1}$ penalizes any state in which any cluster of $k+1$ particles has relative angular momentum less than $m$. Some of the most celebrated FQH states are realizable via the projection Hamiltonians:
\begin{enumerate}
	\item  The Laughlin $2m$-state is the \emph{unique} ground state of $H_{2}^{2m-1}$ \cite{Haldane_Sphere_Pseudo}.
	\item The $\mathbb{Z}_k$ Read-Rezayi state is the \emph{unique} ground state of $H_{k+1}^{1}$ (In fact, since it is impossible to have relative angular momentum equal to one, the Hamiltonian $H_{k+1}^{0}$ suffices) \cite{Read-Rezayi}.
	\item The Gaffnian state is the \emph{unique} ground state of $H_{3}^{2}$ \cite{Gaffnian}.
\end{enumerate}
In all of these examples, the Hamiltonian $H_{k+1}^{r-1}$ has a unique ground state. The uniqueness is a good sign since realistic Hamiltonians (on disk and sphere) must have a unique ground state. Unfortunately, for `large' enough $k$ and $r$, it is impossible for $H_{k+1}^{r-1}$ to have \emph{unique} ground state \cite{Simon_Pseudopotentials, pakatchi2021CFT}.  Thus, we need to eventually modify $H_{k+1}^{r-1}$ to keep the Hamiltonian realistic. Nonetheless, it would be instructive to make a few observations on the null states of $H_{k+1}^{r-1}$. 

We begin by reviewing some aspects of the \emph{pattern of zeros} formalism. Given a ground state sequence $\gv{\Psi}=(\Psi_n)$, the quantity $S_a$ is the minimal angular momentum that a cluster of $a$ particles can have in $\Psi_n$ in the thermodynamic limit $n\to \infty$ (i.e., for all $\psi_n$ with at least $a$ variables). Alternatively, $S_a$ is the minimal total degree of variables $z_1, \cdots, z_a$ in $\Psi_n$. One calls the infinite sequence $S=(S_1=0, S_2, \cdots, S_a, \cdots)$ the \emph{pattern of zeroes} of $\gv{\Psi}$ \cite{Pattern_of_Zeros}. By definition of pattern of zeros, $\gv{\Psi}$ is a zero-energy state of $H_{k+1}^{S_{k+1}-1}$ (in fact, a zero-energy state of $H_a^{S_a-1}$ for all $a$ with $S_a>0$.). In Ref. \cite{Pattern_of_Zeros}, the authors obtain various consistency conditions on the pattern of zeros as an abstract sequence. They argue that since FQH phases are classifiable, only the `first few' $S_a$ should contain new information about the phase. As such, they propose focusing on states satisfying the following:
\begin{equation}
	S_{pk+q}=\left\{\frac{p(p-1)}{2}k+pq\right\}(S_{k+1}-S_k)+pS_k+S_q, \qquad (0\leq q<k)
	\label{kcluster}
\end{equation}
Given the circumstances, $k$ and $(S_1=0, S_2, \cdots, S_k, S_{k+1})$ would partially classify the respective FHQ phase. We should also mention that, with $m=(S_{k+1}-S_k)$, filling fraction and shift are $\nu = k/m$ and $\mathcal{S}=m- 2S_k/k\in \mathbb{Z}$ in this formulation. Now, as far as $(k,r)$-clustering states are concerned, we have $S_1=S_2=\cdots=S_k=0$. It is tempting to write $S_{k+1}=r$, but that would be incorrect. We will shortly provide a $(k,r)$ clustering example that has $S_{k+1}< r$. 
Nevertheless, if a $(k,r)$ clustering state is consistent with Eq. \eqref{kcluster}, then it should satisfy the additional condition $m=S_{k+1}=r$. One can see this by comparing the filling fraction of the clustering state ($k/r$) with the one specified by the pattern of zeroes ($k/m$). In short, this discussion motivates the study of $(k,r)$ clustering states with the following pattern of zeros:
\begin{equation}
	S_{pk+q}=\left\{\frac{p(p-1)}{2}k+pq\right\}r, \qquad (0\leq q<k)
	\label{kcluster_last}
\end{equation}
We call such a $(k,r)$ clustering sequence a \emph{periodic} one. We explore periodicity more in section \ref{periodicity_sec}.  Periodic sequences, in some sense, satisfy a generalized Pauli exclusion principle: ``no more than $k$ particles can occupy $r$ consecutive orbitals in the LLL.'' In that section, we will obtain additional conditions on the intensive graphs to make the extension sequence periodic.

In general, if $\Psi$ is a zero-energy state of $H_{k+1}^{r-1}$, then it would satisfy \eqref{fusion_eqs}. Following Ref. \cite{Simon_projection}, we say a zero-energy state of $H_{k+1}^{r-1}$ is \emph{proper}, if it does not vanish when $k$ particles are fused. The discussion in subsection \ref{clustering_sec} shows that the densest proper zero-energy states of $H_{k+1}^{r-1}$ are $(k,r)$ clustering. However, there are many cases where the densest zero-energy state is improper. For example, as noted in Ref. \cite{Simon_projection}, consider $k=2$ and $r=6$. The Laughlin $2$-state and $(2,6)$ clustering states are both zero-energy states of $H_{3}^{5}$. The Laughlin state is improper, while clustering states are proper. In both cases, any cluster of $3$ electrons has relative angular momentum $6$. Now, a $(2,6)$ clustering state has $N=2n$ particles and $N_\phi^{\text{proper}}=6n-6$ flux quanta. On the other hand, Laughlin $2$-state, with same number of particles, has $N_\phi^{\text{Laughlin}}=4n-2$. Thus the Laughlin state has a higher density for all $n> 2$.

A secondary subtlety is that not every $(k,r)$ clustering state is a null state of the projection Hamiltonian $H_{k+1}^{r-1}$. In general, $S_{k+1}\neq r$ and $H_{k+1}^{r-1}$ rejects any state with $S_{k+1}< r$. For example, consider the intensive graph drawn in Fig. \ref{fig:nonexample}.(a). This leads to a $(2,2p+2)$ clustering sequence $\gv{\Psi}^{(g)}$, where $\Psi^{(g)}_n = \mathrm{SGP}(G_n)$ and $G_n=\mathrm{Ext}_n(g)$. For illustration purposes, we have drawn $G_3$ in Fig. \ref{fig:nonexample}.(b). As can be seen in the figure, when $p>2$, the minimal angular momentum of a $3$-cluster is equal to $S_3=6$, not $r=2p+2$. As a result, $\Psi_n^{(g)}$ cannot be the zero-energy state of $H_{3}^{2p+1}$. However, if a $(k,r)$ intensive graph $g$ is such that $\gv{\Psi}^{(g)}$ is periodic, then it will be a zero-energy state of $H_{k+1}^{r-1}$. Similarly, Jacks are null states of $H_{k+1}^{r-1}$ because they are \emph{periodic} clustering states.

\subsection{Separability and Modified Hamiltonians $H_\chi$}
The modified Hamiltonian presented here was pioneered by us elsewhere \cite{pakatchi2021CFT}. In Ref. \cite{pakatchi2021CFT}, the subject of study are the $\mathbb{Z}_k^{(r)}$-algebras and their wavefunctions. The Jacks $\mathcal{J}_n^{(k,r)}$ are examples of such wavefunctions \cite{estienne1, estienne2}. We have shown that $\mathbb{Z}_k^{(r)}$ wavefunctions satisfy a new property called \emph{separability}. In turn, we modify to $H_{k+1}^{r-1}$ relying on this local property. In the other paper, we also provide evidence that $H_\chi$ is \emph{likely} to have a unique ground state at least when $r=4$. While Ref. \cite{pakatchi2021CFT} is the source for these arguments, \ref{Zk2_appendix} provides an overview. In this paper, we take separability as an abstract property and study its implications on graphs. In the end, to stay as close to Jacks as possible, we filter intensive graphs so that the wavefunctions are periodic and separable. In particular, any such construction will be realizable by a model Hamiltonian $H_\chi$ (hopefully, uniquely). 
We now delve into the definition of separability.

We begin by introducing some terminology. Denote by $\mathcal{T}_{k+1}^{m}$ the space of symmetric homogeneous translational invariant polynomials in $k+1$ variables and degree $m$. We sometimes use an alternative notation $\ket{P; z_1, \cdots, z_{k+1}}$ for a polynomial $P(z_1, \cdots, z_{k+1})$ in $\mathcal{T}_{k+1}^{m}$. The ket notation is to remind us of the Hilbert space structure of $\mathcal{T}_{k+1}^{m}$: Concretely, the inner product of $\ket{P_1}, \ket{P_2}\in \mathcal{T}_{k+1}^m$ is given by
\begin{equation}
	\braket{P_1}{P_2} = \int \prod_{i=1}^{k+1} dx_idy_i \: \overline{P_1(z_1, \cdots, z_{k+1})}P_2(z_1, \cdots, z_{k+1}) \exp\left(-\frac{1}{2\ell_B^2}\sum_{i=1}^{k+1}|z_i|^2\right)
\end{equation}
This the Bargmann space structure \cite{girvin1984formalism} that $L^2$ inner product of the LLL induces on $\mathcal{T}_{k+1}^m$.

Let $\gv{\Psi}=(\Psi_n)$ be a $(k,r)$ clustering sequence and let $m=S_{k+1}$. We say $\gv{\Psi}$ is separable if there exists a non-zero $\ket{\chi;z_1, \cdots, z_{k+1}}\in \mathcal{T}_{k+1}^{m}$, with normalization $\ket{\chi;1, 0^{\times k}}=1$, such that for all $n$:
\begin{equation}
	(\mathcal{P}_{k+1}^m\Psi_n)(z_1, \cdots, z_{nk}) = \ket{\chi;z_1, \cdots, z_{k+1}}Q(z_{\mathrm{cm}};z_{k+1}, \cdots, z_{nk})
	\label{separable_eq}
\end{equation}
Here, $Q$ is some polynomial and $z_{\mathrm{cm}}=(z_1+\cdots+z_{k+1})/(k+1)$. Given the conditions, we have shown in \ref{separability_consq_appendix} that $\deg \chi=S_{k+1}=m=r$ by necessity. In particular, $\gv{\Psi}$ is a null state of $H_{k+1}^{r-1}$. We also show that Eq. \eqref{separable_eq} takes the following nice form
\begin{equation}
	(\mathcal{P}_{k+1}^r\Psi_{n+1})(z_1, \cdots, z_{nk}) = 
	\ket{\chi;z_1, \cdots, z_{k+1}}\prod_{i\geq k+1}(z_i-z_{\mathrm{cm}})^r\Psi_n(z_{\mathrm{cm}}, z_{k+1}, \cdots, z_{nk})
\end{equation}

We may now describe our modified Hamiltonian $H_\chi$. We use the notation $I=i_1<\cdots<i_{k+1}$ and $z_{I}=z_{i_1}, \cdots, z_{i_{k+1}}$. Let $\ket{\chi}\in \mathcal{T}_{k+1}^{r}$ and $\ket{\chi; 1, 0^{\times k}}=1$. The modified Hamiltonian $H_\chi$, parametrized by such $\chi$, is given by
\begin{equation}
	\label{separable_hamiltonian_eq}
	H_{\chi} = H_{k+1}^{r-1}+
	V_r
	\sum_{I}
	\left(1-\frac{\ketbra{\chi;z_I}{\chi;z_I}}{\braket{\chi}{\chi}}\right)
	\mathcal{P}_{k+1}^r(z_I), \qquad (V_r>0)
\end{equation}
In the new term, one first projects to the sector where $k+1$ particles have relative angular momentum $r$. Then a secondary projection, this time in $\mathcal{T}_{k+1}^{r}$, takes us to the subspace orthogonal to $\chi$. Thus, the zero-energy states $\Psi$ of $H_\chi$ are those null states of $H_{k+1}^{r-1}$ that satisfy
\[
\mathcal{P}_{k+1}^r\Psi(z_1, \cdots, z_{nk}) = \ket{\chi;z_1, \cdots, z_{k+1}}Q(z_{\mathrm{cm}};z_{k+1}, \cdots, z_{nk})
\]
for some polynomial $Q$. We claim that no improper null state of $H_{k+1}^{r-1}$ can be a null state of $H_\chi$. To see this, let $\Psi'$ be an improper null state. Suppose for some $\ket{\chi'}\in \mathcal{T}_{k+1}^r$ and some polynomial $Q'$, we can factorize $\mathcal{P}_{k+1}^{r}\Psi'$ as
\[
\mathcal{P}_{k+1}^r\Psi'(z_1, \cdots, z_{nk}) = \ket{\chi';z_1, \cdots, z_{k+1}}Q'(z_{\mathrm{cm}};z_{k+1}, \cdots, z_{nk})
\]
Now, since $\Psi'$ is improper, we have $\ket{\chi';1, 0^{\times k}}=0$. Therefore, even if the factorization above is possible, we can never have $\chi'=\chi$, and the claim follows. As a corollary, the densest null states of $H_\chi$ are separable (and, in particular, $(k,r)$ clustering). In other words, $\chi$-separable states are realizable by $H_\chi$. If $H_\chi$ has a unique ground state, the realization is unique as well.

We finish this section by providing a family of intensive graphs that lead to periodic and separable extensions. A \emph{$(k,l,m)$ saturated} graph $g$ has $k$ nodes, $l$ loops for each node, $kl=$even, and any ordered pair of distinct nodes is connected with an arrow of multiplicity $0<m\leq l$. In other words, the multiplicity function is the following $k\times k$ matrix
\begin{equation}
	\mu(i,j)=
	\begin{cases}
		l & i=j\\
		m\leq l & i\neq j	
	\end{cases}, \qquad (kl=\text{even})
\end{equation}
Then $\gv{\Psi}^{(g)}$ is periodic (discussed in section \ref{periodicity_sec}, proved in \ref{saturated_proper_appendix}) and separable (discussed in section \ref{Separability_sec}, proved in \ref{last_appendix}). Moreover, if $h=g^c$ is the disjoint union of $c$ copies of $g$, then $\gv{\Psi}^{(h)}$ is also separable and periodic. Let us present a few concrete known examples:
\begin{enumerate}
\item The $(k=2, l=2r, m=-)$ saturated graph $\ell_r$ is the intensive graph of the Laughlin $2r$-state (i.e. a single loop of multiplicity $2r$).
\item Using the representation of Ref. \cite{cappelli}, the $k$-fold disjoint union $\ell_1^{k}$ is an intensive graph for $\mathbb{Z}_k$ Read-Rezayi states.

\item The $(k=2, l=2,m=1)$ saturated graph, denoted $\jmath$, is the intensive graph for Gaffnian.
\item The $(k=2, l=2,m=2)$ saturated graph is the intensive graph for Haffnian.
\item One can show that $\mathcal{J}_2^{(2c,3)}$ is the SGP of $T_2\otimes \jmath^c$ for all $c$ (see Example: Binary Cubics, subsection 7.4, in Ref. \cite{pakatchi2021CFT}). This loosely suggests that the Jacks $\mathcal{J}_n^{(2c,3)}$ are presentable with the intensive graph being $\jmath^{c}$.
\end{enumerate}
The remainder of this paper is a thorough discussion, justification, and generalization of the content provided in this section.

\newpage
\section{Discussion on Periodicity}
\label{periodicity_sec}
In this section, we will expand on the discussion of section \ref{discussion_sec} and take a closer look at periodicity. We begin by interpreting periodicity in terms of the free bosonic states in the LLL and their `orbital' structure. We then translate periodicity into graph theory using two concepts: orientations and cuts. In the end, we find the graphic implications of periodicity and filter the intensive graphs accordingly.

\subsection{Periodicity and Highest Root Partitions}
\label{highestrootpartition_sec}
\subsubsection{Root-Partitions}
It is useful to interpret the $N_\phi+1$ states in the LLL (in the sphere geometry) as orbitals. The one-particle state corresponding to the $m$th orbital ($0\leq m\leq N_\phi$) is $\phi_m(z)\propto z^m$ and has angular momentum $-\frac{1}{2}N_\phi + m$. Knowing the occupation numbers of the orbitals, we may specify a free bosonic state. Concretely, let $n_m$ denote the number of bosons in $m$th orbital, and make a list $n=(n_0,n_1, \cdots, n_{N_\phi})$ which we refer to as occupation (list). We denote the partition associated with an occupation $n$ by $\mathrm{Part}(n)$:
\begin{equation}
	\mathrm{Part}(n) = (
	\underbrace{N_\phi, \cdots, N_\phi}_{\times n_{N_\phi}}, \cdots, \underbrace{2,\cdots, 2}_{\times n_2}, 
	\underbrace{1,\cdots, 1}_{\times n_1}, 
	\underbrace{0,\cdots, 0}_{\times n_0}
	)
\end{equation}
Conversely, given any partition $\lambda=(\lambda_1, \cdots, \lambda_N)$ with $\lambda_1\leq N_\phi$, the frequency list $\mathrm{Freq}(\lambda)$ gives the respective occupation. Now, the free boson state with occupation $n$ is the \emph{symmetric monomial} $m_{\lambda}$ where $\lambda=\mathrm{Part}(n)$:
\begin{equation}
	m_{\lambda}(z_1, \cdots, z_N) = \frac{1}{\prod_{p=0}^{N_\phi}n_p!}\mathscr{S}\Big[z_1^{\lambda_1}z_2^{\lambda_2}\cdots z_N^{\lambda_N}\Big]
\end{equation}

Let $\Psi$ be a wavefunction with $N$ particles on the sphere with $N_\phi$ flux quanta. Additionally, suppose $\Psi$ is homogeneous of degree $M$ (i.e. has angular momentum $-\frac{1}{2}NN_\phi+M$). We denote the Hilbert space of such wavefunctions by $\mathcal{H}(M, N_\phi, N)$. We want to decompose $\Psi$ in terms of symmetric monomials $m_\lambda$. The set of relevant partitions $\pi(M,N_\phi, N)$ consists of $\lambda=(\lambda_1, \lambda_2, \cdots)$ satisfying $\sum_i \lambda_i=M$, $\lambda_1\leq N_\phi$, and $\lambda_{N+1}=0$. The decomposition of $\Psi$ in terms of the free bosonic states is called the \emph{root-decomposition} of $\Psi$:

\begin{equation}
	\Psi(z_1, \cdots, z_N)=\sum_{\lambda\in \pi(M,N_\phi, N)}c_{\lambda}m_{\lambda}(z_1, \cdots, z_N)
\end{equation}
Here, the coefficients $c_\lambda$ are $\mathbb{C}$-numbers. We call $\lambda$ a \emph{root-partition} of $\Psi$ if $c_{\lambda}\neq 0$.

Suppose $\Psi=\sum_{\lambda}c_{\lambda} m_{\lambda}$ is an $(N,N_\phi)$ uniform state. Uniformity leads to relations between the coefficients $c_\lambda$. For one, translational invariance results in the following identities (see \ref{umbral1}): Given any occupation $n=(n_0, \cdots, n_{N_\phi})$ such that $\mathrm{Part}(n)\in \pi(\frac{1}{2}NN_\phi-1, N_\phi, N)$ we have
(Notation: $e^m=\mathrm{Freq}(m)$, i.e., a single boson occupying the $m$th orbital):
\begin{equation}
	\sum_{m=1}^{N_\phi}m \: n_{m-1}\: c_{\mathrm{Part}(n-e^{m-1}+e^m)}=0
	\label{constraint1}
\end{equation}
To describe implications of the self-adjoint property, define: $(n_0, n_1, \cdots, n_{N_\phi})^t=(n_{N_\phi}, \cdots, n_1, n_0)$ (i.e. flipped upside-down). Then we have
\begin{equation}
	c_{\mathrm{Part}(n)}=c_{\mathrm{Part}(n^t)}
	\label{constraint2}
\end{equation}
Although the full Hilbert space $\mathcal{H}(\frac{1}{2}NN_\phi, N_\phi, N)$ is a high-dimensional space, due to the linear constraints \eqref{constraint1} and \eqref{constraint2}, the number of coefficients $c_\lambda$ that we need to determine independently is much smaller. One approach to constructing trial FQH states would be to specify the special root-partitions that will fix the entire wavefunction.

\subsubsection{Dominance and Periodicity}
There is a natural partial order on the free bosonic states in $\mathcal{H}(M, N_\phi, N)$ called \emph{dominance}. Let $m,m'$ be two orbitals with $m-m'>1$. Suppose $n$ is a configuration with $n_m\neq 0$ and $n_{m'}\neq 0$. The two-particle operation which moves a boson from orbital $m$ to orbital $m-1$ and simultaneously moves one from $m'$ to $m'+1$ is called a \emph{squeezing move}. The resulting configuration $\widetilde{n}$ is
\begin{equation}
	\widetilde{n}_l = \begin{cases}
		n_{l}-1 & l=m,m'\\
		n_{l}+1 & l=m-1, m'+1\\
		n_l & \mathrm{otherwise}
	\end{cases}
\end{equation}
In general, given two partitions $\lambda, \mu$ of $M$, if $\mu$ can be obtained from $\lambda$ by a series of squeezing moves, then we say $\lambda$ \emph{dominates} $\mu$ and write $\lambda\succeq \mu$.

A wavefunction $\Psi$ has a \emph{highest root-partition} $\Lambda$ if $\Lambda$ is a root-partition and it dominates all other root-partitions. For example, Jack polynomials $\mathcal{J}_n^{(k,r)}$ \eqref{jack_eq} have a highest root-partition $\Lambda(n,k,r)$. Let us recall the definition for this partition (Notation: $x^{\times y}$ is $x$ repeated $y$ times):
\begin{equation}
	\begin{aligned}
		\Lambda(n,k,r)&=([(n-1)r]^{\times k}, \cdots, [2r]^{\times k}, r^{\times k}, 0^{\times k})\\
		&=\mathrm{Part}(k, 0^{\times r-1}, k, 0^{\times r-1}, \cdots, k, 0^{\times r-1},k)
	\end{aligned}
\end{equation}
In general, if $\gv{\Psi}=(\Psi_n)$ is a $(k,r)$ is a clustering sequence, then $\Lambda(n,k,r)$ is a root-partition of $\Psi_n$. Moreover, as we argue in subsection \ref{ori}, if $\Psi_n$ admits a highest root-partition $\Lambda$, then $\Lambda=\Lambda(n,k,r)$ by necessity.

The partition $\Lambda(n,k,r)$ has a remarkable interpretation. Among $\lambda\in \pi(n(n-1)\:kr/2,\:(n-1)r,\: nk)$, the partition $\Lambda(n,k,r)$ is the \emph{unique} one satisfying the \emph{$(k,r)$ exclusion principle}: ``No consecutive $r$ orbitals can be occupied by more than $k$ bosons.'' 
To elaborate, a general partition $\lambda$ with $\lambda_{N+1}=0$ is called \emph{$(k,r,N)$-admissible} \cite{Feigin} if
\begin{equation}
	\lambda_i - \lambda_{i+k}\geq r, \qquad (1\leq i\leq N-k)
\end{equation}
The frequencies $\mathrm{Freq}(\lambda)$ satisfy the generalized Pauli principle if and only if $\lambda$  is $(k,r,N)$-admissible. Among all $(k,r,nk)$ admissible partitions, $\Lambda(n,k,r)$ is the densest (i.e. $M=\lambda_1+\cdots+\lambda_N$ is the smallest possible).

\subsubsection{Periodicity in Terms of Pattern of Zeroes}
For a partition $\lambda=(\lambda_1,\lambda_2, \cdots, \lambda_N)$ we define the $a$th \emph{partial sum} and \emph{reverse partial} sum as
\begin{subequations}
	\begin{align}
		P_a(\lambda) &= \lambda_1+\cdots+\lambda_a\\ P_a^*(\lambda)&=\lambda_{N}+\lambda_{N-1}+\cdots+\lambda_{N-a+1}
	\end{align}
\end{subequations}
It can be shown that (see \cite{macdonald} (1.15--16)) $\lambda\succeq \mu$ if and only if $P_a(\lambda)\geq P_a(\mu)$ for all $a$. Additionally, since $P_a(\lambda)\geq P_a(\mu)$ is the same as $P^*_{N-a}(\lambda)\leq P^*_{N-a}(\mu)$, an alternative formulation would be $P_a^*(\lambda)\leq P_a^*(\mu)$ for all $a$. We also  note that, with $\Lambda= \Lambda(n,k,r)$, we have ($a=pk+q$, with $0\leq q<k$):
\begin{equation}
	\label{patternZ_eq}
	P_a^*(\Lambda)
	=\left\{\frac{p(p-1)}{2}k+pq\right\}r:=\mathcal{S}_k(a)r
\end{equation}

Recall that, for a wavefunction $\Psi$, the pattern of zeros $S_a(\Psi)$ is the smallest total power that the variables $z_1, \cdots, z_a$ can have in $\Psi$. We may calculate $S_a(\Psi)$ as follows:
\begin{equation}
	S_a(\Psi)=\min_{\lambda\in \mathrm{RP}(\Psi)}P^*_a(\lambda)
\end{equation}
where $\mathrm{RP}(\Psi)$ stands for the set of root-partitions of $\Psi$. Previously we defined periodicity of a $(k,r)$ clustering sequence $\gv{\Psi}$ in terms of its pattern of zeroes: $S_a=\mathcal{S}_k(a)r$. The revelation that $\mathcal{S}_k(a)r=P_a^*(\Lambda(n,k,r))$ now shows that $\gv{\Psi}$ is periodic if and only if it admits a highest root-partition.

Let us summarize why periodicity is a natural property for us to consider. For one, the Jacks are periodic. Moreover, we have $S_1=\cdots=S_k=0$ and $S_{k+1}=r$. Therefore, periodic clustering states are null states of the projection Hamiltonian $H_{k+1}^{r-1}$. Additionally, in the sense of their highest root-partition, periodic states are the densest states satisfying a $(k,r)$ exclusion principle. Next, we will look for additional conditions on the intensive graphs that would result in periodicity. We will discuss the notions of \emph{reorientations} and \emph{cuts} that are central to a graphic treatment of root-decomposition.

\subsection{Reorientations \& Root Decomposition}\label{ori}
\begin{figure}
	\centering
	\includegraphics[scale=.75]{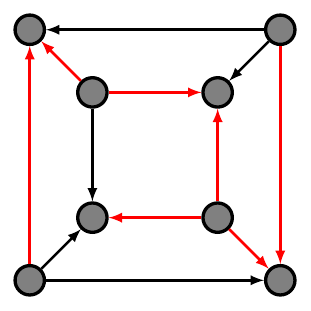} 
	\includegraphics[scale=.75]{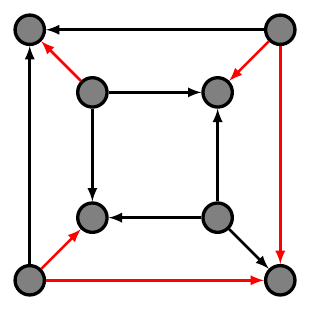} 
	
	\includegraphics[scale=.75]{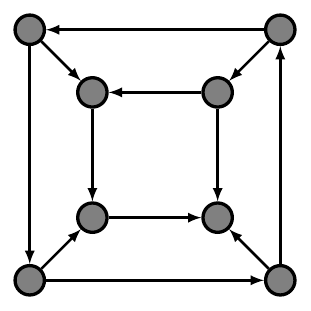} 
	\includegraphics[scale=.75]{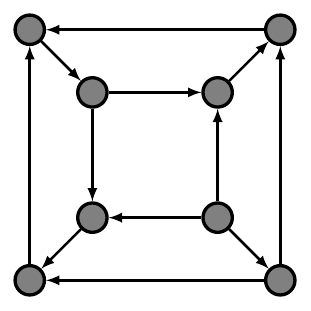} 
	
	\caption{The main graph $G$ in the example is the shard $T_2\otimes \mathfrak{g}(1,0,1,1)$ (Fig. (\ref{extensive_graphs_fig}.f)). In the first row, two reorientations of $G$ are illustrated. Here $\omega$ sends the red arrows to $-1$ and the black ones to $+1$. Directly underneath each symbolic depiction, the corresponding reoriented graph $G_\omega$ is drawn.}
	\label{orifig}
\end{figure}

In this subsection, we relate `reorientations' of the graph $G_n=T_n\otimes g$, to the root-decomposition of $\Psi_n^{(g)}$. Given a loopless graph $G=(V,\mu)$, let us define the \emph{arrow set} $E$ as
\begin{equation}
	E =\bigcup_{i,j\in V} E_{i,j}, \qquad E_{i,j}=\{
	\underbrace{i\to j,\, i\to j,\, \cdots,\, i\to j}_{\times \mu(i,j)}
	\}
\end{equation}
with $E_{i,j}=\emptyset$ if $\mu(i,j)=0$. Given each $e\in E_{i,j}$ we write $s(e)=i$ (source) and $t(e)=j$ (target). For an arrow $e=s(e)\to t(e)$, we define $-e\equiv s(e)\leftarrow t(e)$. 
From the above considerations, we can determine $\mu$ from $E$ and vice versa. As a result, we can also describe $G$ as the pair $G=(V,E)$. Some concepts, like reorientations, are easier to define using arrow sets (compared to multiplicity functions). A \emph{reorientation} is a function $\omega: E\to \{+1,-1\}$. For each reorientation, we define a new arrow set
\begin{equation}
	E^{\omega}=\{\omega(e)\,e\mid e\in E\}
\end{equation}
Intuitively, a reorientation $\omega$ visits each arrow of $G$ and either keeps its direction or reverses it. If $G$ has the data $(V,E)$, we define the \emph{reoriented graph} $G^\omega$ via the data $(V, E^\omega)$ (see Fig. \ref{orifig}). Let us now introduce some terminology regarding reorientations:

\begin{dfn}
	The two constant functions $\omega(E)=+ 1$ and $\omega(E)=-1$ are called the \emph{trivial} and \emph{reversal} reorientations respectively. Note that if $\omega$ is trivial, then $G=G^{\omega}$. If $\omega$ is the reversal orientation, we use the notation $-G:=G^{\omega}$.
\end{dfn}
\begin{dfn}
	The \emph{sign} of a reorientation is defined as
	\begin{equation}
		\mathrm{sgn}(\omega) = \prod_{e\in E} \omega(e)
	\end{equation}
	i.e. $\mathrm{sgn}(\omega)$ is $-1$ only when $G$ and $G^\omega$ disagree on the direction of an odd number of arrows in $\mathcal{F}(G)$.
\end{dfn}

\begin{dfn}
	For any orientation $\omega$, define the partition $T(\omega)$ ($\mathrm{Ord}$ sorts a sequence in non-decreasing fashion)
	\begin{equation}
		T(\omega) = \mathrm{Ord}(d^+_1(\omega),d^+_2(\omega), \cdots, d^+_N(\omega))
	\end{equation}
	where $d_i^+(\omega)$ is the out-degree of node $i$ in the reoriented graph $G^{\omega}$. We call $T(\omega)$ the \emph{orientation type} of $\omega$.
\end{dfn}
\begin{dfn}
	Let $\lambda$ be an orientation type. Define
	\begin{equation}
		\Omega_{\lambda}^{\pm}=\{\omega\mid T(\omega)=\lambda, \mathrm{sgn}(\omega)=\pm1\}
	\end{equation}
	An orientation type $\lambda$ is called \emph{balanced} if $|\Omega_\lambda^+|=|\Omega_\lambda^-|$, and \emph{skew} otherwise.
\end{dfn}

With these definitions, it can be shown that \cite{sabidussi}
\begin{subequations}
	\begin{align}
		\mathrm{SGP}(G) &= \sum_{\lambda}\sigma_{\lambda}m_{\lambda}\\ \sigma_{\lambda}&=
		\Big(|\Omega_\lambda^+|-|\Omega_\lambda^-|\Big) \prod_{p\geq 0}\mathrm{Freq}(\lambda)_p!
	\end{align}
\end{subequations}
Here, the sum runs over the orientation types $\lambda$ of $G$. Therefore, $\sigma_{\lambda}\neq 0$ only if $\lambda$ is a skew orientation type. In other words, the root-partitions of $\mathrm{SGP}(G)$ are the skew orientation types of $G$.

For the rest of this subsection, we will focus on graphs $G=T_n\otimes g$, where $g$ is a $(k,r)$ intensive graph with $c$ components. Since $\Psi_n^{(g)}=\mathscr{N}(g)\mathrm{SGP}(G)$, where $\mathscr{N}(g)=[n!^{c}k!^n]^{-1}$, the root-decomposition is
\begin{subequations}
	\begin{align}
		\Psi_n^{(g)} &= \sum_{\lambda}c_{\lambda}m_{\lambda}\\
		c_{\lambda}&=\frac{\sigma_\lambda}{n!^{c} k!^n}
		=\frac{\prod_{p\geq 0}\mathrm{Freq}(\lambda)_p!}{k!^n}
		\frac{(|\Omega_\lambda^+|-|\Omega_\lambda^-|)}{n!^c}
	\end{align}
\end{subequations}
In particular, since the trivial orientation of $G$ has type $\Lambda:=\Lambda(n,k,r)$, and $\prod_{p\geq 0}\mathrm{Freq}(\Lambda)_p=k!^n$, we find
\[
c_{\Lambda}=\frac{|\Omega^+_{\Lambda}|-|\Omega^-_{\Lambda}|}{n!^c}
\]
To find $c_{\Lambda}$, we need to talk about the reorientations that have type $\Lambda:=\Lambda(n,k,r)$. The proof for what is about to follow can be found in \ref{oriproof_appendix} (for the general case $c\geq 1$). For demonstration purposes, we assume $c=1$. Let $\omega$ be an reorientation with $T(\omega)=\Lambda$. Given any two maximum independent sets $[I], [J]$ with $1\leq I<J\leq n$, we have
\[
G^{\omega}|_{[I]\cup [J]}\simeq \pm G_2
\]
i.e., the bonds of $G^{\omega}$ are either the shard $G_2$ or its reversal $-G_2$. As $g$ is even, meaning number of arrows in the shard (i.e. $kr$) is even, all of these orientation have $\mathrm{sgn}(\omega)=+1$. Moreover, there are a total of $n!$ orientations of type $\Lambda$. Therefore,
\begin{equation}
	\Psi_n^{(g)} = m_{\Lambda(n,k,r)}+\sum_{\lambda\neq \Lambda}c_{\lambda}m_{\lambda}
\end{equation}
This brings us to the discussion on the highest root-partition. Partitions can be equipped with a total order called \emph{lexicographic} order. One writes $\lambda\geq_L \mu$ if either $\lambda=\mu$ or the first non-vanishing difference $\lambda^*_i-\mu_i^*$ is negative. Here, $\lambda_i^*=\lambda_{N-i+1}$. It can be shown that if $\lambda$ is an orientation type of $G$ (skew or balanced), then $\Lambda\geq_L \lambda$. It is a general fact of partitions that $\lambda\succeq \mu$ results in $\lambda\geq_L \mu$. Consequently, if $\Psi_n^{(g)}$ has a highest root-partition $\Lambda$, then $\Lambda=\Lambda(n,k,r)$ by necessity. The same argument works for any clustering state (graphic or not).

Reorientations are a global concept and, thus, generally hard to work with. On the other hand, cuts, which are a local concept, are better suited for crafting a condition on the intensive graphs $g$. As such, we will introduce cuts next. Subsequently, we use them to translate periodicity and separability as conditions on $g$.

\subsection{Cuts \& Periodicity}
\label{cuts}
Let $G=(V,\mu)$ be a graph of order $N$ with the nodes labeled by $1,2,\cdots, N$. An $a$-cut $C$ of a graph $G=(V,\mu)$, denoted by $C\vdash_a G$, is a subset $C\subset V$ with $|C|=a$. It is a `cut' in the sense that it partitions the node set $V$ into $C$ and $V-C$. The number of arrows in the restricted graph $G|_C$ is denoted by
\begin{equation}
	w(C)=\sum_{i,j\in C}\mu(i,j)
\end{equation}
and is called the \emph{weight} of the cut.

Let us rewrite the SGP of a graph in terms of its $a$-cuts. Throughout, we use the shorthand notations $\vv{x}, \vv{y}$ for $x_1, x_2, \cdots x_a$ and $y_1, y_2,\cdots, y_{N-a}$ respectively. For a fixed cut $C\vdash_a G$, we define
\begin{subequations}
	\begin{align}
		\Gamma_C^{(L)}&=
		\prod_{i,j\in C}(x_{i}-x_{j})^{\mu(i,j)}\\
		\Gamma_C^{(R)}&=
		\prod_{I,J\notin C}(y_{I}-y_{J})^{\mu(I,J)}\\
		\Gamma_C^{(\times)}
		&=
		\prod_{i\in C}\prod_{J\notin C}
		(y_{J}-x_i)^{\mu(J,i)}
		(x_i-y_{J})^{\mu(i,J)}
	\end{align}
\end{subequations}
Now with $\widetilde{\Gamma}_C=	\Gamma^{(L)}_C\,
\Gamma_C^{(R)}\,
\Gamma_C^{(\times)}$, and $\Gamma_C=\mathscr{S}_{\vv{x}}\mathscr{S}_{\vv{y}}\widetilde{\Gamma}_C$ ($\mathscr{S}$ being the symmetrization operator), the SGP becomes
\begin{equation}
	\mathrm{SGP}(G)(\vv{x},\vv{y})=
	\sum_{C\vdash_a G}\Gamma_C(\vv{x},\vv{y})
\end{equation}
Let $\Gamma_C^m(\vv{x}, \vv{y})=\mathcal{P}_a^m(\vv{x})\Gamma_C$ be the component of $\Gamma_C$ in which the $\vv{x}$ variables have relative angular momentum $m$. By definition of weight, we have $\Gamma^m_C=0$ for all $m<w(C)$. Moreover, defining $\bar{x}=(x_1+\cdots+x_a)/a$, one computes
\begin{equation}
	\Gamma_C^{w(C)}(\vv{x}, \vv{y})=
	\mathrm{SGP}(G|_C)(\vv{x})\mathscr{S}_{\vv{y}}\left\{
	\Gamma_C^{(R)}(\vv{y})
	\prod_{J\notin C}
	(y_{J}-\bar{x})^{v^+_C(J)}
	(\bar{x}-y_{J})^{v^-_C(J)}
	\right\}
	\label{projection_cut}
\end{equation}
Here, $v^+_C(J)=\sum_{i\in C}\mu(J,i)$, and $v^-_C(J)=\sum_{i\in C}\mu(i,J)$. We say $C$ is a \emph{vanishing cut} if $\mathrm{SGP}(G|_C)(\vv{x})=0$. Although the specifics of the leading projection are not important to the discussions of the pattern of zeros, the details of the leading $(k+1)$-projection are crucial for separability (see section \ref{Separability_sec}).

We aim to find a condition on intensive graph $g$ so that $\gv{\Psi}^{(g)}$ is periodic. Firstly, since $\Lambda:=\Lambda(n,k,r)$ is a skew orientation type of $\mathrm{Ext}_n(g)$, we always have $S_a\leq \mathcal{S}_k(a)r$. We need to design conditions on $g$ so the inequality $S_a\geq \mathcal{S}_k(a)r$ also holds. Interpreting $S_a$ as the minimal relative angular momentum of an $a$-cluster, it suffices to demand that for all $C\vdash_a G$, the following inequality holds:
\begin{equation}
	w(C)\geq \mathcal{S}_k(a)r
	\label{weight_condition_EQ}
\end{equation}
This is not a necessary condition but a sufficient one. For example, the existence of a vanishing cut $C$ with $w(C)<\mathcal{S}_k(a)r$ does not ruin periodicity. Nevertheless, dealing with vanishing cuts is complicated, and we wish to avoid it. We can reformulate \eqref{weight_condition_EQ} in terms of the intensive graph $g$:
\begin{dfn}
	Let $g=(V_g, \mu_g)$ be a $(k,r)$ intensive graph. Let $\mathcal{U}=\{U_1, \cdots, U_\ell\}$ be a set of subsets of $V_g$. Define $||\mathcal{U}||:=\sum_{i=1}^\ell |U_i|$ and
	\begin{equation}
		\mu(\mathcal{U}):=\sum_{1\leq i<j\leq \ell}\sum_{u\in U_i}\sum_{v\in U_j}\mu_g(u,v)
	\end{equation}
	We say $g$ is \emph{proper} if for all $\mathcal{U}$ the following inequality holds:
	\begin{equation}
		\mu(\mathcal{U})\geq \mathcal{S}_k(||\mathcal{U}||)\,r
	\end{equation}
	To see the relation to cuts, note that to each $\mathcal{U}$, with $||\mathcal{U}||=a$, we can associate a cut $C(\mathcal{U})\vdash_a \mathrm{Ext}_n(g)$:
	\begin{equation}
		C(\mathcal{U}) = \bigcup_{J=1}^{n}\{J\}\otimes U_J
	\end{equation}
	We have $w(C(\mathcal{U}))=\mu(\mathcal{U})$, and any cut $C\vdash_a \mathrm{Ext}_n(g)$ is equal to $C(\mathcal{U})$ for some $\mathcal{U}$.
\end{dfn}

Let us finish this section by providing a concrete example of proper intensive graphs. Recall that a \emph{$(k,l,m)$ saturated} graph $g$ has $k$ nodes, $l$ loops for each node, $kl=$even, and any ordered pair of distinct nodes is connected with an arrow of multiplicity $0<m\leq l$. In other words, the multiplicity function is the following $k\times k$ matrix
\begin{equation}
	\mu(i,j)=
	\begin{cases}
		l & i=j\\
		m\leq l & i\neq j	
	\end{cases}, \qquad (kl=\text{even})
\end{equation}
In \ref{saturated_proper_appendix}, we prove the properness of saturated graphs. Moreover, given $c$ saturated graphs $g_i$ with data $(k_i,l_i,m_i)$, and $r=l_i+(k_i-1)m_i$ fixed, the disjoint union $g=g_1\sqcup \cdots\sqcup g_c$ is proper as well. This is a consequence of the following proposition, proven in \ref{disjoint_appendix}:
\begin{prop}
	\label{disjoint_prop}
	Let $g=g_1\sqcup g_2\sqcup \cdots\sqcup g_c$ with $g_i$ a $(k_i, r)$ connected intensive graph.
	\begin{enumerate}
		\item If for all $i$, the sequence $\gv{\Psi}^{(g_i)}$ is periodic, then $\gv{\Psi}^{(g)}$ is periodic as well.
		\item If $g_i$ is proper for all $i$, then $g$ is also proper.
	\end{enumerate}
\end{prop}
\noindent
Beyond the disjoint union of saturated graphs, we will neither attempt to check the properness of any other class of intensive graphs nor classify proper intensive graphs. The classification problem appears to be a formidable combinatorical problem.

\section{Separability in Terms of Graphs}
\label{Separability_sec}
Among $(k,r)$ intensive graphs, those relevant to separability are all \emph{circulant}. Given a graph $g=(U,\mu)$, with $|g|=|U|=k$, we can label the nodes by elements of $\mathbb{Z}_k$, i.e. $0,1,\cdots, k-1$. The multiplicity $\mu$ can then be interpreted as a $k\times k$ matrix called the \emph{adjacency} matrix. The \emph{circulant graph} $\mathfrak{g}(\xi)$, with $\xi=(\xi_0, \xi_1, \cdots, \xi_{k-1})$ being a vector, is defined via the adjacency matrix
\begin{equation}
	\mu = \begin{pmatrix}
		\xi_0 & \xi_1 & \cdots & \xi_{k-2} & \xi_{k-1}\\
		\xi_{k-1} & \xi_0 & \xi_{1} & &\xi_{k-2}\\
		\vdots & \xi_{k-1} & \xi_{0} & \ddots & \vdots\\
		\xi_{2}&&\ddots&\ddots & \xi_1\\
		\xi_1 & \xi_2 &\cdots & \xi_{k-1}& \xi_0
	\end{pmatrix}
\end{equation}
i.e. $\mu(a,b)=\xi_{b-a}\pmod{k}$. Note that both $k,r$ can be read directly from $\xi$; i.e. $\ell(\xi)=k$ (the length) and $\abs{\xi}=r$. Clearly all circulant graphs are flow-regular of degree $|\xi|$ and order $\ell(\xi)$. Note that $\mathfrak{g}(\beta)^{c}$, i.e. the disjoint union of $c$ copies of $\mathfrak{g}(\beta)$, is also circulant. Concretely, defining the $ck$-dimensional vector $\beta^{c}$ via
\[
[\beta^{c}]_{s}=\begin{cases}
	\beta_{s/c} & c\mid s\\
	0 & \mathrm{otherwise}
\end{cases}
\]
we have $\mathfrak{g}(\beta)^{c}=\mathfrak{g}(\beta^c)$. We call $\beta$ a \emph{basic} vector if 
\begin{enumerate}
	\item $\beta_0\neq 0$; i.e. $\mathfrak{g}(\beta)$ is fully looped.
	\item $\beta_{-1}\neq 0$; This results in $\mathfrak{g}(\beta)$ having a (Hamiltonian) cycle $0\to k-1\to k-2\to \cdots\to 2\to 1\to 0$. In particular, $\mathfrak{g}(\beta)$ is connected.
	\item $\ell(\beta)|\beta|=$even; meaning $\mathfrak{g}(\beta)$ is even.
\end{enumerate}
For any $c\geq 1$, $\beta$ basic and $\xi=\beta^c$ the graph $\mathfrak{g}(\xi)$ is $(\ell(\xi),\abs{\xi})$ intensive (with $c$ connected components).

\begin{figure}
	\centering
	\includegraphics[scale=.65]{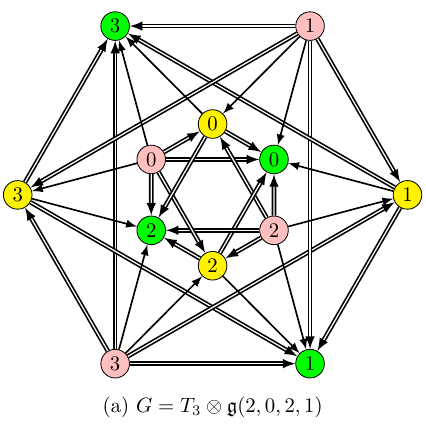} 
	\hspace{10pt}
	\includegraphics[scale=.65]{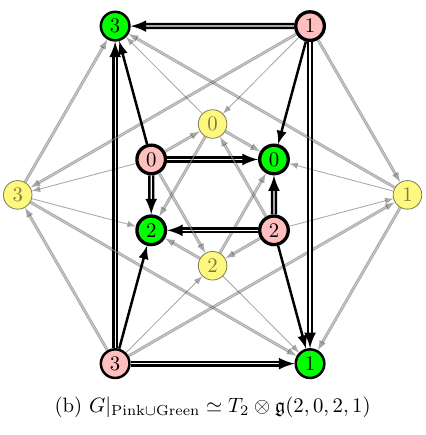} 
	
	\includegraphics[scale=.22]{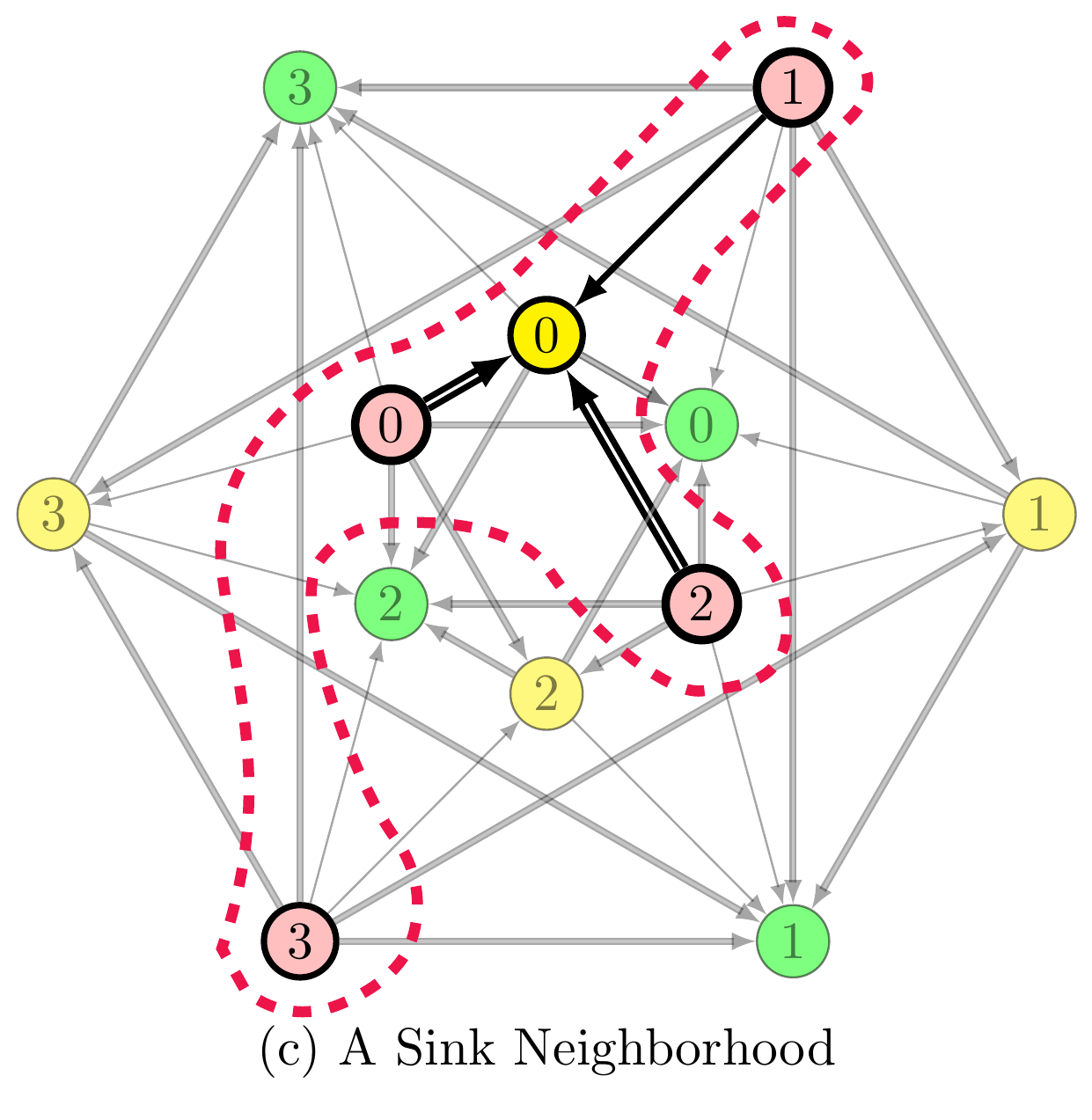} 
	\hspace{10pt}
	\includegraphics[scale=.22]{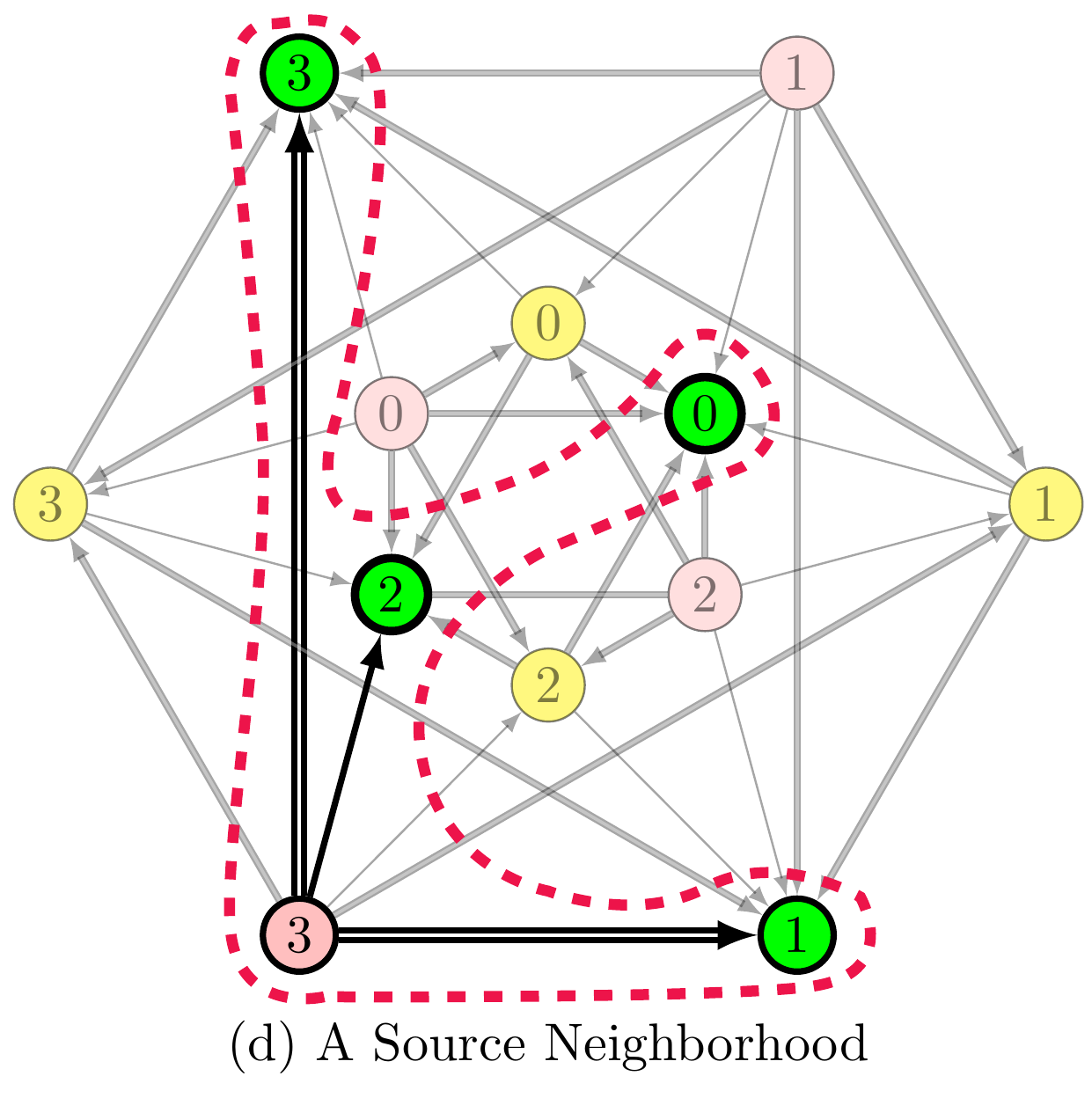} 
	\caption{And example depicting the local structure of graphs $T_n\otimes \mathfrak{g}(\xi)$. (a) The graph $G=T_3\otimes \mathfrak{g}(2,0,2,1)$ (see Fig. \ref{tensorEXP}). (b) The induced subgraph of $G$ restricted to two maximum independent set (Pink \& Green). The resulting subgraph is isomorphic to the shard $T_2\otimes \mathfrak{g}(2,0,2,1)$. (c) The neighborhood of the node Yellow-0 (in a shard). The in-degrees are a permutation of $\xi=(2,0,2,1)$ (no out-degrees). (d) The neighborhood of the node Pink-3 (in a shard). The out-degrees are a permutation of $\xi=(2,0,2,1)$ (no in-degree).}
	\label{localfig}
\end{figure}
The extensions $\mathrm{Ext}_n(g)$, with $g=\mathfrak{g}(\xi)$, have a finer local structure compared to extensions $\mathrm{Ext}_n(g)$ with $g$ a generic intensive graph. Let us demonstrate this using the language of neighborhoods. Given a graph $G=(V,\mu)$ and a node $x\in V$, define the \emph{neighborhood} $N(x)$ as the quadruple
\begin{subequations}
	\begin{align}
		N_{+}(x)&=\{y\in V\mid \mu(x,y)\neq 0\}\\
		N_{-}(x)&=\{y\in V\mid \mu(y,x)\neq 0\}\\
		\delta_+(x)&=\{\mu(x,y)\mid y\in N_+(x)\}\\
		\delta_-(x)&=\{\mu(y,x)\mid y\in N_-(x)\}
	\end{align}
\end{subequations}
It is straightforward to see that for a shard $G_2=T_2\otimes \mathfrak{g}(\xi)$, there are only two possibilities for $N(x)$:
\begin{enumerate}
	\item $N_+(x)$ is an independent set of size $k$, $N_-(x)=\emptyset$, and $\delta_+(x)=$ a permutation of $\xi$. This is called a \emph{source neighborhood}.
	\item $N_-(x)$ is an independent set of size $k$, $N_+(x)=\emptyset$, and $\delta_-(x)=$ a permutation of $\xi$. This is called a \emph{sink neighborhood}.
\end{enumerate}
This local structure is depicted in Fig. \eqref{localfig}. Note that if we think of $N(v)$ as a graph, then the local structure translates to (proportionality up to a positive integer)
\begin{equation}
	\mathrm{SGP}(N(v))(\vv{x})\propto \epsilon\ket{\chi_\xi; \vv{x}}
\end{equation}
where $\epsilon=1$ if $N(v)$ is a source neighborhood, $\epsilon=(-)^{|\xi|}$ if $N(v)$ is a sink neighborhood, and
\begin{equation}
	\ket{\chi_\xi;\vv{x}}=\frac{1}{b(\xi) k!}
	\sum_{\sigma\in S_{k+1}}
	\prod_{m\in Z_k}(x_{\sigma(1)}-x_{\sigma(m+2)})^{\xi_m}\in \mathcal{T}_{k+1}^{r}
\end{equation}
where if $\xi=(2m, 0^{\times(k-1)})$ then $b(\xi)=2$, while for all other $\xi$ we have $b(\xi)=1$. Note that $\chi_{\xi}(1, 0^{\times k})=1$.

The effects of this local property on $\mathrm{Ext}_n(\mathfrak{g}(\xi))$ are seen through the $(k+1)$-cuts. For brevity, from now on, the term ``cut'' will mean a $(k+1)$-cut. Let $g$ be any intensive graph. We say $C\vdash \mathrm{Ext}_n(g)$ is a \emph{standard cut} if $C$ includes a maximum independent set. For the simple case where $g$ is connected, there are $n(n-1)k$ such cuts:
\begin{equation}
	C_{J,x} = [J]\cup\{v\}, \qquad (v\notin [J])
\end{equation}
with $[J]={J}\otimes V_g$ a color class of $\mathrm{Ext}_n(g)$ ($1\leq J\leq n$). If $v\in [J_v]$, for $J_v\neq J$, due to patching isomorphism $\mathrm{Ext}_n(g)|_{[J_x]\cup[J]}\simeq \mathrm{Ext}_2(g)$, a standard cut $C_{J,x}$ is nothing but the neighborhood of $v$ in the shard $\mathrm{Ext}_n(g)|_{[J_v]\cup[J]}$ (see Fig. \eqref{localfig}). Using our formula for projections of cuts up to leading order \eqref{projection_cut}, it is now straightforward to show that
\begin{equation}
	(\mathcal{P}_{k+1}^r\Psi_{n+1}^{(g)})(\vv{x},\vv{y})=\ket{\chi_{\xi};\vv{x}}\prod_{i=1}^{nk-1}(y_i-\bar{x})^r\Psi_n(\bar{x},\vv{y})+R
\end{equation}
where the \emph{reminder} term is given by
\begin{equation}
	R(\vv{x},\vv{y})=\sum_{C\text{ non-standard}}(\mathcal{P}_{k+1}^r\Gamma_C)(\vv{x},\vv{y})
\end{equation}
Two features will stop $\gv{\Psi}^{(g)}$ from being separable with minimal polynomial $\ket{\chi_\xi}$:
(1) $S_a(\Psi_n)<r$ for some $n$, and (2) $R\neq 0$.
We will now design a set of conditions to avoid these two situations.
\begin{dfn}
	Suppose $\beta$ is a basic vector, $c\geq 1$, and $\xi=\beta^c$. Let $g=\mathfrak{g}(\xi)=(V_g,\mu_g)$. Let $\mathcal{U}=\{U_1, \cdots, U_\ell\}$ be a set of subsets of $V_g$, with $||\mathcal{U}||=k+1$. We call $g$ \emph{separative} if
	\begin{enumerate}
		\item $\mu(\mathcal{U})\geq r$ for all $\mathcal{U}$.
		\item If $\mu(\mathcal{U})=r$, then the associated cut $C(\mathcal{U})$ is either vanishing or standard.
	\end{enumerate}
\end{dfn}
\noindent
Some examples of presentable separable states are given by the following proposition (proven in \ref{last_appendix}).

\begin{prop}
	\label{sep_prop}
	Let $\beta$ be one of these basic vectors:
	\begin{enumerate}
		\item $\beta = (2m)$.
		\item $\beta = (l, m^{\times(k-1)})$ with $l\geq m> 0$, $kl=$even.
		\item $\beta = (1, 0^{\times(k-2)}, 1)$.
		\item $\beta = (1, 0^{\times(2K-3)},1, 1)$.
	\end{enumerate}
	Then with $\xi=\beta^{c}$ ($c\geq 1$ arbitrary), 
	the sequence $\gv{\Psi}^{(\xi)}$ is $\chi_\xi$-separable.
\end{prop}

\paragraph{Acknowledgements}
I am thankful for insightful discussions with my research supervisor Xiao-Gang Wen and my friend Michael DeMarco. This work is supported by NSF FRG grant:  DMS-1664412.

\appendix
\section{Haldane Sphere}
\label{HaldaneSphere_app}
Haldane \cite{Haldane_Sphere_Pseudo} initiated the
study of fractional quantum Hall states on the sphere. The spheres have the advantage of being genus zero, compact and 
having no boundary. In contrast, while $\mathbb{R}^2$ is genus zero and without boundary, it is not compact. A defect of $\mathbb{R}^2$ being unbounded is infinitely degenerate Landau levels. The natural solution to this issue is to compactify $\mathbb{R}^2$ by adding a point at infinity. The resulting topological space is a sphere $\mathbb{S}^2$. In the context of quantum Hall states, one should always realize a sphere with curvature $1/R$ (i.e. radius $R$) as being embedded in $\mathbb{R}^3$. In other words, our electron gas is \emph{confined} to move on a sphere but obeys the electromagnetic laws of $\mathbb{R}^3$. 

The only option for generating a uniform perpendicular magnetic field over the sphere is to put a Dirac monopole \cite{dirac1931quantised} at the center of the sphere:
\begin{equation}
	\vv{B} = B\gv{\Omega}, \qquad B=\frac{\hbar S}{eR^2}>0
\end{equation}
where $\gv{\Omega}$ is the unit normal vector (i.e. $\vv{r}=R\gv{\Omega}$). With the magnetic flux quantum being $\Phi_0=h/e$, a total of $N_\phi=2S$ magnetic flux quanta pierces through the sphere. Dirac has shown that $N_\phi$ must be an integer \cite{dirac1931quantised}.

Let $\vv{A}$ be a vector potential for the magnetic field, i.e., $\nabla\times \vv{A}=B\gv{\Omega}$.
The kinetic angular momentum is defined as
\begin{equation}
	\gv{\Lambda} = \vv{r}\times \gv{\pi} = \vv{r}\times (\vv{p}+e\vv{A})
\end{equation}
Computing the commutations of the components of $\vv{\Lambda}$, one finds $[\Lambda^{\alpha}, \Lambda^\beta]=i\hbar \epsilon^{\alpha\beta \gamma }(\Lambda^{\gamma}-\hbar S \Omega^\gamma)$, where $\alpha,\beta, \gamma=x,y,z$, and $\epsilon$ is the Levi-Civita symbol. These commutations suggest that $\vv{L}=\gv{\Lambda}+\hbar S \gv{\Omega}$ is the generator for rotations. This is confirmed via the computation $[L^\alpha, \Omega^\beta]=i\hbar \epsilon^{\alpha \beta\gamma}\Omega^\gamma$. In terms of these operators, the Hamiltonian governing the dynamics of a single electron is
\begin{equation}
	H = \frac{\Lambda^2}{2mR^2}=
	\frac{\hbar\omega_c}{2S}\left(\frac{L^2}{\hbar^2}-S^2\right)
\end{equation}
and has eigenvalues $E_l=\frac{\hbar\omega_c}{2S} [l(l+1)-S^2]$ with $l$ an integer ($\omega_c=eB/m$ is the cyclotron frequency). For energies to be positive, we need $l=S+n$, with $n=0,1,2\cdots$. The collection of states with energy $E_{S+n}$ are called the $n$th Landau level. The $n$th Landau level is the irreducible representation of the angular momentum algebra (given by $[L^-, L^+]=2\hbar L^z, [L^z, L^\pm]=\pm \hbar L^{\pm}$) with highest weight $l=S+n$ and has a degeneracy of $2(S+n)+1$. In particular, the lowest Landau level (LLL) has an energy $\hbar\omega_c/2$ and a degeneracy $2S+1$.

We will be exclusively interested in the lowest Landau level in this paper. Due to the nature of the Hamiltonian, the LLL ``wavefunctions'' $\Phi_{S, m}(\theta, \phi)$ ($-S\leq m\leq S$) are  the simultaneous eigenfunction of $L^2, L_z$; i.e.
\begin{equation}
	L^2 \Phi_{S,m}(\theta, \phi)=\hbar^2 S(S+1)\Phi_{S,m}(\theta, \phi), \qquad
	L_z \Phi_{S,m}(\theta, \phi)=m\hbar\Phi_{S,m}(\theta, \phi)
\end{equation}
To find these functions, we need to fix a gauge. However, irrespective of the choice for the vector potential $\vv{A}$, there will be some singularities since there exists no continuous tangent field over the sphere. We choose $\vv{A}$ so that there is only one singularity, and that singularity is at the north pole (the point ``at infinity''):
\begin{equation}
	\vv{A} = - \frac{\hbar S}{eR}\cot \frac{\theta}{2}\vv{e}_\phi
\end{equation}
This will allow us to find the wavefunctions $\Phi_{S,m}(\theta,\phi)$ everywhere on sphere except for the north pole (i.e. $\mathbb{S}^2-\{N\}$). If we were to instead choose $\vv{A}'=+(\hbar S/eR)\tan(\theta/2)\vv{e}_\phi$, a secondary solution $\Phi'_{S,m}(\theta, \phi)$ would be found. The latter solution is valid on the the whole sphere except for the south pole (i.e. $\mathbb{S}^2-\{S\}$). A true wave-``function'' on the sphere is actually the pair $(\Phi_{S,m},\Phi'_{S,m})$ and is a \emph{section} (rather than a function) over the sphere \cite{wu1976dirac}. On the overlap of the charts (i.e. $\theta\neq 0, \pi$), the two vector potentials are related via $\vv{A}'=\vv{A}+\frac{\hbar}{e}\nabla (2S\phi)$. Thus, the two wavefunction $\Phi, \Phi'$ are related by a gauge transformation:
\begin{equation}
	\label{eq:gauge_transform}
	\Phi_{S,m}'(\theta, \phi)=\Phi_{S,m}(\theta, \phi)e^{-2iS\phi}
\end{equation}
We now put our attention towards finding the explicit solutions $\Phi_{S,m}(\theta, \phi)$. Using the vector potential $\vv{A}$, the operators $L^2, L_z$ can be written as
\begin{subequations}
	\begin{align}
		L^2&=
		-\hbar^2\left[
		\frac{1}{\sin\theta}\pd{}{\theta}\left(\sin \theta \pd{}{\theta}\right)+
		\frac{1}{\sin^2\theta} \pdd{}{\phi}
		-\frac{S}{\sin^2(\theta/2)}
		\left(S+i\pd{}{\phi}\right)
		\right]\\
		L_z &= \frac{\hbar}{i}\pd{}{\phi} -  \hbar S
	\end{align}
\end{subequations}
Introducing the spinor coordinates $u=\cos (\theta/2)e^{i\phi}$ and $v=\sin(\theta/2)$, the (unnormalized) simultaneous eignefunctions $\Phi_{S,m}(\theta,\phi)$ are $u^{S+m}v^{S-m}$.

The LLL wavefunctions are almost always written in terms of their complex coordinates, which is synonymous with stereographic projection. It is this reparametrization that reveals the holomorphic nature of the LLL wavefunctions.  As the figure \ref{stereo_fig} illustrates, the complex coordinate $z$ of a point on the sphere (except for the north pole) is defined as $z=2R\cot(\theta/2)e^{i\phi}$. To reparametrize the other chart (i.e. $\mathbb{S}^2-\{S\}$), we do a stereographic projection with the roles of $S$ and $N$ reversed. This leads to a reparametrization $w=2R\tan (\theta/2)e^{-i\phi}$. Consequently, on the overlap (i.e. $\mathbb{S}^2-\{S,N\}$) the two coordinates are related via $z/(2R)=(2R)/w$. The measure (volume $2$-form) on the sphere can now be written as
\begin{equation}
	d\omega = R^2\sin\theta d\theta\wedge d\phi = -\frac{i}{2}\frac{dz\wedge d\bar{z}}{[1+z\bar{z}/(2R)^2]^2}=
	-\frac{i}{2}\frac{dw\wedge d\bar{w}}{[1+w\bar{w}/(2R)^2]^2}
\end{equation}
In other words, the inner product of two states $\Psi_1, \Psi_2$ over the sphere is
\begin{equation}
	\braket{\Phi_1}{\Phi_2}=\int d\omega \: \overline{\Phi_1(
		z, \bar{z})}\Phi(z,\bar{z})
\end{equation}
Due to their crucial role in our analysis of uniform states, it is of utmost importance to us to find the generators of angular momentum algebra $L^z, L^{\pm}$ in terms of complex coordinates $z,\bar{z}$. These are computed to be \cite{de1997analytic}:
\begin{subequations}
	\label{eq_angular_single}
	\begin{align}
		\hbar^{-1}L_- &= 2R\partial + \frac{\bar{z}^2\bar{\partial}+S\bar{z}}{2R}\\
		\hbar^{-1}L_z &= z\partial - \bar{z}\bar{\partial} - S\\
		\hbar^{-1}L_+ &= \frac{-z^2\partial +Sz}{2R}
		-2R\bar{\partial}
	\end{align}
\end{subequations}
Now, either by direction transforming $u^{S+m}v^{S-m}$, or using the above algebra, one can find the simultaneous eigenstates of $L^2, L_z$:
\begin{subequations}
	\label{eq_spherical_solution}
	\begin{align}
		\Phi_{S,m}(z,\bar{z}) &=\sqrt{\frac{2S+1}{4\pi R^2}\binom{2S}{S+m}} \frac{(z/2R)^{S+m}}{(1+z\bar{z}/4R^2)^{S}}\\
		\Phi'_{S,m}(w,\bar{w})&=
		\sqrt{\frac{2S+1}{4\pi R^2}\binom{2S}{S+m}}
		\frac{(w/2R)^{S-m}}{(1+w\bar{w}/4R^2)^{S}}
	\end{align}
\end{subequations}
In obtaining $\Psi'$, we have utilized the gauge transformation \eqref{eq:gauge_transform} in the form:
\begin{equation}
	\Phi'(w,\bar{w})=
	\frac{w^{2S}}{(w\bar{w})^S}\Phi\left(\frac{4R^2}{w}, \frac{4R^2}{\bar{w}}\right)
\end{equation}
This shows that, aside from the the universal kernel $K(z,\bar{z})=(1+z\bar{z}/4R^2)^{-S}$, the single-particle wavefunctions in the lowest Landau level are polynomials $P(z)$ with $\deg P\leq 2S$.

We now move on to bosonic many-body wavefunctions on the sphere that are confined to the lowest Landau level. Aside from normalization, the most general form for such a wavefunction is (recall that $2S=N_\phi$)
\begin{subequations}
	\begin{align}
			\label{spherical_wavefunction_eq}
		&\Phi(z_1, \bar{z}_1, \cdots, z_N, \bar{z}_N)=\Psi(z_1, \cdots, z_N)\prod_{i=1}^N (1+z_i\bar{z}_i/4R^2)^{-S}\\
		&\Phi'(w_1, \bar{w}_1, \cdots, w_N, \bar{w}_N)=\Psi'(w_1, \cdots, w_N)\prod_{i=1}^N (1+w_i\bar{w}_i/4R^2)^{-S}\\
		&\Psi'(w_1, \cdots, w_N)=\frac{1}{(4R^2)^{NN_\phi/2}}\prod_{i=1}^N w_i^{N_\phi}\Psi\left(\frac{4R^2}{w_1}, \cdots, \frac{4R^2}{w_N}\right)
	\end{align}
\end{subequations}
where $\Psi$ is a symmetric polynomial with local degree $\leq 2S=N_\phi$. Note the similarity between the polynomial $\Psi'$ (which is the `wavefunction near infinity') and definition of the adjoint polynomial $\Psi^\dagger$ \eqref{adjoint_eq}. 

We finish this appendix by an important comment about FQH ground states on the sphere. If $\Phi$ is to represent an FQH ground state, then it must be an $SO(3)$-singlet; i.e. $L^2\Phi=0$. Alternatively, we must have $L_-\Phi = L_z\Phi = L_+\Phi = 0$. Acting the $N$-body extension of the operators in \eqref{eq_angular_single} on the wavefunction \eqref{spherical_wavefunction_eq}, we find that $\Psi$ needs to satisfy the following system of partial differential equations:
\begin{subequations}
	\begin{align}
		L_-\Psi &:= \sum_i \partial_i\Psi=0\\
		L_z\Psi &:= \left(-\frac{NN_\phi}{2}+\sum_i z_i\partial_i\right)\Psi=0\\
		L_+\Psi &:= 
		\left(\sum_i N_\phi z_i-z_i^2\partial_i\right)\Psi=0
	\end{align}
\end{subequations}
The conditions $L_-\Psi =0$ (resp. $L_+\Psi =0$) are sometimes called the \emph{lowest} (\emph{highest}) weight condition. That being said, sometimes authors use a different convention for the signs: $L_+'=L_-$, 
$L_-'=L_+$ and $L_z'=-L_z$. Of course, if one uses $\vv{L}'$ generators instead of $\vv{L}$, then it is more natural to defined `highest'/`lowest' weight  condition the other way around.

\section{Uniform States: Four Equivalent Formulations}
\label{uniform_appendix}
This appendix will state and prove the equivalence for four different formulations of the uniform states. We begin by proving a fact about translational invariant polynomials. 
\begin{lem}
Let $P(z_1, \cdots, z_N)$ be translational invariant (not necessarily symmetric). Let $d=\deg_{z_1}P$ and $D=\sum_{j=2}^N \partial_j$. Let $Q(z_2, \cdots, z_N)=P(0, z_2, \cdots, z_N)$. Then $D^{d+1}Q = 0$ and $P = \exp (-z_1D)Q$. 
\end{lem}
\begin{proof}
By translational invariance, we have $P(z_1, \cdots, z_N)=Q(z_2-z_1, \cdots, z_N-z_1)$. Moreover, given any function $F(z_2, \cdots, z_N)$, we have $F(z_2+t, \cdots, z_N+t)= \exp(t D)F(z_2, \cdots, z_N)$ (Taylor expand $f(t):=F(z_i+t)$). Combining the two observations, the lemma is proved.
\end{proof}

\begin{thm}
	\label{uniform_equivalent}
Let $P(z_1, \cdots, z_N)$ be a non-zero symmetric polynomial in $N$ variables and $N_\phi$ be some non-negative integer. The following are equivalent
\begin{enumerate}
\item $P$ is translational invariant, homogeneous, and $\deg_{\ell}P\leq N_\phi$. Moreover, $P$ is self-adjoint:
\[
P(z_1, \cdots, z_N) = \prod_{i=1}^N z_i^{N_\phi} P(-1/z_1, \cdots, -1/z_N)
\]
This is what we called an $(N,N_\phi)$ uniform state.

\item $\deg_\ell P\leq N_\phi$ and for any conformal mapping $f(z)= (az+b)/(cz+d)$ ($\det f:=ad-bc\neq 0$), $P$ satisfies
\begin{equation}
P\left(f(z_1), \cdots, f(z_N)\right)=
\prod_{i=1}^N\left[f'(z_i)\right]^{\frac{1}{2}N_\phi}
P(z)
\label{mobius}
\end{equation}
One says $P$ is conformally covariant.
\item $\deg_\ell P\leq N_\phi$ and $P$ is a solution to the following system of partial differential equations
\begin{subequations}
\label{HLWC}
\begin{align}
L_-P&:=\sum_i \partial_i P =0\\
L_zP&:=\sum_i (z_i\partial_i-\frac{N_\phi}{2})P\\
L_+P&:=\sum_i (N_\phi z_i -z_i^2\partial_i)P=0
\end{align}
\end{subequations}
These are the angular momentum operators corresponding to the induced representation of $SO(3)$ on a sphere with $N_\phi$ flux quanta. The conditions amount to $P$ being $SO(3)$-singlet.

\item $P$ is translational invariant, $\deg_\ell P = N_\phi$ and $P$ is homogeneous of degree $M=\frac{1}{2}NN_\phi$.
\end{enumerate}
\end{thm}
\begin{proof}
The proof of $4\Rightarrow 1$ will be provided in \ref{cayley_appendix}. Here, we prove $1\Rightarrow 2 \Rightarrow 3 \Rightarrow 4$.

$(1\Rightarrow 2)$ The three criteria of an $(N,N_\phi)$ uniform state is the same as covariance of $P$ (in the sense of \eqref{mobius}) under the three transformations $f(z)=z+c$ (translations), $\lambda z$ (scaling), and $-1/z$ (inversion). Moreover, a simple chain rule shows that if $P$ is covariant under $f$ and $g$, it will be covariant under $f\circ g$ as well. Therefore, it suffices to write an arbitrary M\"{o}bius transformation as compositions of translations, scalings and inversion. If $c=0$, then $z\mapsto (az+b)/d$ ($ad\neq 0$) is a scaling followed by a translation. If $c\neq 0$ and $ad-bc\neq 0$, define
\begin{enumerate}
\item $f_1(z)=z+d/c$ (translation).
\item $f_2(z)=-1/z$ (inversion).
\item $f_3(z)=[(ad-bc)/c^2] z$ (scaling).
\item $f_4(z)=z+a/c$ (translation).
\end{enumerate}
We have $f(z)=(f_4\circ f_3\circ f_2\circ f_1)(z)$.

($2\Rightarrow 3$) Consider the M\"{o}bius transformations
\[
f_-(z;\tau) = z+\tau, \quad f_+(z;\tau)=\frac{z}{1+\tau z}
\]
Assuming $\tau$ is small, if we expand Eq. \eqref{mobius} for $f_{\pm}$ up to $O(\tau^2)$, the desired relations $L_{+}P=L_-P=0$ follow. In fact, given any polynomial $P(z_1, \cdots, z_N)$ of degree no higher than $N_\phi$ in any variable $z_i$ (not necessarily symmetric), one can directly check that
\begin{subequations}
	\begin{align}
		\label{ap_trans_eq}
		(e^{\tau L_-}P)(z_1, \cdots, z_N)&=
		P\left(z_1+\tau, \cdots, z_N+\tau\right)\\
			(e^{\tau L_+}P)(z_1, \cdots, z_N)&=
		\prod_{i=1}^N(1+\tau z_i)^{N_\phi}P\left(\frac{z_1}{1+\tau z_1}, \cdots, \frac{z_N}{1+\tau z_N}\right)		
	\end{align}
\end{subequations}
Note that since $[L_+, L_-]=2L_z$, checking $L_zP=0$ separately is redundant.

($3\Rightarrow 4$) First of all, $L_-P=0$ is equivalent to translational invariance. This follows from \eqref{ap_trans_eq} (which is a fancy Taylor expansion). Secondly, it is a general fact that (see \ref{cayley_appendix}) if $P(z_1, \cdots, z_n)$ is a symmetric homogeneous translational invariant polynomial, then $\deg P\leq N\deg_\ell P/2$. Combined with the hypothesis $\deg_\ell P\leq N_\phi$, we find the following inequality
\[
\deg P\leq \frac{N\deg_\ell P}{2}\leq \frac{NN_\phi}{2}
\]
Now, note that $L_zP=0$ is Euler's homogeneous function theorem and states that $\deg P = NN_\phi/2$.  It only remains to show that $\deg_\ell P=N_\phi$. But this follows from the above inequality.
\end{proof}

\section{Cayley Decomposition}
\label{cayley_appendix}
In this appendix we will prove Theorems \ref{cayley_1} and \ref{cayley_2}. Almost all of the content of this appendix can be found in Elliot\cite{elliott}. Let us start by proving Theorem \ref{cayley_1}:
\emph{If $G$ is an $(N,N_\phi)$ extensive graph, then $SGP(G)$ is either zero or an $(N,N_\phi)$ uniform state.}
\begin{proof}
Suppose $\mathrm{SGP}(G)\neq 0$. Using the definition,
\[
P=\mathrm{SGP}(G) = \sum_{\sigma}\prod_{i,j}(z_{\sigma(i)}-z_{\sigma(j)})^{\mu(i,j)}
\]
Translational invariance and homogeneity are immediate. Moreover,
\begin{align*}
&P\left(-\frac{1}{z_1}, \cdots, -\frac{1}{z_N}\right)
\\&\qquad=
\sum_{\sigma}
\prod_{i,j}
(z_{\sigma(i)}-z_{\sigma(j)})^{\mu(i,j)}
\prod_{i,j} (z_{\sigma(i)}z_{\sigma(j)})^{-\mu(i,j)}\\
&\qquad=
\sum_{\sigma}
\prod_{i,j}
(z_{\sigma(i)}-z_{\sigma(j)})^{\mu(i,j)}
\prod_{i}z_{\sigma(i)}^{-(d_i^++d_i^-)}
\end{align*}
So far, this is true for all graphs. If $G$ is $N_\phi$-regular, however, $d_i^++d_i^-=N_\phi$, and we recover
\[
P(z_1, \cdots, z_n) = \prod_{i=1}^N z_i^{N_\phi}
P\left(-\frac{1}{z_1}, \cdots, -\frac{1}{z_N}\right)
\]
\end{proof}
We now move on to prove Theorem \ref{cayley_2}, aka Cayley decomposition. This is done is several steps. Throughout $\deg_i P$ stands for the degree of $P$ in $z_i$.

\begin{lem}[Elliot \cite{elliott}, \S 88\emph{(bis)}] Let $P(z_1, \cdots, z_n)$ be a homogeneous translational invariant polynomial of degree $\deg P$. Then
\begin{equation}
2\deg P\leq \sum_i \deg_i P
\end{equation}
\end{lem}
\begin{proof}
We prove this by induction on $n$. If $n=2$, then $P(z_1,z_2)=C(z_1-z_2)^{w}$, for some $w$, and the equality holds ($d_1+d_2=2w$). Now suppose the assertion is true for any translational invariant homogeneous polynomial $Q$ in $n-1$ variables. 
We will proceed to prove by contradiction; i.e. we assume
\[
2\deg P>\sum_i \deg_i P
\]
Suppose $P$, as a polynomial in $z_1$, has a root at $z_2$ of degree $r$, for some $0\leq r< \deg P$ ($r=\deg {P}$ is excluded since it violates $2\deg P>\sum_i \deg_i P$). Then
\[
P(z_1, \cdots, z_n)=(z_1-z_2)^r Q(z_1, \cdots, z_n)
\]
with $\lim_{z_1\to z_2}Q\neq 0$. Now $\deg Q = \deg P - r$, $\deg_{j}Q=\deg_{j}P-r$ for $j=1,2$, and $\deg_i Q=\deg_i P$ for $i>2$. Therefore,
\[
2\deg Q>\sum_{i=1}^{n} \deg_iQ, \qquad (\star)
\]
i.e. $Q$ has the same relation on degrees as $P$. Define 
$$Q'(z_2,z_3, \cdots, z_n)=Q(z_2, z_2,z_3, \cdots, z_n)\neq 0$$
note that $\deg Q'=\deg Q$, $\deg_2 Q'\leq \deg_1 Q+\deg_2 Q$ and $\deg_iQ'=\deg_iQ$ for $i>2$. Using relation $(\star)$
\[
2\deg Q' > \deg_{1}Q+\deg_2 Q+\sum_{i>2}\deg_iQ\geq \sum_{i=2}^{n}\deg Q'_i
\]
As $Q'$ is a translational invariant homogeneous polynomial in $n-1$ variables, this is a contradiction with the induction hypothesis.
\end{proof}

\begin{thm}[Elliot \cite{elliott}, \S 89]
Let $P(z_1, \cdots, z_n)$ be a homogeneous translational invariant polynomial. Then we can write $P$ as a finite sum of the form
\begin{equation}
P(z_1, \cdots, z_n) = \sum_{\alpha} C_{\alpha}\prod_{i<j}(z_i-z_j)^{m_{\alpha}(i,j)}
\end{equation}
where for all $\alpha$ (which is some index),  $m_{\alpha}(i,j)=m_{\alpha}(j,i)$, $m_{\alpha}(i,i)=0$ and
\[
D_i(\alpha):=\sum_{j=1}^{n} m_{\alpha}(i,j)\leq \deg_i P
\]
In other words, $P$ is a linear combination of products of differences in which no variable $z_i$ occurs more than $\deg_i P$.
\end{thm}
\begin{proof}
We prove by induction on $n$. For $n=2$, $P=C(z_1-z_2)^w$ for some $w$, and the assertion is trivially true. Suppose the assertion is true for translational invariant homogeneous polynomials in $n-1$ variables. Let $d_i=\deg_i P$ for $1\leq i\leq n$ and $w=\deg P$. We can write
\[
P(z_1, \cdots, z_n)=z_1^{d_1}Q(z_2, \cdots, z_n)+R(z_1, \cdots, z_n)
\]
with $Q,R$ polynomials so that $\deg_1 R<d_1$. Comparing the coefficient of $z_1^{d_1}$ in the equality $P(z_1+c, \cdots, z_n+c)=P(z_1, \cdots, z_n)$, one finds that $Q$ is also translational invariant. Henceforth, $Q$ satisfies the induction hypothesis; i.e.
\[
Q = \sum_\alpha C_\alpha Q_\alpha
\]
with $C_\alpha$ some numeric factors, and
\[
Q_{\alpha}(z_2, \cdots, z_n) = 
\prod_{2\leq i<j\leq n}(z_i-z_j)^{m_{\alpha}(i,j)}
\]
where $\sum_{j=2}^n m_{\alpha}(i,j)=\deg_i Q_\alpha\leq \deg_i Q\leq d_i$ for $i\geq 2$.
Note that, the way $Q_\alpha$ is defined,
\[
2\deg Q = 2\deg Q_\alpha=\sum_{i=2}^n \deg_iQ_\alpha
\]
and $\deg Q = w-d_1$. Let $s_i(\alpha)\geq 0$ be such that $\deg_{i}Q_{\alpha}=d_i-s_i(\alpha)$ for $i\geq 2$. Then
\[
\sum_{i\geq 2} s_i(\alpha) =\sum_{i\geq 2}d_i-2(w-d_1)=
d_1+\left(\sum_{i=1}^n d_i-2w\right)
\]
Using the lemma, we find that $\sum_{i\geq 2} s_i(\alpha) \geq d_1$. Therefore, there exists $0\leq \nu_i\leq s_i$ such that $\sum_{i\geq 2}\nu_i(\alpha) = d_1$. Define
\begin{align*}
P_\alpha(z_1, \cdots, z_n) &= \prod_{i=2}^{n}(z_1-z_i)^{\nu_i(\alpha)}Q_{\alpha}(z_2, \cdots, z_n)\\
N(P)(z_1, \cdots, z_n)&=\sum_{\alpha}C_{\alpha}P_\alpha(z_1, \cdots, z_n)
\end{align*}
Note that $N(P)$ is of the desired form. Let us define $P_1=P$, and $P_m=P_{m-1}-N(P_{m-1})$ for $m>1$ (the significance of which will be clear momentarily). Note that $P_2$ is either zero (in which case we are done), or it is a translational invariant homogeneous polynomial with $\deg_1 P_2<\deg_1 P_1=d_1$. Repeating the procedure over and over, eventually we either end up with $P_r=0$ or $\deg_1 P_r=0$. In the former case we are done. In the latter case $P_r$ is independent of $z_1$, so it is in $n-1$ variables. Therefore it can be written in the desired form by induction hypothesis.
\end{proof}

\begin{cor}
Let $P$ be a translational invariant homogeneous polynomial of degree $w$ satisfying $2w = \sum_{i=1}^n \deg_i P$. Then we can write $P$ as a finite sum of the form
\begin{equation}
P(z_1, \cdots, z_n) = \sum_{\alpha} C_{\alpha}\prod_{i<j}(z_i-z_j)^{m_{\alpha}(i,j)}
\end{equation}
where for all $\alpha$,  $m_{\alpha}$ is a symmetric matrix with zero diagonal such that $\sum_{j=1}^{n} m_{\alpha}(i,j)= \deg_i P
$. In other words, $P$ is a superposition of products of differences in which $z_i$ appears exactly $d_i=\deg_iP$ times.
\end{cor}
\begin{proof}
Using the theorem,
\[
P = \sum_{\alpha}C_\alpha \prod_{i<j}(z_i-z_j)^{m_\alpha(i,j)}
\]
with $D_i(\alpha)\leq d_i$. At the same time, $2w = \sum_i d_i = \sum_i D_i(\alpha)$ for all $\alpha$. Therefore, $\sum_{i}(d_i-D_i(\alpha))=0$. The only possibility would then be $D_i(\alpha)=d_i$ for all $\alpha$.
\end{proof}

\begin{cor}[Cayley Decomposition]
\label{cayd}
Let $P$ be a translational invariant homogeneous symmetric polynomial of local degree $N_\phi$, and total degree $M=NN_\phi/2$. Then there exists $(N,N_\phi)$ extensive graphs $G_1, \cdots, G_p$, and $\mathbb{C}$-numbers $C_1, \cdots, C_p$ so that
\begin{equation}
P = \sum_{\alpha=1}^p C_\alpha \mathrm{SGP}(G_\alpha)
\end{equation}
\end{cor}
\begin{proof}
By previous corollary, we can write $P$ as
\[
P=\sum_{\alpha}C'_\alpha\prod_{i<j}(z_i-z_j)^{m_\alpha(i,j)}
\]
Define $\mu_{\alpha}(i,j)=m_\alpha(i,j)$ if $i<j$, and $\mu_{\alpha}(i,j)=0$ otherwise. Also define $V_\alpha=\{1,2,\cdots, N\}$. Then the data $G_\alpha=(V_\alpha, \mu_{\alpha})$ is an $(N,N_\phi)$ extensive graph. Finally, as $P$ is symmetric
\[P(z_1, \cdots, z_N) = \frac{1}{N!}\sum_{\sigma\in S_N}P(z_{\sigma(1)}, \cdots, z_{\sigma(N)})\\
= \sum_{\alpha} \frac{C'_\alpha}{N!}\mathrm{SGP}(G_\alpha)
\]
\end{proof}
\begin{cor}
A polynomial $P$ is an $(N,N_\phi)$ uniform state if  it is symmetric, translational invariant, of local degree $N_\phi$, and homogeneous of degree $M=NN_\phi/2$. (This is $4\Rightarrow 1$ in Theorem \ref{uniform_equivalent})
\end{cor}
\begin{proof}
Use the previous corollary to write $P=\sum_{\alpha} C_\alpha \mathrm{SGP}(G_\alpha)$. Then Theorem \ref{cayley_1} finishes the job.
\end{proof}

\section{Extensions of Intensive Graphs}
\label{graph_appendix}
In this appendix we discuss some of the graph theoretic properties of graphs $G_n=T_n\otimes g$, with $g$ being a $(k,r)$ intensive graph. The following graph theoretic notions will be needed:
\begin{dfn}
	We call two nodes $x,y$ in a graph $G=(V,\mu)$ \emph{adjacent} if $\mu(x,y)+\mu(y,x)\neq 0$.
\end{dfn}
\begin{dfn}
	Let $G=(V,\mu)$ be a graph and $S\subset V$ a subset of nodes. The pair $G|_{S}:=(S,\mu|_S)$ is called the \emph{induced subgraph of $G$ by $S$} ($\mu|_S$ is the restriction of $\mu$ to $S\subset V$).
\end{dfn}
\begin{dfn}
	Let $G_1=(V_1, \mu_1)$ and $G_2=(V_2, \mu_2)$. We say $G_1$ is \emph{isomorphic} to $G_2$, and write $G_1\simeq G_2$, if there exists a bijection $\phi:V_1\to V_2$ so that $\mu_1(x,y)=\mu_2(\phi(x), \phi(y))$ for all $x,y\in V_1$.
\end{dfn}

\begin{dfn}
	Let $G_1=(V_1, \mu_1)$ and $G_2=(V_2, \mu_2)$. We define the \emph{disjoint union} $G_1\sqcup G_2=(V, \mu)$ as follows. The node set $V$ is the disjoint union of $V_1, V_2$; i.e. $V=V_1\sqcup V_2$. We define
	\[
	\mu(x,y)=\begin{cases}
		\mu_1(x,y) & x,y\in V_1\\
		\mu_2(x,y) & x,y\in V_2\\
		0 & \mathrm{otherwise}
	\end{cases}
	\]
\end{dfn}

\begin{dfn}
	Let $G=(V,\mu)$. A path is a sequence of nodes $[i_1,i_2,\cdots ,i_\ell]$ such that $i_m, i_{m+1}$ are adjacent for all $1\leq m<\ell$. Given two elements $x,y\in V$ we write $x\sim y$ if there is a path between $x,y$ (we assume $x\sim x$ by default). This relation is clearly an equivalence on $V$. Suppose $x_1, \cdots, x_c\in V$ are such that
	\[
	V=[x_1]\sqcup [x_2]\sqcup \cdots \sqcup [x_c]
	\]
	where $[x_i]$ is the equivalence class of $x_i$. Then $G_i:=G|_{[x_i]}$ is called a \emph{(connected) component} of $G$, and $c(G)=c$ is called the \emph{number of components} of $G$. Note that
	\[
	G=G_1\sqcup G_2\sqcup\cdots\sqcup G_c
	\]
\end{dfn}

\subsection{Terminology: Horizontal and Vertical Sets}
Let $g=g_1\sqcup g_2\sqcup\cdots \sqcup g_c$ with each $g_i$ connected. We use the notations $g_i=(U_i,\nu_i)$, and $g=(U,\nu)$ for the graph data. Let $G_n=T_n\otimes g$, and its data be $G_n=(V, \mu)$.
\paragraph{(Vertical Sets)}
\label{vertical_sets}
Given $1\leq J\leq n$ define
\begin{equation}
	\mathcal{V}_J^{(i)}=\{J\}\otimes U_i=\{J\otimes x\mid x\in U_i\}
\end{equation}
This is a subset of nodes in $T_n\otimes g_i$. Now, given a sequence $\vv{J}=(J_1, \cdots, J_c)$ (wit $1\leq J_i\leq n$) we define
\begin{equation}
	\mathcal{V}_{\vv{J}}=\bigcup_{i=1}^c \mathcal{V}_{J_i}^{(i)}
\end{equation}
We call $\mathcal{V}_{\vv{J}}$ the \emph{$\vv{J}$th vertical set} of $G$. Clearly $\mathcal{V}_{\vv{J}}$ is an independent set of size $|g|$.

Given a fixed $m$, define a $c\times m$ matrix $\vv{S}$
\[
\vv{S}=\begin{pmatrix}
	J_{1}^{(1)} & J_{1}^{(2)} & \cdots & J_{1}^{(m)}\\
	J_{2}^{(1)} & J_{2}^{(2)} & \cdots & J_{2}^{(m)}\\
	\vdots\\
	J_{c}^{(1)} & J_{c}^{(2)} & \cdots & J_{c}^{(m)}\\
\end{pmatrix}
\]
where, for each $1\leq t\leq c$ we have $J_t^{(i)}<J_t^{(i+1)}$, and all entries of $\vv{S}$ are in integers in the interval $[1,n]$. Note that each column of $\vv{S}$ is a $\vv{J}$-vector. The $i$th row of $\vv{S}$ is denote by $S_i$. Note that
\[
T_n\otimes g_i|_{S_i\otimes U_i}\simeq T_{m}\otimes g_i
\]
via the isoomorphism $\phi_i(J_i^{(t)}\otimes x)=t\otimes x$ for all $x\in g_i$. Defining, $\phi=\phi_1\sqcup \phi_2\sqcup \cdots \sqcup \phi_c$, we find that
\[
\left.T_n\otimes g\right|_{\bigcup_{j=1}^{m} \mathcal{V}_{\vv{J}^{(j)}}}=
\left.T_n\otimes g\right|_{\bigcup_{i=1}^{c} S_i\otimes U_i}\simeq T_m\otimes g
\]
Two special cases are important to us: 
\begin{enumerate}
	\item Let $\vv{J}, \vv{J}'$ be such that for all $1\leq i\leq c$ we have $J_i\neq J'_i$. Then
	\begin{equation}
		G_n\mid_{\mathcal{V}_{\vv{J}}\cup \mathcal{V}_{\vv{J}'}} \simeq G_2
	\end{equation}
	In a manner of speaking, $T_n\otimes g$ is ``locally'' the same as the shard $T_2\otimes g$.
	\item For any sequence $\vv{J}$ we have
	\begin{equation}
		G_n|_{\mathcal{V}^c_{\vv{J}}}\simeq G_{n-1}
	\end{equation}
	with $\mathcal{V}^c_{\vv{J}}$ the complement of $\mathcal{V}_{\vv{J}}$.
\end{enumerate}
The isomorphisms discussed in this section are natural byproducts of how $T_n$ and the tensor product are defined. We do not need $g$ to be intensive for these isomorphisms. The role of intensivity is to make it so that the vertical sets are the only maximum independent sets.

\paragraph{(Horizontal Sets)}
\label{horizontal_sets}
Given $x\in U$, we define the \emph{$x$th horizontal set} of $G$ as
\begin{equation}
	H_x = T_n\otimes \{x\} = \{J\otimes x\mid J\in T_n\}
\end{equation}
Take two arbitrary elements of $H_x$, e.g. $X=I\otimes x$ and $Y=J\otimes x$, say with $I<J$. Then $\mu(Y,X)=0$, while $\mu(X,Y)=\nu(x,x):=l(x)$ with $l(x)$ the number of loops on $x$. In other words, any pair of nodes in $H_x$ are connected with an arrow of multiplicity $l(x)$ ($\neq 0$ since $g$ is fully looped).

\subsection{Maximum Independent Sets}
\label{maxind_appendix}
\subsubsection*{Maximum Independent Sets of $T_n\otimes g$}
Let $g$ be a $(k,r)$ intensive graph. Say $g=g_1\sqcup g_2\sqcup \cdots \sqcup g_c$ with each $g_i$ connected. Let $G_n=T_n\otimes g$. We want to find the independence number $\alpha(G_n)$. Note that if we take any $k+1$ nodes (or more) in $G_n$, then by pigeonhole principle there exists a horizontal set $H_x$ containing at least two of those nodes. As any two nodes in $H_x$ are connected with an arrow of multiplicity $l(x)\neq 0$, it is impossible to have an independent set of size $k+1$. As each $\mathcal{V}_{\vv{J}}$ is an independent set of size $k$, we find $\alpha(G_n)=k$. Note that it was the fully looped condition of $g$ that ensured this.

Next, we want to show that $G_n$ has no maximum independent sets other than the vertical sets $\mathcal{V}_{\vv{J}}$. Let $S$ be an arbitrary maximum independent set. With $G_n^{(i)}=T_n\otimes g_i\subset G_n$, let $S^{(i)}$ be the maximum independent set $S$ induces on $G^{(i)}_n$. If we show that $S^{(i)}$ is necessarily of the form $\mathcal{V}_J^{(i)}$ for some $1\leq J\leq n$ then \emph{all} maximum independent sets of $G_n$ are of the form $\mathcal{V}_{\vv{J}}$. In other words, we need

\begin{prop}
	\label{indie}
	Suppose $g=(U,\mu)$ is connected, fully looped, and flow-regular of degree $d$. Then the graph $G_n=T_n\otimes g$ has exactly $n$ independent sets of size $|g|$; namely
	\begin{equation}
		\mathcal{V}_J=\{J\otimes x\mid x\in U\}, \qquad 1\leq J\leq n
	\end{equation}
\end{prop}
\noindent
To prove this, consider a function $\mathfrak{J}:U\to \{1,2, \cdots, n\}$. Associated to $\mathfrak{J}$, we define a set
\[
[\mathfrak{J}]=\{\mathfrak{J}(x)\otimes x\mid x\in U\}
\]
Any maximum independent set of $T_n\otimes g$ is necessarily of this form (as $g$ is fully looped). 
The idea is to find what $\mathfrak{J}$ can be if $[\mathfrak{J}]$ is an independent set of size $|g|$. Define $J_{\min}=\min_{x\in g}\mathfrak{J}(x)$ and $x_{\min}\in \mathfrak{J}^{-1}(J_{\min})$. We want to show that $\mathfrak{J}(x)=J_{\min}$ for all $x$.
\begin{lem} Let $y\in U$ be such that $y\neq x_{\min}$ and $\mu(x_{\min},y)\neq 0$. Then $\mathfrak{J}(y)=J_{\min}$.
\end{lem}
\begin{proof}
	Suppose $\mathfrak{J}(y)\neq J_{\min}$. Then, by minimality of $J_{\min}$, we have an arrow $J_{\min}\to \mathfrak{J}(y)$ in $T_n$. By hypothesis $\mu(x_{\min},y)\neq 0$ as well. Therefore there exists an arrow $J_{\min}\otimes x_{\min}\to \mathfrak{J}(y)\otimes y$. This is a contradiction with $X$ being an independent set.
\end{proof}
\noindent
Now let $y\in U$ be arbitrary. Note that if there is a directed path $x_{\min}\to y_1\to y_2\to \cdots \to y_\ell=y$, using the lemma inductively, we find that $J_{\min}=\mathfrak{J}(y_1)=\cdots=\mathfrak{J}(y)$. So it suffices to prove that such a path always exists.

\begin{lem} Let $g=(U,\mu)$ be a connected flow-regular graph (say of degree $d$). Then for any two $x,y\in U$, there exists a directed path $x=x_0\to x_1\to\cdots \to x_\ell=y$. (One says $g$ is \emph{strongly connected}).
\end{lem}
\begin{proof}
	Let $S\subset V$. Given any $s\in S$ define
	\[
	d_{\mathrm{ext}}^+(s)=\sum_{t\in V-S}\mu(s,t), \qquad 
	d_{\mathrm{ext}}^-(s)=\sum_{t\in V-S}\mu(t,s)
	\]
	and furthermore, $d_{\mathrm{int}}^\pm(s)=d-d_{\mathrm{ext}}^\pm(s)$ (we are implicitly using flow-regularity here). The number $d^+_{\mathrm{int}}$ counts how many arrows go out of $s$ and land back in $S$. Similarly, $d^-_{\mathrm{int}}$ is the number of arrows coming into $s$ from another node in $S$. As every arrow/loop has a start and a finish
	\[
	\sum_{s\in S} (d_{\mathrm{int}}^+(s)-d_{\mathrm{int}}^-(s))=0
	\]
	Defining $D_S^\pm=\sum_{s\in S}d_{\mathrm{ext}}^\pm(s)$, the above equality results in $D_S^+=D_S^-$.

	Now we consider a very specific $S$. Suppose $S=\{x\}\cup S'$ where $S'$ is the set of all $z\in U$ so that there is a directed path from $x$ to $z$. By definition of $S$, we have $D^+=0$. Since $D^+=D^-$ we find $D^-=0$ as well. However, since $g$ is connected, this is impossible unless $S=U$. Therefore, there exists a directed path from $x$ to all other nodes in $g$.
\end{proof}

\subsection{Proof of $(k,r)$ Clustering Property of $\Psi_n^{(g)}$}
\label{proof_cluster_appendix}
Let $G=(V,\mu)$ be a loopless graph with $|G|=N$ and $\alpha(G)=k$. Let $P=\mathrm{SGP}[G]$. We want to compute the specialization $P(w^{\times k}, z_{k+1}, \cdots, z_{N})$. Let $f: V\to\{z_1, \cdots, z_N\}$ be a bijections. Define 
$$P_f=\prod_{i,j\in V}(f(i)-f(j))^{\mu(i,j)}$$
Then $P=\sum_f P_f$ with the sum over all bijections. For a given $f$, define $S_f=f^{-1}\{z_1, \cdots, z_k\}$. Now if $S_f$ is \emph{not} an independent set, then $P_f(w^{\times k}, z_{k+1}, \cdots)=0$.  We call such an $f$ a \emph{vanishing bijection}. Therefore, suppose $f$ is non-vanishing. Then
\[
P_f(z_1, \cdots, z_k, z_{k+1}, \cdots, z_N) = \prod_{i\in S}\prod_{j\notin S_f}(f(i)-f(j))^{\mu(i,j)}
(f(j)-f(i))^{\mu(j,i)}
\prod_{j_1, j_2\notin S_f}(f(j_1)-f(j_2))^{\mu(j_1, j_2)}
\]
Consequently, we have
\[
P_f(w^{\times k}, z_{k+1}, \cdots, z_N) = \prod_{j\notin S_f}(-1)^{\sum_{i\in S_f}\mu(i,j)}
(f(j)-w)^{\sum_{i\in S_f} \mu(i,j)+\mu(j,i)}
\prod_{j_1, j_2\notin S_f}(f(j_1)-f(j_2))^{\mu(j_1, j_2)}
\]
At this point, it is useful to introduce some general terminology. Given any maximum independent set $S$, for any $i\in V-S$, define 
\begin{subequations}
	\begin{align}
		\nu^+_S(i)&=\sum_{s\in S}\mu(i,s)\\
		\nu^-_S(i)&=\sum_{s\in S}\mu(s,i)\\
		\nu_S(i) &= \nu_S^+(i)+\nu^-_S(i)\\
		\mathrm{sgn}(S)&=\prod_{i\in V-S}(-1)^{\nu_S^-(i)}
	\end{align}
\end{subequations}
Moreover, note that for any maximum independent set $S$, there are $k!(n-k)!$ bijections $f$ such that $S_f=S$. Concretely, such $f$ sends $S$ (resp. $V-S$) to a permutation of $\{z_1, \cdots, z_{k}\}$ (resp. $\{z_1, \cdots, z_{N-k}\}$). Combining all of our observations, we conclude that

\newpage

\begin{equation}
		\mathrm{SGP}(G)(w^{\times k},z_1, \cdots, z_{N-\alpha})=
		k!\sum_{S}\mathrm{sgn}(S)\sum_{\sigma\in \mathfrak{S}_{N-a}}\prod_{i\notin S}(z_{\sigma(i)}-w)^{\nu_S(i)}\prod_{i,j\notin S}(z_{\sigma(i)}-z_{\sigma(j)})^{\mu(i,j)}
	\label{SGPw}
\end{equation}
with the $S$-sum being over all maximum independent sets of $G$ (we use the notation $\mathfrak{S}_n$ for the symmetric group).

Specializing now to the graphs $G_n=T_n\otimes g$, 
consider any node $X\in V-\mathcal{V}_\vv{J}$ (with $\vv{J}=(J_1, \cdots, J_c)$). If $X$ is a node in the component $T_n\otimes g_i$, we write $\mathfrak{c}(X)=i$. Suppose $X=J\otimes x$ for some $1\leq J\leq n$ and $x\in U_i$. Then
\begin{align}
	\nu_{\vv{J}}^+(X)&
	=\theta(J-J_{\mathfrak{c}(X)})r\\
	\nu_{\vv{J}}^-(X)&=
	\theta(J_{\mathfrak{c}(X)-J})r
\end{align}
where $\theta(x)=1$ for $x>0$ and $\theta(x)=1$, and $\theta(x)=0$ if $x<0$. To obtain the above relation we have used the fact that $g$ is flow-regular of degree $r$. Note that, since $J\neq J_{\mathfrak{c}(X)}$, we always have $\nu_{\vv{J}}^+(X)\nu_{\vv{J}}^-(X)=0$ and
\begin{align}
	\nu_{\vv{J}}(X)&:=\nu_{\vv{J}}^+(X)+\nu_{\vv{J}}^-(X)=r \qquad (\forall X)\\
	\mathrm{sgn}(\mathcal{V}_\vv{J})&:=(-1)^{\sum_{X}\nu_{\vv{J}}^-(X)}=
	(-1)^{\sum_{i=1}^c(J_i-1)|g_i|r}=1
\end{align}
In the latter, we are using the fact that each component of $g$ is even (i.e. $|g_i|r=$even for all $i$). Plugging all of this information into Eq. \eqref{SGPw}, together with the fact that all maximum independent sets of $T_n\otimes g$ are vertical (and in particular, there are $n^c$ such sets), we find
\begin{equation}
	\mathrm{SGP}(G_n)(w^{\times k},Z)=n^c k!\prod_{i=1}^{(n-1)k}(z_i-w)^r
	\mathrm{SGP}(G_{n-1})(Z)
\end{equation}
Here, we have used the reduction isomorphism $G_{n-1}\simeq G_n|_{V-\mathcal{V}_{\vv{J}}}$. As a result, with the particular normalization we chose for $\Psi_n^{(g)}$, we have
\begin{equation}
	\Psi^{(g)}_{n}(w^{\times k},Z)=\prod_{i=1}^{(n-1)k}(z_i-w)^r
	\Psi^{(g)}_{n}(Z)
\end{equation}

\subsection{Reorientations of Type $\Lambda(n,k,r)$}
\label{oriproof_appendix}
In this section we will study reorientations with type $\Lambda(n,k,r)$ in graphs $T_n\otimes g$. Let us start our discussion by some general terminology.

Let $G=(V,E)$ is a graph, and $\omega$ is a reorientation of $G$. To make the notation as clear as possible, we denote the out-degree of $x\in V$ in the graph $G^{\omega}$ by $d^+_x(G,\omega)$. Usually, the reference to $G$ is clear from the context, but keeping track of $G$ will prove useful in this section.

The second remark is about isomorphisms. Previously we have defined graph isomorphisms (\ref{graph_appendix}) in terms of the multiplicity function. We will reformulate isomorphisms in terms of the arrow sets. Let $G=(V_G,E_G)$ and $H=(V_H, E_H)$. An isomorphism $\phi:G\xrightarrow{\sim} H$ is a pair of bijections $\phi_V: V_G\xrightarrow{\sim} V_H$ and 
$\phi_E: E_G\xrightarrow{\sim} E_H$ such that
\begin{equation}
	\phi_E(e)=\phi_V(s(e))\to \phi_V(t(e))
\end{equation}
Let $\phi=(\phi_V, \phi_E)$ be an isomorphism of $G\simeq H$, and $\omega$ be a reorientation of $G$. We define the \emph{pushforward} reorientation $\phi_*\omega$ as
\begin{equation}
	\phi_*\omega = \omega\circ \phi^{-1}_E
\end{equation}
which is a reorientation on $H$. Note that $(\phi_*\omega)(\phi_E(e))=\omega(e)$. If we define $\phi^\omega=(\phi_V, \phi_E^\omega)$ as follows
\[
\phi^{\omega}_E(\omega(e)e)=
\omega(e)\phi_E(e)= 
(\phi_*\omega)(\phi_E(e))\phi_E(e)
\]
Then $\phi^\omega: G^{\omega}\xrightarrow{\sim}H^{\phi_*\omega}$ is an isomorphism. So in particular, the pushforward respects the type: $T(\omega)=T(\phi_*\omega)$. As for the sign, we have
\[
\mathrm{sgn}(\phi_*\omega)=\prod_{e\in E_G}(\phi_*\omega)(\phi_E(e))=\prod_{e\in E_G}\omega(e)=\mathrm{sgn}(\omega)
\]
So pushforward also respects the sign.

Throughout, let $g$ be a $(k,r)$ intensive graph, $G_n=T_n\otimes g$, and $V_n$ the vertex set of $G_n$.  Furthermore, let $\omega_n$ be an arbitrary reorientation of $G_n$ of type $\Lambda(n,k,r)$. Note that all parts of $\Lambda(n,k,r)$ are of the form $mr$, with $0\leq m<n$. Define
\begin{equation}
	V^{[m]}(G_n,\omega_n) := \{x\in V\mid d^+_x(G_n, \omega_n)=mr\}
\end{equation}
\begin{prop}
	\label{ori_1}
	$V^{[m]}(G_n,\omega_n)$ is an independent set for all $0\leq m<n$. Moreover, if $x\in V^{[m]}(G_n,\omega_n)$ and $y\in V^{[m']}(G_n,\omega_n)$ with $m>m'$ then no arrow $y\to x$ exists in $G_n^{\omega_n}$.
\end{prop}

\begin{proof}
	Let us start with $U=V^{[0]}(\omega_n)$. As no two nodes with out-degree zero can be adjacent, $U:=V^{[0]}(\omega_n)$ is a maximum independent set of $G_n^{\omega_n}$ (and therefore also of $G_n$). Define $H=G_n|_{V_n-U}$, and let
	$\phi: H\xrightarrow{\sim} G_{n-1}$ be the reduction isomorphism \eqref{reduction_graph_iso}. Let $\omega_H=\omega_n|_H$ be the reorientation $\omega_n$ induces on $H\subset G_n$, and $\omega_{n-1}=\phi_*\omega_H$. We make the following observations:
	\begin{enumerate}
		\item Take any node $x\notin U$. There exists no arrow in $G_n^{\omega_n}$ with source in $U$. Moreover, as $U$ is a maximum independent set, there are exactly $r$ arrows with source $x$ and target a node in $U$. Therefore,
		\[
		d_x^+(H, \omega_H)=d_x^+(G_n,\omega_n)-r
		\]
		for all $x\in V_H$.
		\item Consequently, $\omega_H$ is of type $\Lambda(n-1,k,r)$. So, $T(\omega_{n-1})=T(\phi_*\omega_H)=\Lambda(n-1,k,r)$.
		\item Given any $m\neq 0$, since $V^{[m]}(G_n, \omega_n)\subset V_H$ ($V_H$ being the node set of $H$), we find
		\[
		\phi\left(V^{[m]}(G_n, \omega_n)\right)=V^{[m-1]}(G_{n-1}, \omega_{n-1})
		\]
		\item Let $m>m'>0$. Suppose $x\in V^{[m]}$ and $y\in V^{[m']}$. By the previous observation, 
		$$
		\begin{aligned}
			\phi(x)&\in V^{[m-1]}(G_{n-1},\omega_{n-1})\\
			\phi(y)&\in V^{[m'-1]}(G_{n-1},\omega_{n-1})
		\end{aligned}
		$$
		So for the secondary part of the proposition, it suffices to show that there is no arrow $\phi(y)\to \phi(x)$ in $G_{n-1}^{\omega_{n-1}}$.
		
	\end{enumerate}
	Since independent sets map to one another under isomorphisms, if $V^{[m-1]}(G_{n-1},\omega_{n-1})$ is a maximum independent set of $G_{n-1}$ then $V^{[m]}(G_n, \omega_n)$ is a maximum independent set of $G_n$. An induction on $n$ now finishes the job (base of induction is $n=1$, for which $G_1=T_1\otimes g$ is a graph with $|g|$ nodes and no arrows).
\end{proof}
\begin{cor}
	When $g$ is connected, each maximum independent set $[J]$ of $G_n$ is equal to $V^{[m_J(\omega_n)]}(G_n, \omega_n)$ for some $0\leq m_J(\omega_n)<n$.
\end{cor}
\begin{proof}
	There are exactly $n$ maximum independent $[J]$ sets for the connected case, and there are $n$ sets $V^{[m]}(G_n, \omega_n)$. So $[J]=V^{[m_J(\omega_n)]}(G_n, \omega_n)$ for some $0\leq m_J(\omega_n)<n$.
\end{proof}
\noindent
Suppose $g$ is connected, and $m_J:=m_J(\omega_n)$ as in the corollary. Define a reorientation $\pi:=\pi(\omega_n)$ on $T_n$ as follows
\begin{equation}
	\pi(I\to J) = \begin{cases}
		+1 & m_I>m_J\\
		-1 & m_I<m_J
	\end{cases}
\end{equation}
Using proposition \ref{ori_1}, and patching isomorphism \eqref{patching_iso}, it is now clear that

\begin{cor}
	Let $g$ be connected, and $1\leq I<J\leq n$. Then $G_n^{\omega_n}|_{[I]\cup [J]}\simeq \pi(I\to J)\: G_2$.
\end{cor}

\begin{cor}
	Let $g$ be connected and even. Then $\mathrm{sgn}(\omega_n)=+1$.
\end{cor}
\begin{proof}
	Define $\omega_{I,J}=\omega_n|_{[I]\cup [J]}$ say with $I<J$. By previous corollary, $\mathrm{sgn}(\omega_{I,J})=(\pi(I\to J))^{kr}$, as there are $kr$ arrows in $G_2$. Since $kr=$even, then $\mathrm{sgn}(\omega)_{I,J}=+1$. Now, $\mathrm{sgn}(\omega_n)=\prod_{I<J}\mathrm{sgn}(\omega)_{I,J}=1$.
\end{proof}

\begin{cor}
	Let $g=g_1\sqcup \cdots\sqcup g_c$ with $g_i$ connected and even. Then $\mathrm{sgn}(\omega_n)=+1$.
\end{cor}
\begin{proof}
	The sign of $\omega_n$ is the product of the sign of the induced reorientation on each component. Since the latter is always $+1$, we have $\mathrm{sgn}(\omega_n)=+1$.
\end{proof}
\noindent
As for the question of how many orientations of type $\Lambda(n,k,r)$ exist, let us start with a connected $g$. In this case, we have $n$ maximum independent sets. Given any permutation $\sigma\in \mathfrak{S}_n$ we can build an orientation $\omega_{\sigma}$ of type $\Lambda(n,k,r)$ with $V^{[m]}(G_n, \omega_{\sigma})=[\sigma^{-1}(n-m)]$. Concretely, if there is an arrow $u\to v$ in $g$, we define
\[
\omega_{\sigma}(I\otimes u\to J\otimes v) =\begin{cases}
	+1 & \sigma(I)<\sigma(J)\\
	-1 & \sigma(I)>\sigma(J)
\end{cases}
\]
As such we have $n!$ orientation of type $\Lambda(n,k,r)$. When $g$ has $c$ connected components, the number of such orientations becomes $n!^c$.

\begin{cor}
	\begin{equation}
		\Psi^{(g)}_{n}=m_{\Lambda(n,k,r)}+\sum_{\lambda\neq \Lambda}c_{\lambda} m_{\lambda}
	\end{equation}
\end{cor}
\begin{proof}
	By the considerations so far, if $g$ has $c$ connected components, then there are $n!^c$ orientations of type $\Lambda=\Lambda(n,k,r)$. Moreover, $\prod_{p\geq 0}\mathrm{Freq}(\Lambda)_p!$ is equal to $k!^n$ for $\Lambda(n,k,r)$. Therefore, with $\mathrm{SGP}[G]=\sum_{\lambda} c_\lambda m_\lambda$, we have
	\[
	c_{\Lambda(n,k,r)} = k!^n\sum_{\tau(\omega)=\Lambda(n,k,r)}\mathrm{sgn}(\omega)=
	n!^c k!^n
	\]
	Moreover, due to the particular normalization for $\Psi_n^{(g)}$, we have
	\[
	\Psi^{(g)}_n = \frac{1}{n!^c k!^n}\mathrm{SGP}(G)=m_{\Lambda(n,k,r)}+\cdots
	\]
\end{proof}
Finally, let us prove that if $\Psi_n^{(g)}$ is squeezable, then highest partition in necessarily $\Lambda(n,k,r)$. More precisely:
\begin{prop}
	$\Lambda(n,k,r)$ is the largest orientation type of $G_n$ in lexicographic order.
\end{prop}
\begin{proof}	
	
	For brevity, let $\Lambda_n:=	\Lambda(n,k,r)$. Let $\omega$ be any reorientation and let $\lambda=T(\omega)$. Define $\lambda^*=(\lambda_N, \lambda_{N-1}, \cdots, \lambda_2, \lambda_1)$. We have
	\[
	[\Lambda_n]^* = (0^{\times k}r^{\times k}\cdots [(n-1)r]^{\times k})
	\]
	If $\lambda^*_k>0$, then $\Lambda_n\geq_L \lambda$, and we are done. If, however $\lambda^*_k=0$, then $G_n^\omega$ possesses a maximum independent set $U$ with all nodes of out-degree zero. Therefore, similar to the argument in the proof of Prop. \ref{ori_1}, if $\omega'$ is pushfroward of the orientation induced on $H=G_n|_{V_n-U}$ (i.e. $\omega'$ is an orientation on $G_{n-1}$), and $\mu=T(\omega')$, we find
	\[
	\lambda^*_{k+l} = r+\mu^*_l
	\]
	This essentially reduces the problem to showing $\Lambda_{n-1}\geq_L \mu$, where $\mu=T(\omega')$ for any orientation $\omega'$ of $G_{n-1}$. An induction on $n$ now finishes the job.
\end{proof}

\section{Immediate Consequences of Separability}
\label{separability_consq_appendix}
Let $\Psi$ be a wavefunction in $N$ variables that does not vanish when fusing $k$ particles, but vanishes of order $r$ when $k+1$ particles are fused. Let us use the shorthand notation $\vv{z}=z_1, \cdots, z_{N-k-1}$. We denote the $k$-fusion as $\Phi(w; z_0, \vv{z}):=\Psi(w^{\times k}, z_0, \vv{z})$. Specializing \eqref{kfusion_eq} to $z_0=w-\zeta$, we compute:
\begin{equation}
\Phi(w;w-\zeta, \vv{z})=(-\zeta)^r\prod_{i=1}^{N-k-1}(z_i-w)^rR(w;w-\zeta, \vv{z})=(-\zeta)^r \prod_{i=1}^{N-k-1}(z_i-w)^rR(w;w, \vv{z})+O(\zeta^{r+1})
	\label{ap1}
\end{equation}
Since $R(w;w,\vv{z})\neq 0$, to leading order $\Phi(w;w-\zeta, \vv{z})\sim \zeta^r$.

We will also use the shorthand $\vv{w}=w_1, \cdots, w_{k+1}$. Let $\mathcal{P}_{k+1}^l$ be projection with respect to \emph{first} $k+1$ variables. Suppose $\mathcal{P}_{k+1}^{l}\Psi=0$ for $l\leq m=S_{k+1}$ and 
\begin{equation}
	\label{D2}
	(\mathcal{P}_{k+1}^{m}\Psi)(\vv{w}, \vv{z}) = 
	\chi(\vv{w})Q\left(\frac{w_1+\cdots+w_{k+1}}{k+1};\vv{z}\right)
\end{equation}
for some $\chi\in \mathcal{T}_{k+1}^{l}$ and some polynomial $Q$. Define $\mathscr{N}_\chi=\chi(1,0^{\times k})$. We may assume $\mathscr{N}_\chi=0,1$, without loss of generality. The key point here is that if $(\mathcal{P}_{k+1}^{l}\Psi)(w^{\times k}, w-\zeta, \vv{z})\neq 0$, then
\[
(\mathcal{P}_{k+1}^{l}\Psi)(w^{\times k}, w-\zeta, \vv{z})=
O(\zeta^l)
\]
Moreover, since $
\Psi(\vv{w},\vv{z})=\sum_{m\geq l}(\mathcal{P}_{k+1}^m\Psi)(\vv{w}, \vv{z})$, the condition \eqref{D2} results in
\begin{equation}
	\Phi(w; w-\zeta, \vv{z})=\mathscr{N}_\chi(-\zeta)^{m}
	Q\left(w;\vv{z}\right)+O(\zeta^{m+1})
	\label{ap2}
\end{equation}
Comparing \eqref{ap1} and \eqref{ap2}, there are only two possibilities:
\begin{enumerate}
	\item $\mathscr{N}_\chi=1$ if and only if $m=r$.
	\item $\mathscr{N}_\chi=0$ if and only if $m<r$
\end{enumerate}
In particular, as $\mathscr{N}_\chi=1$ is part of the definition of separability, separable $(k,r)$ clustering states have $S_{k+1}=m=r=\deg \chi$. 

Now, let $\gv{\Psi}$ be a $\chi$-separable (in particular, $(k,r)$ clustering) sequence. Then, using the clustering property, we have
\[
\Psi_{n+1}(w^{\times k}, w-\zeta, \vv{z}) = (-\zeta)^{r}\prod_{i=1}^{nk-1}(z_i-w)^r
\Psi_n(w, \vv{z}) + O(\zeta^{r+1})
\]
Combined with \eqref{ap2} with $\mathscr{N}_\chi=1$ and $m=r$, we find that
\[
	Q(w;\vv{z})=\prod_{i=1}^{N-k-1}(z_i-w)^r\Psi_n(w,\vv{z})
\]
which means, with $w_{\mathrm{cm}}=(w_1+\cdots+w_{k+1})/(k+1)$, 
\begin{equation}
	(\mathcal{P}_{k+1}^{r}\Psi_{n+1})(\vv{w}, \vv{z})=\chi(\vv{w})\prod_{i=1}^{N-k-1}(z_i-w_{\mathrm{cm}})^r\Psi_n(w_{\mathrm{cm}}, \vv{z})
\end{equation}

\section{Motivations Behind the Design of $H_\chi$}
\label{Zk2_appendix}
In this subsection, we take a look at $H_{\chi}$ Hamiltonians for $\chi\in \mathcal{T}_{k+1}^{r}$  ($\chi(1, 0^{\times k})$) for a few values of $k,r$. In each case, we inquire if $H_{k+1}^{r-1}$ and $H_{\chi}$ have a unique ground state. We will also give a lightning-quick review of the $\mathbb{Z}_k^{(r)}$-algebras where the $H_\chi$ Hamiltonians originate.

Before we get to the case-by-case study, let us quickly review the space $\mathcal{T}_{k+1}^{r}$. Recall that $\mathcal{T}_{k+1}^{r}$ is the Bargmann space of symmetric homogeneous translational invariant polynomials in $k+1$ variables and of degree $r$. Let us describe a basis for this space. Let $\lambda = (\lambda_1, \lambda_2, \cdots, \lambda_{\ell})$ be a partition such that $2\leq \lambda_i\leq a$ (for all $i$) and $\lambda_1+\cdots+\lambda_\ell = m$. We use the notation $\pi^*(r,k+1)$ for the set of these partitions. Let $z_{\mathrm{cm}}=(z_1+\cdots+z_{k+1})/(k+1)$ be the center-of-mass. Define
\begin{equation}
	\label{pbasis_EQ}
	p_n(z_1, \cdots, z_{k+1})=\sum_{i=1}^{k+1}(z_i-z_{\mathrm{cm}})^n, \qquad p_{\lambda}=p_{\lambda_1}p_{\lambda_2}\cdots p_{\lambda_\ell}
\end{equation}
Then the set $\{p_{\lambda}\mid \lambda\in \pi^*(r,k+1)\}$ is a basis for $\mathcal{T}_m^a$ \cite{liptrap2008translation}. Thus, the dimension of this space is given by $d(r,k+1)=|\pi^*(r,k+1)|$ \cite{Simon_Pseudopotentials}. Throughout, $d(r,k+1)$ is the dimension of $\mathcal{T}_{k+1}^r$. The condition $\chi(1, 0^{\times k})=1$ is referred to as `fixing the normalization'.

\paragraph{Laughlin States} We begin by $k=1$ and $r=2m$. In this case, $\chi(z_1, z_2)=(z_1-z_2)^{2m}$ by necessity and any null state of $H_{k+1}^{r-1}$ is $\chi$-separable. Thus, the null states of $H_{k+1}^{r-1}$ and $H_\chi$ coincide.

\paragraph{Paired States} Consider $k=2$ and $r=2,3,4$. These correspond to Pfaffian, Gaffnian, and Haffnian. In all of these cases, $d(r,k+1)=1$. After fixing the normalization, the respective minimal polynomial $\chi$ can only be one thing: $\chi_{\text{Pf}}=(9/5) p_2$, $\chi_{\text{Gf}}=(27/7)p_3$, and 
$\chi_{\text{Hf}}=(81/25) p^2_2$. In all of these examples, the densest null state of $H_{k+1}^{r-1}$ is also the densest null state of $H_\chi$.

\paragraph{Read-Rezayi States} The next set of examples is the Read-Rezayi family. Let $r=2$ and $k$ arbitrary. Again, we have $d(2, k+1)=1$ and after fixing the normalization condition,  we are left with $\chi(z_1, \cdots, z_{k+1}) = \frac{1}{k}\sum_{i<j}(z_i-z_j)^2$. Again, the densest zero-energy state of $H_{k+1}^{1}$ and $H_\chi$ are the same.

\paragraph{$\mathbb{Z}_k^{(2)}$-States}
The example to differentiate between $H_{k+1}^{r-1}$ and $H_\chi$ would be $r=4$ and $k\geq 3$. In this case, $d(4,k+1)=2$, so $H_{k+1}^3$ does not possess a unique ground state. In contrast, we will present an argument that if $\chi\in \mathcal{T}_{k+1}^4$ (with $\chi(1, 0^{\times k})=1$), then $H_\chi$ is  \emph{likely} to have a \emph{unique} ground state. These CFT-based examples are borrowed from another work of ours \cite{pakatchi2021CFT}, where they are thoroughly analyzed. However, as far as the current paper is concerned, we only use these examples to motivate the study of $H_\chi$ Hamiltonians and highlight the importance of separability.

A $\mathbb{Z}_k^{(m)}$-algebra is a special type of parafermionic conformal field theory with charge conjugation and $\mathbb{Z}_k$ symmetry. It has $k$ fields $\psi_0\equiv 1, \psi_1, \psi_2, \cdots, \psi_{k-1}$, with scaling dimensions $h_a\equiv h(\psi_a)=a(k-a)/2\nu$ where $\nu=k/2m$. The $\mathbb{Z}_k$ symmetry manifests through the simple fusion rules $\psi_a\star \psi_b = \psi_{a+b\pmod{k}}$. The charge conjugate of $\psi_a$ is denoted by $\psi_{a}^+=\psi_{k-a}$. The operator product algebra, which is invariant under charge conjugation, is given by
\begin{subequations}
	\begin{align}
		\psi_a(z)\psi_b(w)(z-w)^{ab\gamma} &= C_{a,b} \psi_{a+b}(w) + \cdots &&(a+b< k)\\
		\psi_a(z)\psi^+_a(w)(z-w)^{2h_a} &= 1 + \frac{2h_a}{c}(z-w)^2 T(w)+\cdots 
	\end{align}
\end{subequations}
Here, $T$ is the energy-momentum tensor, $c$ is the central charge and $C_{a,b}$ are the structure constants (which we calculate in Ref. \cite{pakatchi2021CFT}). As an example, the case $m=1$ yields the Zamolodchikov-Fateev parafermions \cite{Zamolodchikov_Fateev_Parafermion}. 

Each associative $\mathbb{Z}_k^{(m)}$-algebra leads to a trial FQH ground state sequence $\mathbf{\Psi}=(\Psi_n)$, with $N(n)=nk$ and $N_\phi(n)=2m(n-1)$, filling fraction $\nu=k/2m$, and shift $\mathcal{S}=2m$:
\begin{equation}
	\Psi_n(z_1, \cdots, z_{nk})=\frac{1}{\prod_{a=1}^{k}C^{n}_{1,a}}\avg{\psi_1(z_1)\cdots \psi_1(z_{nk})}
	\prod_{i<j}(z_i-z_j)^{1/\nu}
	\label{example_EQ}
\end{equation}
For example, the wavefunctions in the special case $m=1$ are the $\mathbb{Z}_k$ Read-Rezayi states \cite{Read-Rezayi}. For a general $m$, these state are $(n k, 2(n-1)m)$ uniform. Moreover, the wavefunction $\Psi_n$ is $(k,2m)$-clustering \cite{estienne2}. In addition, we have shown that $\gv{\Psi}$ is  separable with minimal polynomial given by
\begin{equation}
	\chi(z_1, \cdots, z_{k+1}) = 
	\lim_{w\to \infty} w^{-2m(k-1)}\Psi_2(w^{\times (k-1)}, z_1, \cdots, z_{k+1})\in \mathcal{T}_{k+1}^{2m}
\end{equation}
Consequently, $\Psi_n$'s are a densest zero-energy state of $H_\chi$. From now on, we will specialize to the case $m=2$ (i.e., $r=2m=4$) and $k\geq 3$.

We now present the argument that, when $\chi\in \mathcal{T}_{k+1}^4$ (with $\chi(1, 0^{\times k})=1$) and $k\geq 3$,  $H_\chi$ is likely to have a unique ground state. When $k>2$, the space $\mathcal{T}_{k+1}^4$ is two-dimensional, and the following is a convenient basis for it:
\[
F_1 =\frac{1}{2k!}\mathrm{SGP}\Big[
\vcenter{
	\hbox{
		\includegraphics[scale=.5]{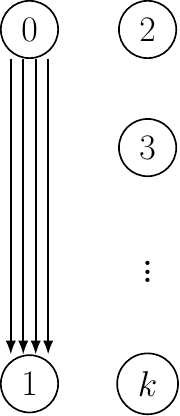}
	}
}
\Big], \qquad F_2=
\frac{1}{4 k (k-3)!}
\mathrm{SGP}\Big[
\vcenter{
	\hbox{
		\includegraphics[scale=.5]{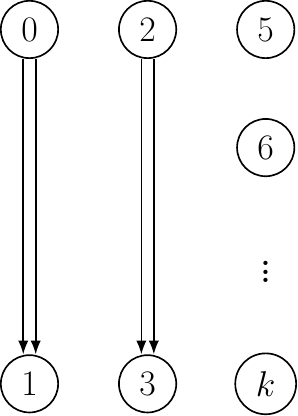}
	}
}
\Big]
\]
Since $F_1(1, 0^{\times k})=1$ and $F_2(1, 0^{\times k})=0$, any $\chi\in \mathcal{T}_{k+1}^{4}$  satisfying the normalization $\chi(1, 0^{\times k})=1$ is of the form $\chi = F_1+ F_2/\beta$, for some parameter $\beta$. As a result, $H_\chi$ is a one-parameter family of Hamiltonians, with $\beta$ as its parameter. Let us denote this Hamiltonian as $H_\beta$ for convenience. In a parallel analysis, one discovers that the $\mathbb{Z}_k^{(2)}$-algebras are a one-parameter family of CFTs, parametrized by $t$. For example, the central charge is given by
\[
c = \frac{4(k-1)t(t+k-1)}{(k+2t-2)(k+2t)}
\]
The square of the structure constants $C_{a,b}$ is also a rational function in $t$ \cite{Zamolodchikov_Fateev_Parafermion, pakatchi2021CFT}. Additionally, if $\chi$ is the minimal polynomial of the $\mathbb{Z}_k^{(2)}$-algebra with parameter $t$, then $\beta = t(t+k-1)$. Therefore, unless if the CFT parameter $t$ satisfies $\beta = t(t+k-1)$, the Hamiltonian $H_\beta$ rejects the wavefunction coming from that $\mathbb{Z}_k^{(2)}$-algebra. In other words, the Hamiltonian fixes the $\mathbb{Z}_k^{(2)}$-algebra and selects its sequence $\mathbf{\Psi}=(\Psi_n)$ \eqref{example_EQ} as a ground state. Thus, \emph{at least among the wavefunctions coming from $\mathbb{Z}_k^{(2)}$-algebras}, the Hamiltonian has a unique densest zero-energy state. Admittedly, this is no proof that $H_\chi$ has a unique ground state. However, the apparent Hamiltonian/CFT duality is suggestive of such uniqueness.

\section{Umbral Tricks}
\label{umbral_appendix}
The symmetric monomials $m_\lambda$ are not a multiplicative basis for space of symmetric polynomials. As such working with root-decomposition can prove difficult. Fortunately, there is a trick to make such calculations easier. Throughout, 
\begin{itemize}
	\itemsep -2pt
\item[--] $R_N=\mathbb{C}[z_1, \cdots, z_N]$ is the (graded) ring of polynomials.
\item[--] $\Lambda_N = \mathbb{C}[z_1, \cdots, z_N]^{S_N}$ is the (graded) ring of symmetric polynomials. 
\item[--] $U = \mathbb{C}[b_0, b_1, b_2,\cdots]$ the so-called \emph{umbral} space. Define $U^N$ as the subspace of polynomials of degree $N$.
\end{itemize}
We define the \emph{umbral (linear) map} $\mathcal{U}:R_N\to U^N$ via
\begin{equation}
	\mathcal{U}(z_1^{\alpha_1}z_2^{\alpha_2}\cdots z_{N}^{\alpha_N})=\frac{1}{N!} b_{\alpha_1}b_{\alpha_2}\cdots b_{\alpha_N}
\end{equation}
and extend it linearly to the entire $R_N$. This naturally induces a linear map $\mathcal{U}:\Lambda_N\to U^N$. The action of $\mathcal{U}$ on augmented symmetric monomial $\widetilde{m}_{\lambda}:=\mathscr{S}(\prod_{i}z_i^{\lambda_i})$ is 
\begin{equation}
\mathcal{U}(\widetilde{m}_{\lambda})=\prod_{p\geq 0} b_p^{\mathrm{Freq}(\lambda)_p}
\end{equation}
Clearly $\mathcal{U}$ is an isomorphism $\Lambda_N\simeq U^N$. Due to the basis of the umbral space being multiplicative (in contrast to $\widetilde{m}_{\lambda}$), working in the umbral space is often simpler. We will demonstrate this with two useful examples.

\subsection{Constraint on Root-Decomposition Imposed by Translation Invariance}
\label{umbral1}
Let us find the constraint imposed by translational invariance puts on the root-decomposition (\ref{constraint1}). A symmetric polynomial $\Psi$ is translational invariant if and only if $L_-\Psi=0$ with $L_-=\sum_i \partial_i$. We need to find how $L_-$ acts on the umbral transform. Suppose $\alpha$ is a permutation of a partition $\lambda$, and $n=\mathrm{Freq}(\lambda)$ (Notation: $\vv{b}^n = b_0^{n_0}b_1^{n_1}\cdots$).
\begin{align*}
	&\mathcal{U}\left(L_- \prod_{i=1}^N z_i^{\alpha_i}\right)=
	\mathcal{U}\left(\sum_{i=1}^N \alpha_i 
	z_i^{-1}\prod_{j=1}^N z_i^{\alpha_i}
	\right)
	\\
	&\qquad=
	\frac{1}{N!}\sum_i \alpha_i \frac{b_{\alpha_i-1}}{b_{\alpha_i}}
	\vv{b}^{n}=
	\frac{1}{N!}\sum_{s\geq 0} sn_s \frac{b_{s-1}}{b_{s}}
	\vv{b}^{n}
	\\
	&\qquad =\left(\sum_{s\geq 0}sb_{s-1}\pd{}{b_s}\right)\mathcal{U}\left( \prod_{i=1}^N z_i^{\alpha_i}\right)
\end{align*}
Therefore if we define
\begin{equation}
	\mathcal{L}_-=\sum_{s\geq 0}sb_{s-1}\pd{}{b_s}
\end{equation}
then $L_-\Psi=0$ if and only if $\mathcal{L}_-\mathcal{U}(\Psi)=0$. Let $c^*_{n}=c_{\mathrm{Part}(n)}$. The umbral transformation of the root-decomposition $\Psi=\sum_{n}c_n^*m_{\mathrm{Part}(n)}=\sum_{n}\prod_{s\geq 0} (n_s!)^{-1} c_n^*\widetilde{m}_{\mathrm{Part}(n)}$ is given by
\[
\mathcal{U}[\Psi]
=\sum_{n} \frac{c_n^*}{\prod_{s\geq 0}n_s!}\vv{b}^{n}
\]
It is now straightforward to check that $\mathcal{L}_-\mathcal{U}[\Psi]=0$ results in \eqref{constraint1}.

\subsection{Root-Decomposition of the SGP of Disjoint Unions}
Let $G_1, G_2, \cdots, G_c$ be $c$ graphs, and write
\begin{align*}
\mathrm{SGP}(G_i) &=\sum_{\nu^{(i)}}C'^{(i)}_{\nu^{(i)}}\widetilde{m}_{\mathrm{Part}(\nu^{(i)})}\\
C'^{(i)}_{\nu^{(i)}}&=\sum_{T(\omega)=\mathrm{Part}(\nu^{(i)})}\mathrm{sgn}(\omega)
\end{align*}
where the first sum is over occupations $\nu^{(i)}$ and the second sum is of orientations $\omega$ of type $\mathrm{Part}(\nu^{(i)})$. We also define $C^{(i)}_{\lambda}=C'^{(i)}_{\mathrm{Freq}(\lambda)}$. We want to find a formula for $\mathrm{SGP}(G)$ with $G=G_1\sqcup G_2\sqcup \cdots\sqcup G_c$ in terms of the $C_{\lambda}^{(i)}$. Given a graph $H=(V,\mu)$ define
$$
P_H = \prod_{i,j=1}^{N}(z_i-z_j)^{\mu(i,j)}
$$
Then $\mathrm{SGP}(H)=\mathscr{S}[P_H]$. Now, note that
\[
\mathcal{U}[P_{G_i}] = \frac{1}{|G_i|!}
\sum_{\nu^{(i)}}C'^{(i)}_{\nu^{(i)}}
\vv{b}^{\nu^{(i)}}
\]
Using the notation  $\vv{z}^{(i)}=(z^{(i)}_1, \cdots, z^{(i)}_{N_i})$, where $N_i=|G_i|$, we have
\[
P_{G}(\vv{z}^{(1)}, \cdots, \vv{z}^{(c)})=\prod_{i=1}^c P_{G_i}(\vv{z}^{(i)})
\]
The umbral transform of product of polynomials with \emph{distinct} variables, is (other than the $N!^{-1}$ in front) the product of the umbrals; i.e.
\[
\mathcal{U}[P_G]=\frac{1}{|G|!}\sum_{\nu^1,\cdots, \nu^{c}}
\left\{\prod_{i=1}^c C'^{(i)}_{\nu^{(i)}}\right\}
\vv{b}^{\nu_1+\nu_2+\cdots+\nu_c}
\]
If we now use the inverse umbral, we can find the root-decomposition of $\mathrm{SGP}(G)$:
\begin{equation}
	\begin{aligned}
	\mathrm{SGP}(G) &= \sum_{\nu^1,\cdots, \nu^{c}}
	\left\{\prod_{i=1}^c C'^{(i)}_{\nu^{(i)}}\right\}
	\widetilde{m}_{\mathrm{Part}(\sum_{i=1}^c \nu^{(i)})}\\
	&= \sum_{\lambda^1,\cdots, \lambda^{c}}
	\left\{\prod_{i=1}^c C^{(i)}_{\lambda^{i}}\right\}
	\widetilde{m}_{\lambda^{(1)}\cup\cdots \lambda^{(c)}}
	\end{aligned}
\label{disjoint_root_decomp}
\end{equation}
In particular, without loss of generality, we may assume $\lambda^{(i)}$ is a skew orientation type of $G_i$ (so that $C^{(i)}_{\lambda^{i}}\neq 0$).

\section{Properness of Saturated Graphs}
\label{saturated_proper_appendix}
Recall that a $(k,l,m)$ saturated graph $g=(V_g, \mu_g)$ is defined as $\mathfrak{g}(l, m^{\times(k-1)})$ with $(l\geq m)$. Let $\mathcal{U}=(U_1, U_2, \cdots, U_n)$ be set of subsets of $V_g$. Without loss of generality, suppose none of $U_i$ are empty. Say $||\mathcal{U}||=a=pk+q$ ($p\geq 0$, $0\leq q<k$). Note that as $|U_i|\leq k$, one necessarily has $n\geq p+1$. Let $G=(V,\mu)=T_n\otimes g$. Define
\[
C_J = \{J\}\otimes U_J,\qquad C=\bigcup_{J=1}^n C_J
\]
Then $C$ is an $a$-cut and $\mu(\mathcal{U})=w(C)$. We call $C_J$ the $J$th \emph{vertical slice}. We are looking for the minimal possible value $w(C)$ can be.

For the purposes of this problem it is easier to work with labels $1,2,\cdots, k$, rather than $0,1,2\cdots, k-1$. So we define the horizontal set $H_k$ as $H_0$.  Given any $1\leq b\leq k$, define the $b$th horizontal \emph{slice} as
\[\mathfrak{h}_b = C\cap H_b\]
where $H_b$ is the $b$th horizontal set. Given any two nodes in $C$, if they are in the same horizontal slice, then they connect via an arrow of multiplicity $l$. Any two nodes in distinct horizontal slices connect with an arrow with multiplicity $m$. Therefore, with $h_b=|\mathfrak{h}_b|$ and $u_i=|U_i|$ we have
\[
w(C)=\mu(\mathcal{U})=m\sum_{i<j}u_iu_j+(l-m)\sum_{b=1}^{k}\frac{h_b(h_b-1)}{2}
\]
Due to the symmetry of the problem, without loss of generality, suppose $h_1\geq h_2\geq \cdots\geq h_{k}$, and call the partition $h = (h_1, h_2, \cdots, h_{k})$ the \emph{shape} of $C$. Note that $h$ is a partition of $a$ of length at most $k$. Define $\rho = ((p+1)^{\times q}, p^{\times(k-q)})$. This is also a partition of $a$ of length at most $k$. Note that both $h_1<\rho_1$ and $h_k>\rho_k$ are impossible for any cut shape $h$. Moreover, if $h_1=\rho_1$ or if $h_k=\rho_k$ then $h=\rho$. So if $h\neq \rho$, we define $Y$ (resp. $y$) as the largest (resp. smallest)  $b$ such that $h_b-\rho_b>0$ (resp. $h_b-\rho_b<0$).

\begin{lem}$h_Y-h_y> 1$. Moreover, $h_{Y}>h_{Y+1}$ and $h_{y}<h_{y-1}$.
\end{lem}
\begin{proof}
Define $s,t>0$ such that $h_Y = \rho_Y+s$ and $h_y = \rho_y-t$. Thus, $h_Y-h_y = (\rho_Y-\rho_y)+s+t$. Since $Y<y$, the first assertion follows. Moreover, by definitions of $Y$ and $y$, we have $h_Y>\rho_{Y}\geq \rho_{Y+1}\geq h_{Y+1}$ and $h_y<\rho_{y}\leq \rho_{y-1}\leq h_{y-1}$.
\end{proof}
\noindent
Let $e^i$ be a $k$-dimensional vector defined as $(e^i)_j=\delta_{ij}$. Due to the lemma, the sequence
$$h'= h-e_Y+e_y$$ is also a partition. We will now construct an $a$-cut $C^{(1)}$ with $w(C^{(1)})\leq w(C)$ of shape $h'$. Given each node in $x\in C$, we denote its \emph{coordinates} as $(J,b)$ (i.e. $x\in C_J$ and $x\in \mathfrak{h}_b$). Now as $h_{Y}-h_y>1$ there exists in node $(J,Y)$ in $C$ so that $(J,y)\notin C$. Define 
\[
C^{(1)} = (C-\{(J,Y)\})\cup\{J,y\}
\]
Let us now classify the nodes of $C$ with respect to this pair:
\begin{enumerate}
	\item Nodes in the vertical slice $C_J$. Such a node connects to neither $(J,Y)$ nor $(J,y)$.
	
	\item Nodes in $\mathfrak{h}_Y-C_J$. These nodes connect to $(J,Y)$ and $(J,y)$ with $l$ and $m$ arrows respectively. Define
	\[d_1=(h_Y-1)l, \qquad d'_1 = (h_Y-1)m\]
	\item Nodes in $\mathfrak{h}_y-C_J$. These nodes connect to $(J,Y)$ and $(J,y)$ with $m$ and $l$ arrows respectively. Define
	\[d_2=(h_y-1)m, \qquad d'_2 = (h_y-1)l\]
	\item Any other node connects to both $(J,Y)$ and $(J,y)$ with $m$ arrows.
\end{enumerate}
As a result,
\[
\begin{aligned}
w(C)-w(C^{(1)})&=d_1+d_2-d'_1-d'_2\\
&=(h_Y-h_y-1)(l-m)\geq 0
\end{aligned}
\]
So $w(C^{(1)})\leq w(C)$. Repeating the process we find a sequence of cuts $C=C^{(0)}, C^{(1)}, \cdots, C^{(l)}=C'$ with $w(C^{(i)})\geq w(C^{(i+1)})$ and $C'$ being of shape $\rho$. As we are looking for a cut of minimal weight, without loss of generality, let us assume $C$ is of shape $\rho$.

For a cut $C$ of shape $\rho$, using $\mathcal{S}_k(a)=kp(p-1)/2 +pq$, we have
\[
w(C) = m\sum_{i<j}u_iu_j + \mathcal{S}_k(a)(l-m)
\]
Moreover, 
\[
\sum_{i<j}u_iu_j = \frac{1}{2}\left[\left(\sum_i u_i\right)^2-\sum_{i}u_i^2\right]
\]
as $\sum_i u_i = a$, minimizing $\sum_{i<j}u_iu_j$ is the same as maximizing $\sum_i u_i^2$ (constrained to $u_i\leq k$). Up to permutation of $i$'s, the solution is $u_1=u_2=\cdots =u_p=k$ and $u_{p+1}=q$ and $u_J=0$ for $J>p+1$. This results in
\[
\min\sum_{i<j}u_iu_j=
\mathcal{S}_k(a)k
\]
Combined with $r=l+(k-1)m$, we now find
\[
\min w(C)=\min \mu(\mathcal{U})=\mathcal{S}_k(a)(l+(k-1)m)=\mathcal{S}_k(a)r
\]
This minimum value occurs for $\mathcal{U}=\{U_1, U_2, \cdots, U_{p+1}\}$ with $U_1=U_2=\cdots=U_p = V_g$ and $U_{p+1}$ any subset of $V_g$ with $|U_{p+1}|=q$. Hence $g$ is proper.

\section{Pattern of Zeroes \& Disjoint Unions}
\label{disjoint_appendix}
Let us start this appendix by giving a combinatorical interpretation to the function
\begin{equation}
	\mathcal{S}_{k}(a)=\frac{p(p-1)}{2}k+pq
\end{equation}
Throughout, $a=pk+q$, $p\geq 0$ and $0\leq q<r$. 
\begin{itemize}
	\item[--] Define the $\pi_k(a)$ as the set of partitions of $a$ with largest part at most $k$. The most important example of such partitions is
	\begin{equation}
		D_k(a)=(k^{\times p}, q, 0^{\times (a-p-1)})
	\end{equation}
	Note that $D_k(a)\succeq \lambda$ for all $\lambda\in \pi_k(a)$.
	
	\item[--] Given any partition $\lambda$ of $a$, we can define the quantity $s(\lambda)$ via
	\begin{equation}
		s(\lambda)=\sum_{i=1}^{a}(i-1)\lambda_i=\sum_{i=1}^a (a-P_i(\lambda))
	\end{equation}
	where $P_i(\lambda)$ is the $i$th partial sum. Clearly, if $\lambda\succeq \mu$, then $s(\lambda)\leq s(\mu)$. In particular, since $D_k(a)\succeq \lambda$ for any $\lambda\in \pi_k(a)$, we have $s(D_k(a))\leq s(\lambda)$. 
\end{itemize}
Concretely computing $s(D_k(a))$, we find the alternative interpretation of $\mathcal{S}_k(a)$ since
\begin{equation}
	s(D_k(a))=\mathcal{S}_k(a)
\end{equation}
This combinatorical quantity has the following property:

\begin{prop}
	\label{pp2}
	\begin{equation}
		\mathcal{S}_{k_1+\cdots+k_c}(a_1+\cdots+a_c)\leq
		\sum_{i=1}^c\mathcal{S}_{k_i}(a_i)
	\end{equation}
\end{prop}
\begin{proof}
	Let $\lambda^{(i)}\in \pi_{k_i}(a_i)$ for $1\leq i\leq c$. Define the partition $\lambda=\sum_i\lambda^{(i)}$ in the obvious way via $\lambda_j=\sum_i\lambda^{(i)}_j$. Define $a=a_1+\cdots+a_c$ and $k=k_1+\cdots+k_c$. Now $\lambda\in \pi_k(a)$ and $s(\lambda)=\sum_{i=1}^c s(\lambda^{(i)})$. Since $\sum_i \lambda^{i}\preceq D_{k}(a)$, we have
	\[
	\mathcal{S}_k(a)=s(D_k(a))\leq \sum_{i=1}^c s(\lambda^{(i)})
	\]
	In particular, choosing $\lambda^{(i)}=D_{k_i}(a_i)$  we obtain conclude $\mathcal{S}_{k}(a)\leq \sum_{i=1}^c \mathcal{S}_{k_i}(a_i)$.
\end{proof}
\noindent
Using Prop. \ref{pp2} we can now prove Prop. \ref{disjoint_prop}.
\begin{prp}
	Let $g=g_1\sqcup g_2\sqcup \cdots\sqcup g_c$ with $g_i$ a $(k_i, r)$ connected intensive graph.
	\begin{enumerate}
		\item If for all $i$, the sequence $\gv{\Psi}^{(g_i)}$ is periodic, then $\gv{\Psi}^{(g)}$ is periodic as well.
		\item If $g_i$ is proper for all $i$, then $g$ is also proper.
	\end{enumerate}
\end{prp}
\begin{proof}
	For (1), by definition of pattern of zeros, we have $S_a=\min_{\lambda}P_a^*(\lambda)$, with $\lambda$ running over root-partitions of $\Psi_n^{(g)}$. Using Eq. \eqref{disjoint_root_decomp}, if $\lambda^{(i)}$'s are the root partitions of $\Psi_n^{(g_i)}$, then root-partitions of $\Psi_n^{(g)}$ are $\lambda^{(1)}\cup \lambda^{(2)}\cup\cdots \cup\lambda^{(c)}$. Now if $\Psi_n^{(i)}$ is periodic, then $P_{a_i}^*(\lambda^{(i)})\geq \mathcal{S}_{k_i}(a_i)$. At the same time,
	\[
	P^*_a(\lambda^{(1)}\cup \lambda^{(2)}\cup\cdots \cup\lambda^{(c)})=\sum_i P_{a_i}^*(\lambda^{(i)})
	\]
	for some $a_i$'s so that $a_1+a_2+\cdots+a_c=a$. By Prop. \ref{pp2}, we have (let $k=k_1+\cdots+k_c$)
	\[
	\sum_i P_{a_i}^*(\lambda^{(i)})\geq 
	\sum_i \mathcal{S}_{k_i}(a_i)\geq \mathcal{S}_{k}(a)
	\]
	As a result, reverse partial sum of any root partition of $\Psi_n^{(g)}$ is greater or equal to $\mathcal{S}_k(a)$. Equality is also possible since $\Lambda(n,k,r)$ is a root partition of $\Psi_n^{(g)}$. Therefore, $\Psi_n^{(g)}$ is perodic.

	For (2), the argument is similar. Take any $a$-cut  $C\vdash_a T_n\otimes g$, and let $C^{(i)}$ be the induced cut on $T_n\otimes g_i$. Since $g_i$ is proper, $w(C^{(i)})\geq \mathcal{S}_{k_i}(a_i)$. Therefore, we find that $w(C)\geq \mathcal{S}_k(a)$. To see that the equality is possible, let $a=pk+q$ and define $\mathcal{U}$ as
	\[
	\mathcal{U}=\{
	\underbrace{V_g, V_g, \cdots, V_g}_{\times p}, U\}
	\]
	with $V_g$ the entire node set of $g$, and $U$ any subset of $V_g$ with $|U|=q$. The associated cut $C(\mathcal{U})$ is then of weight $\mathcal{S}_k(a)$.
	
\end{proof}

\section{Proof of Proposition \ref{sep_prop}}
\label{last_appendix}
Throughout, let $g=\mathfrak{g}(\beta)$ and $k$ be the number of nodes in $g$. The graph $\mathfrak{g}(\xi)=\mathfrak{g}(\beta^c)=g^c$ will have $kc$ nodes. We use the notation $G=T_n\otimes g$ and $G^c=(T_c\otimes g)^c=T_n\otimes g^c$. By a cut $C$, we will always mean a $kc+1$ cut of $G^c$. Since there is much opportunity for confusion, we will always remain faithful to these conventions even if it costs us brevity.

\begin{proof}[Proof for $\beta=(l, m^{\times(k-1)})$]
	We start with part (4) since this is essentially a corollary of $(k,l,m)$ saturated graphs being proper. Let $C$ be a cut of $G^c$. Define $C^{(1)}, C^{(2)},\cdots, C^{(c)}$ as the induced cut on each component. Since $g=\mathfrak{g}(\beta)$ is $(k,l,m)$ saturated, it is proper. Therefore, if for some $i$ we have $|C^{(i)}|>k+1$, then $w(C)>r$. Moreover, if there exists $i\neq j$ with $|C^{(i)}|=|C^{(j)}|=k+1$, we again have $w(C)>r$.  So the only possibility is, up to permutation of components, $|C^{(1)}|=k+1$ and $|C^{(i)}|=k$ for $i=2, \cdots, c$. Using the proof for properness of saturated graphs, we have $w(C^{(1)})=r$ if and only if $C^{(1)}$ is standard in $G$. Moreover, $w(C^{(i\neq 1)})=0$ only when $C^{(i)}$ is a maximum independent set of $G$. Hence, if $w(C)=r$, then $C$ is a standard cut of $G^c$.
\end{proof}

To prove the other cases, we need more terminology. Let $H_a$ be the $a$th horizontal set of $G^c$. Define $h_a=|C\cap H_a|$, i.e. the number of nodes $C$ has in the $a$th horizontal set. Define $f_s$ as the frequency of $s$ in the sequence $h=(h_0, h_1, h_2, \cdots, h_{kc-1})$. Let $K$ be the largest number appearing in $h$. We call the sequence $\varphi(C)=(f_2, f_3, f_4, \cdots, f_K)$ the \emph{horizontal clique profile (HCP)} of $C$. Note that we are not listing $f_1$ in the HCP. The \emph{clique weight} of $C$ is defined as
\begin{equation}
	\widehat{w}(C):=\xi_0\sum_{s=2}^{K}\frac{s(s-1)}{2}f_s 
\end{equation}
The clique-weight is a lower bound for $w(C)$, i.e., $w(C)\geq \widehat{w}(C)$. Note that for a standard cut $C$, we have $\varphi(C)=(1)$. We will build on the terminology as we go along. However, we have enough to prove the statement for $\beta=(2m)$. 

\begin{proof}[Proof for $\beta=(2m)$]
	Let $\xi=\beta^k=(2m,0^{\times(k-1)})$. Consider a $(k+1)$-cut $C\vdash G$. All arrows are horizontal arrows in this example and $w(C)=\widehat{w}(C)$. Suppose $\varphi(C)=(f_2, \cdots, f_K)$, then
	\[
	w(C)=\widehat{w}(C)=2m\sum_{s=2}^{K}
	\frac{s(s-1)}{2}
	f_s
	\]
	Clearly the only possibility for $w(C)=2m$ is $\varphi(C)=(1)$. For this special case, $\varphi(C)=(1)$ is clearly the same as $C$ being standard.
\end{proof}

It remains to prove the statement for $\beta=(1,0^{\times k-2}, 1)$ and $\beta=(1, 0^{\times 2K-3}, 1,1)$. 
It would be useful to classify non-standard cuts by their clique weight. We have three different possibilities:
\begin{enumerate}
	\item A fat cut: This means $\widehat{w}(C)>r$.
	\item An ambiguous cut: A non-standard cut with clique weight equal to $r$.
	\item A delicate cut: These are non-standard cuts with $\widehat{w}(C)<r$.
\end{enumerate}
We call any cut with $w(C)>r$ an irrelevant cut, as they have no relevance to separability. Fat cuts are all irrelevant. 

Next, we discuss relevant ambiguous cuts $C$. In other words,  $w(C)=\widehat{w}(C)=r$. If this is the case, all the arrows in $C$ are horizontal. Let us describe the restrictions $G^c|_C$.
Each connected component of $G^c|_C$ that is not an isolated node is a so-called ``$(s,l)$-clique'' ($s>1$): An $(s,1)$-clique is a graph $(V,\mu_1)$ with $V=\{1,2,\cdots, s\}$ and
\[
\mu_1(i,j)+\mu_1(j,i)=1, \qquad \text{for all }i<j
\]
An $(s,l)$-clique is graph $(V,\mu)$, with $V$ the same as before, and $\mu = l\mu_1$ for some $\mu_1$ characterizing a $(s,1)$-clique. For each $2\leq s\leq K$, the graph $G|_C$ has $f_s$ components that are $(s,\beta_0)$-cliques. Say $H$ is one of those components, i.e. an $(s,\beta_0)$-clique. Then $\mathrm{SGP}(H)=0$ if $\beta_0=$odd. This in turn results in $\mathrm{SGP}(G^c|_C)=0$. In this case, we say the cut has an odd clique and is thus vanishing. Now, we have $\beta_0=1$ for both of our remaining examples. Therefore, the ambiguous cuts are irrelevant or vanishing in both cases.

It remains to investigate the delicate cuts. These are all of the possibilities:
\begin{enumerate}
	\item $\beta=(1,0^{\times k-2}, 1)$. The only HCP possibility for $\widehat{w}(C)<2$ is $\varphi(C)=(1)$. 
	\item $\beta=(1,0^{\times 2K-3}, 1,1)$. The only HCP possibilities for $\widehat{w}(C)< r=3$ are $\varphi(C)=(1), (2)$.
\end{enumerate}
We want to show that any cut with $\varphi(C)=1$ is standard, irrelevant, or vanishing. We will prove this in a more general setting shortly. Moreover, for the latter case, we need to show that any cut with $\varphi(C)=(2)$ is either irrelevant or vanishing. Let us begin by proving this last statement.
\begin{lem}
	Let $\beta=(1,0^{\times 2K-3}, 1,1)$, $g=\mathfrak{g}(\beta)$ and $G=T_n\otimes g$. Suppose $C$ is a $2Kc+1$-cut of $G^c$ with HCP $\varphi(C)=(2)$. Then $C$ is either vanishing or irrelevant
\end{lem}
\begin{proof}
	Given a graph $\Gamma$, let $\overline{\Gamma}$ be the \emph{underlying undirected graph}. An undirected (simple) $n$-path is the following graph
	\[
	P_n = \underbrace{\bullet -\bullet - \bullet\cdots \bullet-\bullet}_{\text{with }n \text{ edges}}
	\]
	$P_0$ is understood as an isolated node.
	
	Now suppose $\varphi(C)=(2)$. If $w(C)=\widehat{w}(C)=2$, then $G^c|_C$ contains an odd clique and is thus vanishing (that said, no such cut actrually exists). If $w(C)=r=3$, then we have only two possibilities for $\overline{G}^c|_C$: Either
	$
	\overline{G}^c|_C\simeq P_3\sqcup P_0^{\sqcup (2Kc-3)}, $
	or $
	\overline{G}^c|_C\simeq P_2\sqcup P_1\sqcup P_0^{\sqcup (2Kc-4)}$. In the latter case, $C$ is a vanishing cut since $G|_C$ contains an odd clique. In the former case, the problem essentially reduces to $c=1$.
	
	Without loss of generality, we may take the four nodes of $C\vdash_{2K+1} T_n\otimes g$ as
	\begin{itemize}
		\itemsep -2pt
		\item $A=I_1\otimes 0$ and $B=I_2\otimes 0$, with $I_1<I_2$.
		\item $C=J_1\otimes b$ and $D=J_2\otimes b$, with $J_1<J_2$ and $0\neq b\in \mathbb{Z}_k$ to be determined shortly.
		
		\item $2K-3$ isolated nodes $X_0, \cdots, X_{2K-4}$.
	\end{itemize}
	We would like $\overline{G}|_{\{A,B,C,D\}}\simeq P_3$. We already have two arrows $A\to B$ and $C\to D$. After some labor, we are reduced to one of 4 possibilities:
	\begin{enumerate}
		\item $I_1\leq J_1<I_2\leq J_2$ and $b$ is either $1$ or $2$.
		\item $J_1\leq I_1<J_2\leq I_2$ and $b$ is either $2K-1$ or $2K-2$.	
	\end{enumerate}
	All four cases lead to $w(C)>3$ and are irrelevant. The approach is always the same, so we will only discuss one of four cases; namely $I_1\leq J_1<I_2\leq J_2$ and $b=1$ (so concretely, $C=J_1\otimes 1$ and $D=J_2\otimes 1$).

	Suppose
	\[
	X_i = N_i\otimes b_i
	\]
	As $\varphi(C)=(2)$, if $i\neq j$ then $b_i\neq b_j$. Without loss of generality, we may take $b_i=i+2$ for $0\leq i\leq m-3$ and $b_i=i+3$ for $i>m-3$, for some $m\geq 2$. To avoid an arrow of type $I\otimes b\to I'\otimes (b-1)$ (which requires $I<I'$), we need to demand
	\[
	J_2\leq N_0\leq N_1\leq\cdots\leq N_{m-3}
	\]
	and
	\[
	N_{m-2}\leq N_{m-1}\leq \cdots \leq N_{2K-4}\leq I_1
	\]
	But then we have an arrow $N_{m-2}\otimes (m+1)\to N_{m-3}\otimes (m-1)$ as $\beta_{-2}=1$ (when $m=3$ we have an arrow $N_0\otimes 3\to J_2\otimes 1$). So we cannot avoid $w(C)>3$ no matter what we do.  
\end{proof}
To finish the discussion, we fully classify the cuts with HCP $\varphi(C)=(1)$ in the most general case.

\subsection*{Cuts with Horizontal Clique Profile $\varphi(C)=(1)$}
Like before, let $\beta$ be a basic vector, $\xi=\beta^c$, $g=\mathfrak{g}(\beta)$, and $G=T_n\otimes g$. Suppose $C$ is a $kc+1$-cut of $G^c$ with $\varphi(C)=(1)$. Without loss of generality, we may take $C$ to be
\[
C = \{J_0\otimes 0, J'_0\otimes 0\}\cup \bigcup_{b=1}^{kc-1}\{J_b\otimes b\}
\]
Here,  $J_0<J'_0$, but there is no assumptions on $J_1, \cdots, J_{kc-1}$.

\begin{lem}
	Let $s\in \{1, \cdots, kc-1\}$ be such $\xi_s\neq 0$. Define $m_s=kc/\mathrm{gcd}(kc, s)$. Then $C$ contains at least one of the following arrows with multiplicity $\xi_s$
	\[
	J_{(a-1)s}\otimes (a-1)s\to J_s\otimes as \quad (1\leq a<m_s), \qquad J_{(m_s-1)s}\otimes (m_s-1)s\to J'_0\otimes 0
	\]
	As a result, $w(C)\geq \xi_0+\xi_1+\cdots + \xi_{kc-1}=r$.  The equality $w(C)=r$ happens only if, for each $s\neq 0$ with $\xi_s\neq 0$, there is only one arrow (of multiplicity $\xi_s$) of the following forms: $J_b\otimes b\to J_{b+s}\otimes (b+s)$ (summations in $\mathbb{Z}_{kc}$) or $J_{(m_s-1)s}\otimes (m_s-1)s\to J'_0\otimes 0$.
\end{lem}
\begin{proof}
	Let us fix  $s\neq 0$ with $\xi_s\neq 0$. For any $1\leq a<m_s$, if $J_{(a-1)s}<J_{as}$, then $J_{(a-1)s}\otimes (a-1)s\to J_s\otimes as$ is an arrow in $C$. So suppose $J_{(m_s-1)s}\leq \cdots\leq J_{2s}\leq J_s\leq J_0<J'_0$. But in this case, $J_{(m_s-1)s}<J'_0$ and $J_{(m_s-1)s}\otimes (m_s-1)s\to J'_0\otimes 0$ is an arrow of $C$. The rest is straightforward.
\end{proof}
We are now almost ready to classify all cuts $C$ with $\varphi(C)=(1)$. In the following proposition $H_{k,c}(l, m)$ is a graph with $kc+1$ nodes, $k>2$ and $c\geq 1$. Labeling the nodes by $0, 1,2,\cdots, kc$, the multiplicity is given by $\mu(0,1)=l$, $\mu(2,3)=m$ and all $\mu(x,y)=0$ if $(x,y)\neq (0,1), (2,3)$. 

\begin{prop}
	Let $\beta=(\beta_0, \beta_1, \cdots, \beta_{k-1})$ be a basic vector and $\xi=\beta^c$ for some $c\geq 1$. Take a $(kc+1)$-cut $C$ of $G=T_n\otimes g^c$ with $\varphi(C)=(1)$. Then $C$ is one of the following:
	\begin{enumerate}
		\item Irrelevant.
		\item Standard.
		\item $k>2$, $\beta=(\beta_0, 0^{\times(k-2)}, \beta_{-1})$, and $G|_C$ is isomorphic to $H_{k,c}(\beta_0, \beta_{-1})$. Therefore, when $\beta_0$ or $\beta_{-1}$ is odd, the cut $C$ is vanishing.
	\end{enumerate}
\end{prop}

\begin{proof}
	Let us write $G^c=G_1\sqcup \cdots \sqcup G_c$, with all $G_i$ isomorphic to $G$. Let $C^{(i)}$ be the cut induced by $C$ on $G_i$. Without loss of generality, suppose $C^{(1)}$ is a $k+1$-cut of $G_1$ with HCP $(1)$, and $C^{(i)}$ ($i>1$) are $k$-cuts of $G_i$ with HCP $(0)$. We have
	\[
	w(C)\geq w(C^{(1)})
	\]
	with equality happening if each $C^{(i)}$ ($i>1$) is a maximum independent set. Therefore, it is enough to show the proposition for $c=1$ (i.e., for the connected case).
	
	So suppose $G=T_n\otimes \mathfrak{g}(\beta)$, with $\beta$ a basic vector, and $C$ a $k+1$-cut of $G$ with $\varphi(C)=(1)$. Define the following sets:
	\begin{align*}
		&C_L = \{J_b\otimes b\mid J_b<J'_0\}, \qquad &B_L&=\{b\in \mathbb{Z}_k\mid J_b\otimes b\in C_L\}\\
		&C_R = \{J_b\otimes b\mid J_b\geq J'_0\}\cup\{J'_0\otimes 0\}, \qquad &B_R&=\{b\in \mathbb{Z}_k\mid J_b\otimes b\in C_R\}\cup\{0\}
	\end{align*}
	Note that since $\beta$ is basic, we have $\beta_{-1}\neq 0$. Using the lemma, if we can find two distinct $b,b'$ in $B_L$ such that $b-1, b'-1\in B_R$, then $w(C)\geq r+\beta_{-1}>r$. Assuming $C$ is relevant, we need to avoid this situation. Let us model this. Suppose we must label each $b\in \{1, \cdots, k-1\}$ by a letter $W_b\in \{L,R\}$. Consider the sequence $LW_{k-1}\cdots W_2W_1R$. We are looking for those sequences that contain the phrase $LR$ at most once. The only possibilities are
	\[
	\underbrace{LL\cdots L}_{\times k-m}\underbrace{RR\cdots R}_{\times (m+1)}, \qquad 0\leq m <k
	\]
	This corresponds to the following
	\begin{align*}
		&C_L = \{J_{0}\otimes 0, J_{m+1}\otimes (m+1), \cdots, J_{k-1}\otimes (k-1)\mid J_i<J'_0\}, \qquad &B_L&=\{0,m+1, \cdots, k-1\}\\
		&C_R = \{J_1\otimes 1, \cdots, J_m\otimes m \mid J_i\geq J'_0\}\cup\{J'_0\otimes 0\}, \qquad &B_R&=\{0, 1,2, \cdots, m\}
	\end{align*}
	No matter how we move $J_i$ around (while respecting the constraints), the arrow $J_{m+1}\otimes (m+1)\to J_m\otimes m$ (or $J_1\otimes 1\to J'\otimes 0$ if $m=0$) is unavoidable. Thus, using the lemma again, if the cut is relevant, we must have
	\[
	J_{m+1}\leq \cdots\leq J_{k-2}\leq J_{k-1}\leq J_0<J'_0\leq J_1\leq J_2\leq \cdots \leq J_m
	\]
	Let us denote the number of arrows of the form $C_L\to C_R$ by $w_{LR}(C)$. To count it, let $0\leq a\leq k-m-1$, and note that we have
	$\beta_{a}+\beta_{a+1}+\cdots+\beta_{a+m}$ arrows of the form $J_{k-a}\otimes (k-a)\to C_R$ (we understand $J_k$ as $J_0$). Let $L_{k,m}=\min(m,k-1-m)$ and define the function $f_{k,m}$ as
	\[
	f_{k,m}(x) = \begin{cases}
		x & 0\leq x\leq L_{k,m}\\
		L_{k,m} & L_{k,m}\leq x\leq k-1-L_{k,m}\\
		k-1-x & k-1-L_{k,m}\leq x\leq k-1
	\end{cases}
	\]
	Then $w_{LR}(C)=r+\sum_{a=0}^{k-1}f_{k,m}(a)\beta_a\leq w(C)$.

	We start with the special basic vector $\beta= (\beta_0, 0^{\times k-2}, \beta_{-1})$. If $k=2$, then both of the cuts described above ($m=,0,1$) are standard. If $k>2$, then no further analysis is required. Any of the cuts above leads to $G|_C\simeq H_{k,1}(\beta_0,\beta_{-1})$ and have weight $r=l+m$. This is the third possibility in the proposition. So suppose there exists $0<s<k-1$ such that $\beta_s\neq 0$. Note that unless $L_{k,m}=0$, we have $f_{k,m}(s)>0$ and $C$ is irrelevant. So let us assume $L_{k,m}=0$ as well. In other words, $m=0, k-1$.

	We consider $m=0$ first. In this case, $J_{1}\leq J_2\leq  \cdots\leq J_{k-1}\leq J_0<J'_0$ and $w_{LR}(C)=r$. Therefore, any arrow of the form $C_L\to C_L$ is forbidden if $C$ is relevant. If $\xi_s\neq 0$ for any $0<s<k-1$, the only possibility is $J_0=J_1=\cdots=J_{k-1}$. Thus, the cut $C$ is standard and a sink neighborhood. Similarly, if $m=k-1$, we have $J_0<J'_0\leq J_1\leq \cdots \leq J_{k-1}$. Any arrow of the form $C_R\to C_R$ is forbidden and we are forced to have $J'_0=J_1=J_2=\cdots= J_{k-1}$. Thus, $C$ is standard and a source neighborhood.
\end{proof}

\newpage

\bibliographystyle{elsarticle-harv}
\bibliography{foo} 

\end{document}